\documentclass[preprint]{aastex}

\shorttitle{Observations of Very Metal-Poor Stars with Subaru/HDS}
\shortauthors{Honda et al.}

\begin{document}

\title{Spectroscopic Studies of Extremly Metal-Poor Stars with
Subaru/HDS: II. The r-process Elements, Including Thorium.\altaffilmark{1}}

\author{Satoshi Honda\altaffilmark{2,3}, Wako Aoki\altaffilmark{2,3},
Toshitaka Kajino\altaffilmark{2,3}, Hiroyasu Ando\altaffilmark{2}, Timothy
C. Beers\altaffilmark{4}, Hideyuki Izumiura\altaffilmark{5}, Kozo
Sadakane\altaffilmark{6}, Masahide Takada-Hidai\altaffilmark{7}}

\altaffiltext{1}{Based on data collected at the Subaru Telescope,
which is operated by the National Astronomical Observatory of Japan.}
\altaffiltext{2}{National Astronomical Observatory, Mitaka, Tokyo,
181-8588, Japan; e-mail: honda@optik.mtk.nao.ac.jp, aoki.wako@nao.ac.jp,
kajino@nao.ac.jp, ando@optik.mtk.nao.ac.jp}
\altaffiltext{3}{Department of Astronomy, Graduate University for
Advanced Studies, Mitaka, Tokyo, 181-8588, Japan}
\altaffiltext{4}{Department of Physics and Astronomy, Michigan State
University, East Lansing, MI 48824--1116; beers@pa.msu.edu}
\altaffiltext{5}{Okayama Astrophysical Observatory, National
Astronomical Observatory of Japan, Kamogata-cho, Okayama, 719-0232,
Japan; izumiura@oao.nao.ac.jp}
\altaffiltext{6}{Astronomical Institute, Osaka Kyoiku University,
Kashiwara-shi, Osaka, 582-8582, Japan; sadakane@cc.osaka-kyoiku.ac.jp}
\altaffiltext{7}{Liberal Arts Education Center, Tokai University, 1117
Kitakaname, Hiratsuka-shi, Kanagawa, 259-1292, Japan;
hidai@apus.rh.u-tokai.ac.jp}

\begin{abstract}

We have obtained high-resolution, high S/N near-UV-blue spectra of 22
 very metal-poor stars ([Fe/H] $<-2.5$) with Subaru/HDS, and measured
 the abundances of elements from C to Th. 
The metallicity range of the observed stars is $-3.2
<$ [Fe/H] $< -2.4$. As found by previous studies, the star-to-star scatter in
the measured abundances of neutron-capture elements in these stars is very
large, much greater than could be assigned to observational errors, and in
comparison with the relatively small scatter in the $\alpha$- and iron-peak
elements. In spite of the large scatter in the ratios of the neutron-capture
elements relative to iron, the abundance {\it patterns} of heavy neutron-capture
elements ($56\leq Z\lesssim 72$) are quite similar within our sample stars. The
Ba/Eu ratios in the 11 very metal-poor stars in our sample in which both
elements have been detected are nearly equal to that of the solar system
r-process component. Moreover, the abundance patterns of the heavy
neutron-capture elements (56 $\leq Z \leq$ 70) in seven objects with clear
enhancements of the neutron-capture elements are similar to that of the solar
system r-process component. These results prove that heavy neutron-capture
elements in these objects are primarily synthesized by the r-process.

In contrast, the abundance ratios of the light neutron-capture elements (38
$\leq$ {\it Z} $\leq$ 46) relative to the heavier ones (56 $\leq$ {\it Z} $\leq$
70) exhibit a large dispersion. Our inspection of the correlation between Sr and
Ba abundances in very metal-poor stars reveals that the dispersion of the Sr
abundances clearly decreases with increasing Ba abundance. This trend is
naturally explained by hypothesizing the existence of two processes, one that
produces Sr without Ba, and another that produces Sr and Ba in similar
proportions. This result should provide a strong constraint on the origin of the
light neutron-capture elements at low metallicity.

We have identified a new highly r-process element-enhanced, metal-poor star,
CS~22183--031, a giant with [Fe/H] = $-2.93$ and [Eu/Fe] = +1.2. We also
identified a new, moderately r-process-enhanced, metal-poor star, CS~30306--132,
a giant with [Fe/H] = $-2.42$ and [Eu/Fe] = +0.85.

The abundance ratio of the radioactive element Th ({\it Z} = 90) relative to the
stable rare-earth elements (e.g., Eu) in very metal-poor stars has been used as
a cosmochronometer by a number of previous authors. Thorium is detected in
seven stars in our sample, including four objects for which the detection of Th
has already been reported. New detections of thorium have been made for the
stars HD~6268, HD~110184, and CS~30306--132. The Th/Eu abundance ratios
(log(Th/Eu)), are distributed over the range $-0.10$ to $-0.59$, with typical
errors of 0.10 to 0.15 dex. In particular, the ratios in two stars,
CS~31082--001 and CS~30306--132, are significantly higher than the ratio
 in the well-studied object CS~22892--052 and those of other moderately
r-process-enhanced metal-poor stars previously reported.
 Since these very metal-poor stars are
believed to be formed in the early Galaxy, this result suggests that the
abundance ratios between Th and stable rare-earth elements such as Eu, both of
which are presumably produced by r-process nucleosynthesis, may exhibit real
star-to-star scatter, with implications for (a) the astrophysical sites of the
r-process, and (b) the use of Th/Eu as a cosmochronometer.

\end{abstract}

\keywords{nuclear reactions, nucleosynthesis, abundances -- stars:
abundances --- stars: population II}

\section{Introduction}

The very metal-poor stars, presently found in the halo of the Galaxy, are
believed to have formed at the earliest times, shortly after it became possible
for the Universe to make stars with sufficiently long main-sequence lifetimes
(i.e., masses $\le 0.8 M_\odot$) to survive for $\sim 14$ Gyr. The chemical
compositions of these stars are thus expected to reflect a quite small number of
nucleosynthesis processes, possibly as small as one, while the compositions of
more metal-rich stars like the Sun reflect the cumulative (hence quite complex)
results of the various processes that have been in operation during the entire
history of Galactic chemical evolution. Recent abundance analyses for extremely
metal-poor stars have provided quite valuable information on the
individual nucleosynthesis processes involved
\citep[e.g.,][]{mcwilliam95b,ryan96, burris00, cayrel03}. 
In particular, detailed abundance studies of heavy metals in such stars
have led to important progress in the understanding of the origin of
neutron-capture elements in the early Galaxy.

Sneden et al. (1996, 2000, 2003) have studied the abundances of the extremely
metal-poor star CS~22892--052, the first example of a growing class of
metal-poor stars that exhibit very large excesses of r-process elements relative
to iron ([r-process/Fe] $> +1.0$). Even from the first analysis, it was apparent
that the relative abundance pattern of the heavy neutron-capture elements (56
$\leq$ {\it Z} $\leq$ 76) in this star was identical (within observational
errors) to that of the (inferred) solar system r-process component, which has been strengthened as more and better data have been acquired. This
striking similarity may be surprising, considering that the abundance pattern of
the Solar System has certainly been influenced by the integrated yields from a
variety of nucleosynthesis sites. It should be pointed out that, in contrast,
the abundance pattern of the lighter neutron-capture elements ($38 \leq Z \leq
48$) in CS~22892--052 exhibit clear deviations from that of the solar-system
r-process component \citep{sneden00,sneden03}.

Recent analyses of a small number of additional very metal-poor stars with
moderate excesses of r-process elements ($0.5 \le $ [r-process/Fe] $ < +1.0$)
\citep[e.g., ][]{westin00,johnson01,cowan02} have obtained similar results, that
is, the abundance pattern of heavy neutron-capture elements ($56\leq Z\lesssim
72$) for each star is quite similar to that of the r-process component in
solar-system material. These results imply that the neutron-capture elements in
these very metal-poor stars were produced by the r-process, which produces quite
similar abundance patterns at least for the range of $Z=56 \sim 72$ (Schatz et
al. 2002; Wanajo et al. 2002; Otsuki, Mathews, \& Kajino 2003). By way of
contrast, the abundance patterns of the lighter neutron-capture elements ($Z <
56$) are significantly different from that of the r-process component in the
Solar System. This observational fact is interpreted as a result of the
existence of (at least) two distinct classes of r-process events \citep[e.g., ][
and references therein]{truran02}.

The search for, and subsequent detailed abundance studies of,
r-process-enhanced, very metal-poor stars enable one to make estimates of the
ages of these objects\footnote{More appropriately, the time interval that has
passed since the production of the r-process elements by the progenitor
object(s) of these stars. For convenience, we will use the term ``age'' to refer
to this time interval.} by use of the radioactive species that can be identified
in them. The long-lived radioactive r-process elements, in particular Th and U,
whose half lives (14 Gyr for $^{232}$Th; 4.5 Gyr for $^{238}$U) are comparable
to or shorter than the cosmic age ($\sim$ 15 Gyr), provide a powerful
tool for determination of a lower limit on the age of the Galaxy, and
hence of the universe. 
If Th and/or U is detected in a very metal-poor star, we can estimate
the lapse of time from the era of the nucleosynthesis process that created these
elements to the present. This is accomplished by measuring the present abundance
ratios, either of U/Th, when they are both detected (which only applies to two
stars thus far), or by measurement of the U or Th abundance ratio as compared to
stable r-process elements, such as Eu, and comparing them to predictions of the
initial production ratios from models (both site-independent, e.g., Schatz et
al. 2002, and site-dependent, e.g., Wanajo et al. 2002, 2003; Otsuki,
Mathews, \& Kajino 2003).

The cosmochronometry technique (in stars) was pioneered by Butcher (1987), who
unfortunately only had roughly solar-abundance stars to work with, hence he had
to contend with the difficulties arising from complex continua and line-blending
problems, which are not so severe for very metal-poor stars. One clear advantage
of the cosmochronometric method is that the resulting age estimate is free from
a host of uncertainties encountered by other estimates of the cosmic age, such
as calibration of distance scales, detailed understanding of stellar evolution,
etc. Furthermore, since it is thought that the large overabundances of r-process
elements are likely to have been associated with a single production event,
quite likely a Type II supernova explosion in the early Galaxy, one is not
forced to model the entire complex history of Galactic chemical evolution in
order to estimate the age. The difficulties in this method, aside from the
rarity of the very metal-poor stars with detectable Th and U (presently
estimated to be no more than $\sim 3$\% of giants with [Fe/H] $<
-2.5$\footnote{We use the usual notation [A/B]$\equiv \rm{log}_{10} (N_A/N_B)
_*-\rm{log}_{10}(N_A/N_B)_\odot$ and log$\epsilon(\rm A)\equiv
\rm{log}_{10}(N_A/N_H)+12.0$, for elements A and B. Also, the term
``metallicity'' will be assumed here to be equivalent to the stellar [Fe/H]
value.}; Beers, private communication), arise from the need to accurately
predict the initial production ratios, which in turn depend on having a detailed
understanding of the nucleosynthesis pathways, nuclear mass models, and
cross-sections for species that have not been measured adequately at present
(see Schatz et al. 2002 for details). 

In the seminal study by Fran\c{c}ois, Spite, \& Spite (1993), Th abundances were
reported for the first time for stars with metallicities as low as [Fe/H]
$\simeq$ --2.6. Sneden et al. (1996) first succeeded in obtaining a clear
detection of Th, along with other r-process elements, in CS~22892--052 (see
above), a relatively bright, extremely metal-poor giant ([Fe/H] = --3.1)
discovered during the HK survey of Beers and collaborators (see Beers et al.
1992). From these authors' measurement of the Th/Eu ratio, the age of this star
was estimated to be 15.2 $\pm$ 3.7 Gyr. In their analysis, they assumed the
initial abundance ratio between Th and stable elements to be the same as that of
the calculated initial solar-system abundance ratio. Subsequently, Westin et al.
(2000) estimated the age of another r-process-enhanced metal-poor star,
HD~115444, to be 14.2 $\pm$ 4 Gyr from the Th/Eu ratio. Johnson \& Bolte (2001)
also measured Th abundances for five very metal-poor stars. Most recently, Cowan
et al. (2002) studied the moderately r-process-enhanced, very metal-poor star,
BD+17$\degr$3248. This star, with [Fe/H] = --2.1, and [Eu/Fe] = +0.9, has the
distinction of being among the most metal-rich stars in which the
r-process-enhancement phenomenon has been observed. These authors, based on an
average of a number of chronometer pairs involving U and Th, obtained an age
estimate of $13.8 \pm 4$ Gyr. Sneden et al. (2003) have assembled a definitive
high-S/N set of spectra for CS~22892--052, drawing on space-based and
ground-based observations. Their analysis has led to a revision of the Th/Eu age
estimate for this star, to $12.8 \pm 3$ Gyr. These results are consistent with
the very recent estimate of the age of the Universe by WMAP (Bennett et al.
2003) to within the reported errors.

Recently, Cayrel et al. (2001) and Hill et al. (2002) reported the detection of
U, as well as Th, in a high-quality VLT/UVES spectrum of the extremely
metal-poor star CS~31082--001, a bright [Fe/H] = --2.9 giant discovered in the
HK survey. They also found that the abundance pattern of neutron-capture
elements with $56 \leq Z \leq 72$ closely mimics that of the r-process component
in the Solar System. However, it is of interest that the situation seems to be
different for nuclei near the heaviest elements. Th ($ Z = 90$) and U ($Z = 92$)
show larger deviations from their expected levels, if we adopt $\sim 14$ Gyr as
the age of this object, as was estimated from the U/Th ratio by \citet{hill02}.
This result suggests that either CS~31082--001 is peculiar in some way that has
affected the surface abundances of the actinides, or that the production ratios
of Th and U, as compared to the stable elements (e.g., Eu) exhibit some real
dispersion amongst very metal-poor stars. Further observational work is clearly
required to answer this important question.

The primary purpose of this study is to obtain measurements of the abundance
patterns of neutron-capture elements for very metal-poor stars, and to examine
the age estimation of stars using the Th/Eu chronometer. In Honda et al. (2003;
hereafter Paper I), we reported equivalent width measurements of absorption
lines in high-quality spectra of 22 very metal-poor stars obtained with the
Subaru Telescope High Dispersion Spectrograph \citep[HDS, ][]{nogu02}. This
sample is as large as those in previous studies by McWilliam et al. (1995),
Ryan, Norris, \& Beers (1996), and Johnson \& Bolte (2001), and the quality of
the spectra is quite high ($S/N > 100$ per resolution element) in the blue
region. In particular, the sample was selected to include as many objects with
excesses of neutron-capture elements as possible, excluding the objects affected
by the s-process, to investigate the nature of r-process nucleosynthesis
in the early Galaxy. 
In this paper, we present the abundance analyses for the neutron-capture
elements, and discuss the observed abundance distributions in very metal-poor
stars with excesses of r-process elements. In \S 2 we describe the
determination of atmospheric parameters for our program stars. The abundance
analysis is described in \S 3, where error estimates in our derived abundances,
and comparisons with previous studies, are also discussed. In \S 4 we discuss
the patterns in the abundance ratios of neutron-capture elements produced by
r-process nucleosynthesis, based on the results of our analysis. In this same
section, the abundance patterns of other elements, including the actinide Th, are
considered, as is the suitability of Th-based chronometers for age estimates.

\section{Atmospheric Parameters}

In order to perform an abundance analysis using model stellar atmospheres, we
first need to determine the atmospheric parameters, i.e., the effective
temperatures, $T_{\rm eff}$, the surface gravities, $\log g$, the microturbulent
velocities, $\xi$, and the metallicities, [Fe/H], for each program star,
based on the available photometric data and our high-resolution spectra.  
These are considered in turn in the following subsections.

\subsection{Effective Temperatures}

We estimate $T_{\rm eff}$ from the available photometric data, employing the
empirical $T_{\rm eff}$ scale obtained by \citet{alon99,alon96} for the $V-K$
color indices, and that by \citet{mcwilliam95b} for the $B-V$ index.  

The broadband $B-V$ color has been frequently used for the determination of
$T_{\rm eff}$, because this color has been measured for the majority of the
metal-poor stars discovered to date. However, the $B-V$ color of a giant star
depends not only on $T_{\rm eff}$, but also on metallicity and the presence, or
not, of molecular absorption features, in particular those arising from carbon.
Hence, we give preference to the use of $V-K$. The photometry data used for the
$T_{\rm eff}$ determination are provided in Table \ref{tab:photo}. The $JHK$
data are taken from the Two Micron All Sky Survey Point Source Catalog (2MASS,
Skrutskie et al. 1997). The $B-V$ data are taken from the SIMBAD database for
bright stars, and from the list of Beers et al. (2003, in preparation) for the
fainter stars. Since \citet{alon99} provides a $T_{\rm eff}$ scale for
photometric data measured on the TCS system \citep{alon98}, the 2MASS infrared
photometry data are first transformed to those of the TCS system \citep{alon94}.
We found that the difference between the two photometric systems is negligible
for our purposes.

Before applying the photometric data to the $T_{\rm eff}$ scale, we first need
to correct for the effects of reddening on the measured colors. Schlegel,
Finkbeiner, \& Davis (1998) constructed a full-sky map of the distribution of
Galactic dust based on far-infrared data observed with the Infrared Astronomical
Satellite (IRAS) and the Diffuse Infrared Background Experiment (DIRBE)
instrument onboard the Cosmic Background Explorer (COBE) satellite. We make use
of the reddening estimates they obtained from this map. An exception is the star
HD~140283, for which a rather high value of $E(B-V)= 0.16$ was derived from the
Schlegel, Finkbeiner, \& Davis (1998) map. It should be kept in mind, however,
that this applies to the entire line of sight, which may not be appropriate for
bright, nearby stars. For this star, which is only some 50 pc away, we instead
adopted a reddening of $E(B-V)= 0.01$, as estimated by Ryan et al. (1996). We
note that Arce and Goodman (1999) cautioned that the map of Schlegel,
Finkbeiner, \& Davis (1998) may overestimate the reddening values when the color
excess $E(B-V)$ exceeds about 0.15 mag, but the color excesses of our stars are
lower than this threshold. The extinction at $V$ is evaluated based on the
relation $A_{\rm v} = 3.1E(B-V)$. We assume that the extinction at $K$ is
negligible.  

The $T_{\rm eff}$ estimated from $V-K$ and $B-V$ are listed in Table 2. These
were derived based on the latest photometric data we could obtain, including the
recent 2MASS release. Our abundance analyses were completed before some of these
photometric data became available, and we have not made a re-analysis if the
difference of the $T_{\rm eff}$ estimated from the updated photometry and
that adopted by the analysis is less than 100~K. As a result, the $T_{\rm eff}$
adopted in the present analysis, also given in Table 3, is sometimes slightly
different (by 80~K maximum) from that derived based on $V-K$.

The $T_{\rm eff}$ estimated from $B-V$ for CS~30306--132 is different by more
than 400~K from that obtained from $V-K$. This object turned out to be moderately
carbon-rich in our analysis, and this discrepancy should be, at least partially,
due to the effect of CH molecular absorption in the B band. Except for this
object, the agreement between the two estimates of $T_{\rm eff}$ is fairly good:
the average of the difference ($T_{\rm eff}(V-K) - T_{\rm eff}(B-V)$) for the 21
objects other than CS~30306--132 is 90~K, with a standard deviation of 64~K.

\subsection{Metallicities, Microturbulent Velocities, and Surface Gravities}

Other parameters ($\log g$, $\xi$, [Fe/H]) are determined by the analysis of Fe
lines. Only lines with equivalent widths less than 100 m{\AA} are used in this
analysis. We exclude stronger lines because they are severely affected by
pressure broadening, which may not be well-estimated by our calculations. The
abundance of Fe is evaluated for each Fe {\small I} line using Kurucz model
atmospheres (Kurucz 1993). 

In order to determine the $\xi$ of our program stars, we use a subset of the Fe
{\small I} lines listed in Table 2 of Paper I. The $\xi$ is set by the value for
which iron lines with a variety of strengths yield a consistent iron abundance,
changing the values of this parameter with steps of 0.1 km s$^{-1}$. The $\xi$
estimated for our objects fall within the range 1 to 2.5 km s$^{-1}$. 

We determined the surface gravity, $\log g$, for each star by demanding that the
iron abundance determined from \ion{Fe}{2} lines is equal to that derived from
\ion{Fe}{1} lines. In the first step, abundances are estimated from \ion{Fe}{1}
and \ion{Fe}{2} lines by assuming a $\log g$ value. We repeat the calculation by
changing the $\log g$ value until the iron abundances derived from both
ionization stages are in agreement. The $\xi$ is then re-determined using the
derived value of $\log g$. The $\log g$ value is then determined again on the
basis of the newly obtained $\xi$, and the process is iterated until
convergence.

\begin{figure}[p]
\begin{center}
\includegraphics[width=12cm]{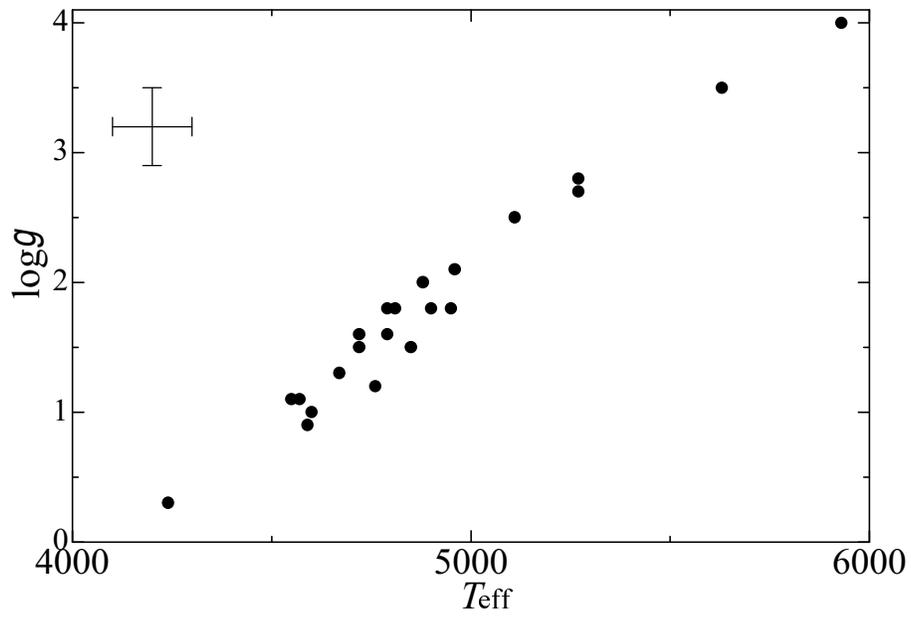}
\caption{Correlation between $T_{\rm eff}$(K) and $\log$ ($g/$cm~s$^{-2}$). Typical uncertainties are shown by error bars.}
\label{fig:tg}
\end{center}
\end{figure}

Figure~\ref{fig:tg} shows the correlation between the estimated $\log g$ and
$T_{\rm eff}$ obtained for our program stars. The scatter is small, indicating
that no extraordinary star is included in our sample, and that the derived $\log
g$ is typical of red giants. Exceptions are the star HD~140283, which we regard
as a main-sequence turnoff star, and BS~17583--100, which is regarded as a
subgiant. 

Our final set of derived atmospheric parameters are summarized in
Table~\ref{tab:param}. We take the \ion{Fe}{1} abundance as the indicator of 
metallicity of each star.

\subsection{Comparisons with Previous Studies}

The atmospheric parameters adopted in the present work are generally in good
agreement with those derived by other recent abundance studies. For example, our
model parameters ($T_{\rm eff}$ = 4570 K, $\log g$ = 1.1, and [Fe/H] $= -2.77$)
for the bright, well-studied giant HD~122563 are in good agreement with those
reported by Westin et al. (2000) ($T_{\rm eff}$ = 4500 K, $\log g$ = 1.3, and
[Fe/H] $= -2.74$) on the basis of a high S/N spectrum of this star. The
parameters of another metal-poor star, HD~115444, derived by the present work
($T_{\rm eff}$ = 4720 K, $\log g$ = 1.5, and [Fe/H] $=-2.85$) also show good
agreement with those obtained for this star by \citet{westin00} ($T_{\rm eff}$ =
4650 K, $\log g$ = 1.5, and [Fe/H] $= -2.99$) . Sneden et al. (2000) derived
model parameters for CS~22892--052 ($T_{\rm eff}$ = 4710 K, $\log g$ = 1.5, and
[Fe/H] = --3.1); these are in good agreement with our values ($T_{\rm eff}$ =
4790 K, $\log g$ = 1.8, and [Fe/H] $ = -2.92$). The small discrepancy in the
derived [Fe/H] values is primarily due to the higher $T_{\rm eff}$ and $\log g$
adopted by this work (see subsection 3.6). Hill et al. (2002) derived model
parameters for CS~31082--001 ($T_{\rm eff}$ = 4825 K, $\log g$ = 1.5, and
[Fe/H]$ = -2.9$). Our estimates for this object ($T_{\rm eff}$ = 4790 K, $\log
g$ = 1.8, and [Fe/H]$ = -2.81$) agree quite well with theirs. Giridhar et al.
(2001) derived model parameters for CS~22169--035 ($T_{\rm eff}$ = 4500 K, $\log
g$ = 1.0, and [Fe/H]$ = -2.8$) and BS~16085--050 ($T_{\rm eff}$ = 4750 K, $\log g$
= 1.0, and [Fe/H]$ = -3.1$). Our estimates for these objects are $T_{\rm eff}$ =
4670 K, $\log g$ = 1.3, and [Fe/H]$ = -2.7$ for CS~22169--035, and
$T_{\rm eff}$ = 4950 K, $\log g$ = 1.8, and [Fe/H]$ = -2.9$ for BS~16085--050.
Giridhar et al. (2001) determined the $T_{\rm eff}$ based on the
requirement that the Fe abundance derived from \ion{Fe}{1} lines be independent
of excitation potential and equivalent width. These differences may arise
because of the alternate methods used to obtain $T_{\rm eff}$.
Differences in $\log g$ and [Fe/H] propagate from that in $T_{\rm eff}$.

\section{Abundance Analyses and Results}

For our quantitative abundance measurements we make use of the analysis program
SPTOOL developed by Y. Takeda (private communication), based on Kurucz's WIDTH6.
SPTOOL calculates synthetic spectra and equivalent widths of lines on the basis
of the given atmospheric parameters, line data, and chemical composition, under
the assumption of LTE. The abundance analyses were made for 33 elements from C
to Th, where we have used the solar-system abundances obtained by
\citet{grevesse96}. The derived abundances for our program stars are listed in
Table~\ref{tab:res1}--\ref{tab:res6}.

\subsection{The Carbon Isotopes ($Z = 6$)}\label{sec:ab6}

Carbon abundances provide an important tool for understanding not only the
evolutionary stage of a given star, but also the nucleosynthesis history of the
early Galaxy. However, since the primary purpose of this work is to investigate
the abundance patterns of the heavy elements, we have selected stars with rather
weak CH G-bands, indicative of low carbon abundances, in order to avoid the
difficulties of line contamination due to CH and CN molecular lines. Even in our
present sample, however, possible contamination from these molecular lines must
be considered. For instance, it is known that contamination from $^{13}$CH lines
can affect the region surrounding the \ion{Th}{2} 4019~{\AA} line (Norris,
Ryan, \& Beers 1997). To ascertain the possible effects of the molecular lines on our
analysis, in this section we obtain estimates of the abundances of the carbon
isotopes ($^{12}$C and $^{13}$C).

\begin{figure}[p]
\includegraphics[width=14cm]{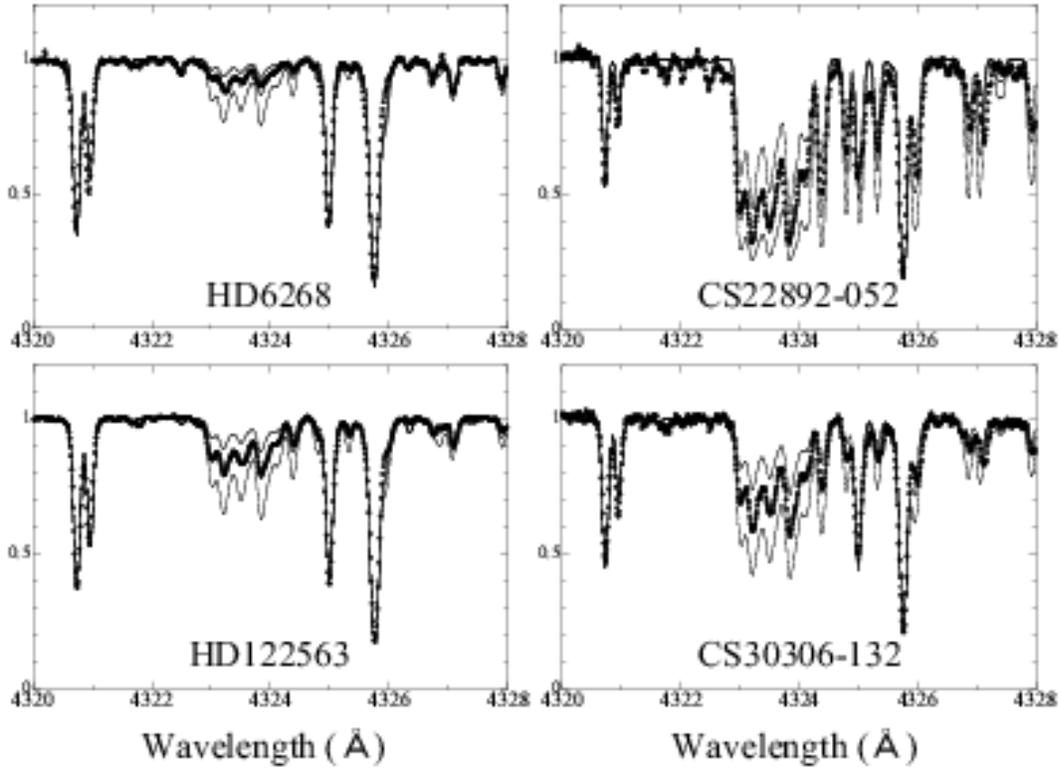}
\caption{Comparison of the observed spectra with the synthetic ones in
 the region near the CH band at 4323 {\AA}. The synthetic spectra were
 computed for three carbon abundances with a difference of 0.3 dex. The
 adopted [C/Fe] are --0.67 for HD~6268, --0.41 for HD~122563, 0.91 for
 CS~22892--052, and 0.34 for CS~30306--132.}
\label{fig:ch}
\end{figure}

The carbon ($^{12}$C) abundance is measured using the spectrum synthesis
technique for the CH 4323 {\AA} band, adopting the CH line data from
\citet{aoki02b}.  We have confirmed that the CH band of the solar spectrum 
is well-reproduced by spectrum synthesis using this line list, and the Kurucz
model atmosphere of the Sun \citep{kurucz93}. Examples of the observed spectra
of the CH band for four of our objects are shown in Figure~\ref{fig:ch}. In this
figure we also show synthetic spectra for three carbon abundances with
differences of 0.3 dex. It is seen that one can certainly specify the best-fit
features to better than 0.3 dex. The CH band is too weak in BS~16085--050 and
BS~16920--017 to measure their carbon abundances, hence for these two objects we
have only estimated upper limits. The derived carbon abundances for our program
stars are listed in Tables~\ref{tab:res1}--\ref{tab:res6}.

\begin{figure}[p]
\includegraphics[width=14cm]{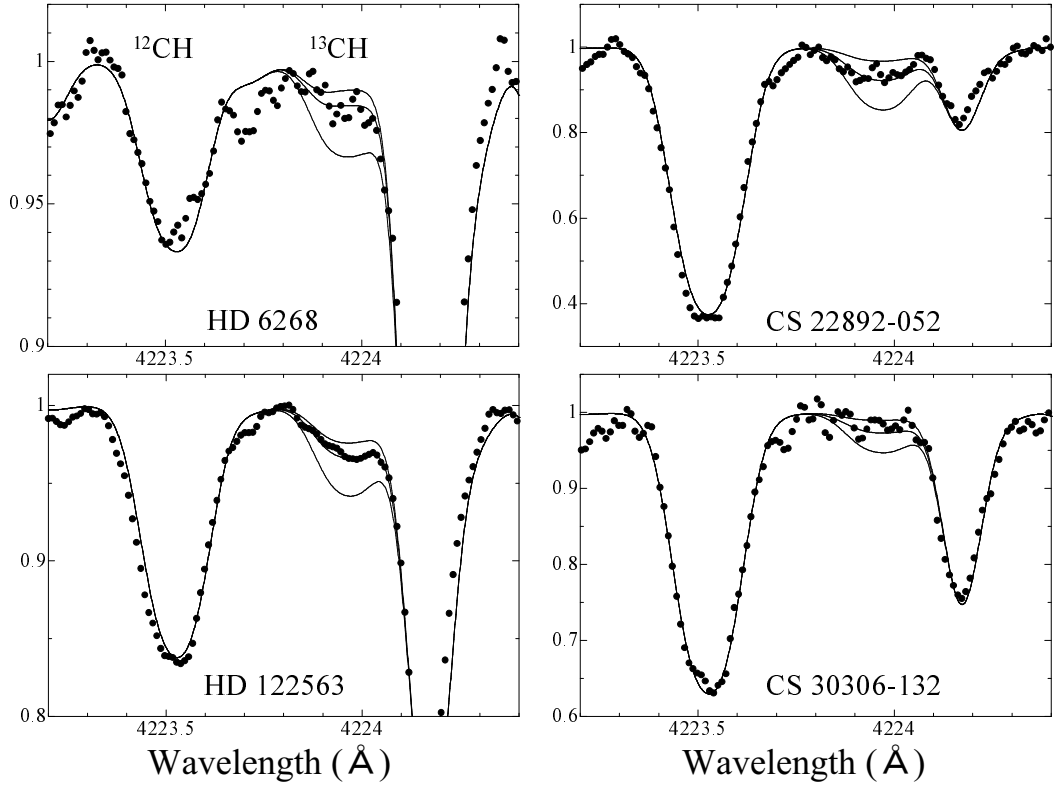}
\caption{Comparison of the observed spectra with the synthetic ones
 around 4224 {\AA} for four objects.
$^{13}$CH lines (4223.9 {\AA}) exist redward of $^{12}$CH lines.
Synthetic spectra are shown for three isotope ratios:
$^{12}$C/$^{13}$C = 2, 4, and 6 for HD~6268, $^{12}$C/$^{13}$C = 3, 5,
 and 7 for HD~122563, and $^{12}$C/$^{13}$C = 10, 20, and 50 for
 CS~22892--052 and CS~30306--132.}
\label{fig:13ch}
\end{figure}

We estimate the carbon isotope ratios ($^{12}$C/$^{13}$C) from the CH lines
around 4200~{\AA} \citep{aoki01}. Examples of the observed spectra for the same
four stars discussed above, along with synthetic spectra for three different
isotope ratios, are shown in Figure~\ref{fig:13ch}. The $^{13}$CH line appears
redward of the $^{12}$CH line in each panel. Note that, in general, the
$^{13}$CH lines are very weak in our objects. As a result, we are able to
estimate the isotope ratios for six of our program stars, but can only obtain
upper limits for 13 objects. Since no $^{12}$CH line appears in this wavelength
region for the other three stars in our sample (CS~22952--015, BS~16085--050,
and BS~16920--017), the isotope ratio is not derived for these objects. The
results of this exercise are also provided in
Tables~\ref{tab:res1}--\ref{tab:res6}. 
The $^{12}$C/$^{13}$C ratios derived by this analysis fall in the range
$4 \le ^{12}{\rm C}/^{13}{\rm C} \le 20$, which are typical values for red
giants.

\subsection{The $\alpha$- and Iron-Peak Elements ($12 \leq Z \leq
28$)}\label{sec:ab12}

The abundances for elements with $12 \leq Z \leq 28$ are derived with a standard
analysis using the equivalent widths reported in Paper I. The transition
probabilities of atomic lines employed in this analysis are adopted from
\citet{westin00}, as mentioned in Paper I.  The results show the behavior of 
typical halo stars for most of our objects \citep[e.g., ][]{cayrel03}, with a
few particularly interesting exceptions. Giridhar et al. (2001) found that
BS~16085--050 is $\alpha$-enhanced, and that CS~22169--035 is $\alpha$-poor,
results which our new data also confirm. We plan to discuss these results in
more detail in a future paper (Honda et al., in preparation). The
Cr abundances for our sample stars also show higher values, as compared with previous
studies \citep[e.g., ][]{cayrel03}. This is because we use \ion{Cr}{2} lines
with high excitation potentials. The Zn abundances derived from the two \ion{Zn}{1}
lines at 4722 and 4810~{\AA} (see Paper I) will be discussed separately, along
with the abundances of other iron-peak elements, in forthcoming paper.

\subsection{The Neutron-Capture Elements ($38 \leq Z \leq 76$)}\label{sec:ab38}

We apply a standard analysis to most lines of the neutron-capture elements using
the equivalent widths reported in Paper I. For some strong lines, as well as
lines contaminated by other absorption features, a spectral synthesis technique
is required. For most lines, the transition probabilities compiled by
\citet{sneden96} and \citet{westin00} are adopted. Line data compiled by
\citet{mcwilliam98} are used for the analysis of \ion{Ba}{2}. For \ion{La}{2}, 
\ion{Eu}{2} and \ion{Tb}{2}, we adopt the recent measurements of transition probabilities by
\citet{lawler01a,lawler01b,lawler01c}, respectively.  Studies of atomic data
have progressed substantially lately, therefore we can also make use of the
newest data for \ion{Nd}{2} and \ion{Yb}{2}, reported by \citet{denhartog03} and
C. Sneden (private communication), respectively.

The effects of hyperfine splitting and isotope shifts are included in the
analysis of the \ion{Ba}{2}, \ion{La}{2}, and \ion{Eu}{2} lines. We assume the
isotope ratios of the r-process fraction in the Solar System in the analysis of
our very metal-poor stars, a convention that has been adopted by others
\citep[e.g., ][]{sneden96}. This assumption is justified by the result of our
analysis that the abundance patterns of stars in our sample for which both Ba
and Eu are detected are in good agreement with that of the r-process component
in the Solar System.

The effect of hyperfine splitting in some La lines is quite significant. In a
previous stage of our analysis, older line data were used, and the effect of
hyperfine splitting was not included. We find, however, that our re-analysis,
which includes this effect and the line data by \citet{lawler01a}, dramatically
reduces the scatter of the derived abundances obtained from individual La lines.
We also find that the scatter of Nd abundances derived from individual lines
significantly decrease by using the line list recently provided by
\citet{denhartog03}, as compared with previous line data. Our analysis underscores
the importance of obtaining accurate line data, including hyperfine splitting,
for use in abundance studies of these heavy elements.

We detect the Yb {\small II} 3694.2 {\AA} line in seven objects, and obtain Yb
abundances by adopting the line data used by Sneden et al. (1996) without
including hyperfine splitting. However, the Yb {\small II} 3694.2 {\AA} line may
indeed be affected by hyperfine splitting, as pointed out by Aoki et al. (2002a) and
\citet{hill02}, because two of the seven isotopes of this element have
odd mass numbers ($^{171}$Yb and $^{173}$Yb). This indicates that the Yb
abundances may be over-estimated by a single line approximation. However, 
from a re-analysis based on new line data (Sneden, private communication),
the Yb abundances appear to be better determined.  

We detect the elements Sr and Ba, both of which have strong resonance lines in
the blue region, for all objects in our sample. Y and Zr are also detected in
most objects. In addition to these four elements, we detect Eu in 11 objects.
Several elements heavier than Eu, including Th (see next subsection), are
detected in seven objects, all of which exhibit overabundances of
neutron-capture elements. We discuss the abundance patterns of these stars in
more detail in \S 4.

\subsection{Thorium ($Z=90$)}\label{sec:ab90}

\begin{figure}[p]
\includegraphics[width=12cm]{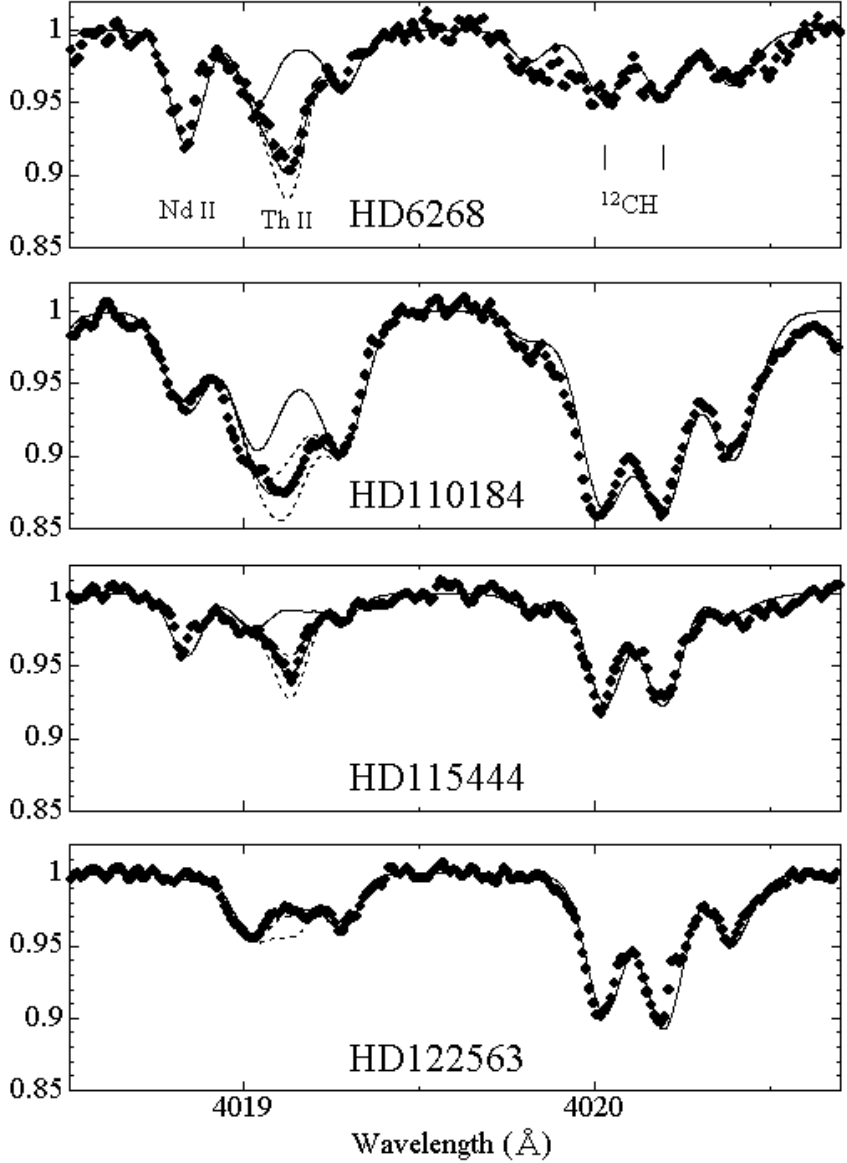}
\caption{Spectral synthesis at the 4019{\AA} region to determine the
 abundances of Th. The filled circles denote the observed data. The
 solid line represents the synthetic spectra with log$\varepsilon$(Th) $=
 - \infty$. Three dashed lines indicate the
 synthetic spectra with three different Th abundances with a step of 0.1
 dex (HD~6268) or 0.15 dex (HD~110184, HD~115444, and HD~122563). 
The adopted Th abundances are log$\varepsilon$(Th) $= -1.93$ for
 HD~6268, and log$\varepsilon$(Th) $= -2.50$ for
 HD~110184, log$\varepsilon$(Th) $= -1.97$ for
 HD~115444.
In the spectrum of HD~122563, the Th line is not detected, and the dashed
 lines show synthetic
 spectra with log$\varepsilon$(Th) $= -2.5$ and $-3.0$.}
\label{fig:th1}
\end{figure}

\begin{figure}[p]
\includegraphics[width=12cm]{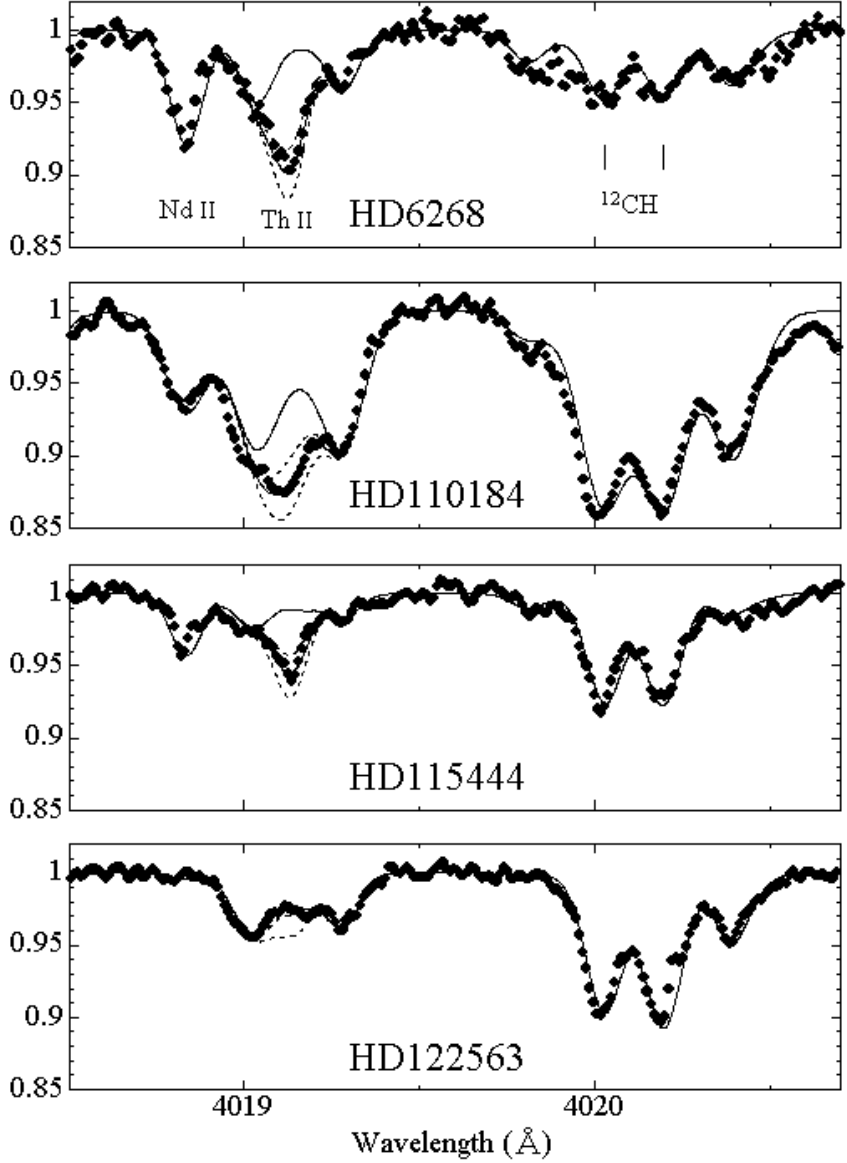}
\caption{The same as Figure 4, but for other four objects. The step of
 the Th abundances adopted in the spectrum synthesis shown by dashed
 lines is 0.1 dex (CS~31082--001) or 0.15 dex (HD~186478, CS~30306--132,
 and CS~22892--052).
The adopted Th abundances are log$\varepsilon$(Th) $= -1.85$ for
 HD~186478, and log$\varepsilon$(Th) $= -1.12$ for
 CS~30306--132, log$\varepsilon$(Th) $= -1.42$ for
 CS~22892--052, and log$\varepsilon$(Th) $= -0.92$ for
 CS~31082--001 . }
\label{fig:th2}
\end{figure}

We detect the Th {\small II} line (4019.1 {\AA}) in seven program stars,
including four stars for which Th abundances have already been reported
in the literature. Figures~\ref{fig:th1} and \ref{fig:th2} show the
observed spectra of this region in these seven stars, as well as the
spectrum of HD~122563, a star with no neutron-capture enhancements, for
comparison.

The \ion{Th}{2} 4019 {\AA} line is known to potentially suffer from blends with
lines of other elements; the impact of the blending of course depends on the
relative abundances of the competing species. The Ce {\small II} line, for
example, is important in stars with super-solar [Ce/Fe] ratios. Sneden et al.
(1996) demonstrated the importance of this line for improving their spectral
syntheses of the \ion{Th}{2} 4019 {\AA} line of CS~22892--052. Norris et al.
(1997) pointed out the significant contribution of $^{13}$CH lines at this
wavelength in $^{13}$C-rich objects. Johnson \& Bolte (2001) compiled a line
list for the analysis of the \ion{Th}{2} 4019 {\AA} line, including the above
contaminating species, which we employ in the present analysis. Before carrying
out our spectral analysis of the Th line, the abundances of Fe, Ni, Nd, Co, and
Ce are obtained first, and fixed, so that only the abundance of the Th is
altered in the final synthesis step. Note that we adopt the partition
function of Th from \citet{morell92}.

In the calculation of the spectra, the contamination of $^{13}$CH lines at
4019.00 {\AA} and 4019.17 {\AA} \citep{johnson01} is also included, adopting the
carbon isotope ratios derived in subsection \ref{sec:ab6}. For CS~31082--001 and
CS~30306--132, for which only an upper-limit of the $^{13}$C abundance was
derived, we use the upper limit as a substitute for the $^{13}$C abundance in
the calculation. We find that the effect of $^{13}$CH is negligible in the
spectra of these two stars.

The detection of the \ion{Th}{2} 4019 {\AA} line has already been reported by
previous authors for CS~22892--052, HD~115444, HD~186478, and CS~30182--001
\citep{sneden96,westin00,johnson01,cayrel01}. We have confirmed this detection
in the spectra obtained with HDS for these four objects. In addition to these
four objects, the Th line is newly detected in three additional stars,
HD~6268, HD~110184, and CS~30306--132. 

\subsection{Upper Limits on U abundances}

The radioactive species uranium is a key element for precision estimation of
stellar ages from abundance studies such as ours, owing primarily to its
relatively ``short'' half-life of 4.5 Gyrs. Errors in its measurement for a
given spectrum lead to smaller errors in the estimated decay age than, for
example, measurements of thorium, with its much longer half-life. However,
detection of U in stars is quite difficult, because of the weakness of its
spectral lines (e.g., \ion{U}{2} 3859 {\AA}), and its low present abundance in
$\sim$ 12--15 Gyr old stars (note that these ages are sufficiently long that
only 10\%--15\% of the originally produced U will survive). Since we found no
evidence of absorption by this element in our spectra, we can only estimate
upper limits on its abundance for the seven stars in which Th is detected
(Tables~\ref{tab:res1}--\ref{tab:res6}). Though \citet{cayrel01} reported the
detection of U in their very high S/N ($\sim 500$) spectrum of CS~31082--001, we
could not find U in this object, probably because of the lower quality of our
spectrum. The upper limits we are able to derive from our present spectra are
rather high, hence no meaningful information is provided for the age estimates
of these stars. Higher quality spectra for these very metal-poor stars are
obviously required, at least if one hopes to detect U. Even without a detection,
improved upper limits to the abundance of U can be used to provide interesting
limits on the ages of these stars (see the discussion in Truran et al. 2002).

\subsection{Error Estimates}

We now estimate the uncertainties in our abundance analysis arising from two
sources. The first source is random error, which might be caused, for example, by
uncertainties in the adopted {\it gf} values and errors in the equivalent width
measurements. The second is the systematic error arising from uncertainties in
our adopted atmospheric parameters.

The size of the random errors are estimated from the standard deviation
(1-$\sigma$) of the abundances derived from individual lines for elements that
had three or more lines available to include in the analysis. For elements based
only on one or two lines, we employ the mean of the random errors estimated from
those elements with multiple lines available. The derived standard deviations
are shown in Table~\ref{tab:res1}--\ref{tab:res6}.

Thorium is a particular case, since the Th abundance is determined from the one
detected absorption line at 4019 \AA\, using the spectrum synthesis technique.
The random error associated with this measurement might be estimated from errors
of the fits themselves (e.g., a $\chi^{2}$ analysis), but we find that this
approach is not useful for our limited quality spectra. In the present analysis,
we have chosen to simply estimate the maximum fitting errors ``by eye''. In
Figures~\ref{fig:th1} and \ref{fig:th2}, synthetic spectra for the adopted Th
abundance and two other possible Th abundances with 0.1~dex or 0.15~dex
differences with respect to our adopted fit are shown (see the figure captions).
We estimate the fitting error in the best cases (for HD~6268 and CS~31082--001)
to be about 0.10 dex, and for those in other cases to be 0.15 dex. We adopt
these fitting errors as estimates of the random error associated with our
derived Th abundance. Note that we neglect the error due to the uncertainty of
the transition probability, which is very difficult to estimate. In any case,
the following discussion pertains primarily to the relative abundances among our
sample stars.

Any ambiguity in the estimated atmospheric parameters for a given star can
result in an inappropriate choice of model atmospheres for use in the abundance
analysis. In order to estimate the typical errors due to this ambiguity, we
evaluate the effects of changes in adopted atmospheric parameters for the case
of HD~115444. The error in $T_{\rm eff}$ is assumed to be 100~K. This should be
a reasonable assumption, because the errors in the photometry data and the
reddening correction is 0.03--0.05~mags, which corresponds to an uncertainty of
about 50~K (for $T_{\rm eff}$ estimated from $V-K$). The difference between the
$T_{\rm eff}$ obtained from $B-V$ and that from $V-K$ is also about
100~K on average (see subsection 2.1). Errors in $\log g$, metallicity, and
$\xi$ are assumed to be 0.3 dex, 0.5 dex, and 0.5 km s$^{-1}$,
respectively. Errors in the resulting abundance estimates due to these
uncertainties are evaluated by varying the individual parameters.

Changes in the final abundances ($\Delta \log \epsilon$) caused by the
above-noted parameter changes are listed for HD~115444 in Table \ref{tab:error}.
As found in previous studies, a higher $T_{\rm eff}$ results in higher derived
abundances, in general. The effect of the difference in metallicity assumed in
the model atmosphere is relatively small. The assumption of a larger $\xi$
results in lower derived abundances. This effect is important in the abundance
analysis for \ion{Al}{1}, \ion{Cr}{1}, \ion{Ni}{1},
\ion{Sr}{2}, and \ion{Ba}{2}, which are based on rather strong lines.

Total systematic errors are derived by the root-sum-square (rss) of the
uncertainties contributed by the four parameters ($\Delta T_{\rm eff}=+100$~K,
$\Delta \log g=+0.3$, $\Delta$ [Fe/H] = +0.5, and $\Delta \xi =+0.5$~km
s$^{-1}$). The typical error due to the uncertainties of the stellar atmospheric
parameters is about 0.14 dex for the neutron-capture elements, with the
exception of Sr and Ba, which are influenced primarily by the adopted
$\xi$.

It should be noted that the systematic errors in the abundance analysis
for neutron-capture elements are small ($\lesssim 0.1$~dex), and the
abundances are affected by the atmospheric parameters in a similar way. 
Therefore, the errors in the {\it relative abundances} among these
elements are smaller than the random errors estimated above.

\subsection{Comparisons with Previous Studies}

The elemental abundances for several of the stars in our sample have already
been studied by previous authors. In this subsection, we compare our results
with those of the previous studies. We focus here on the neutron-capture
elements of the four stars in which Th was previously detected.

\begin{figure}[p]
\includegraphics[width=15cm]{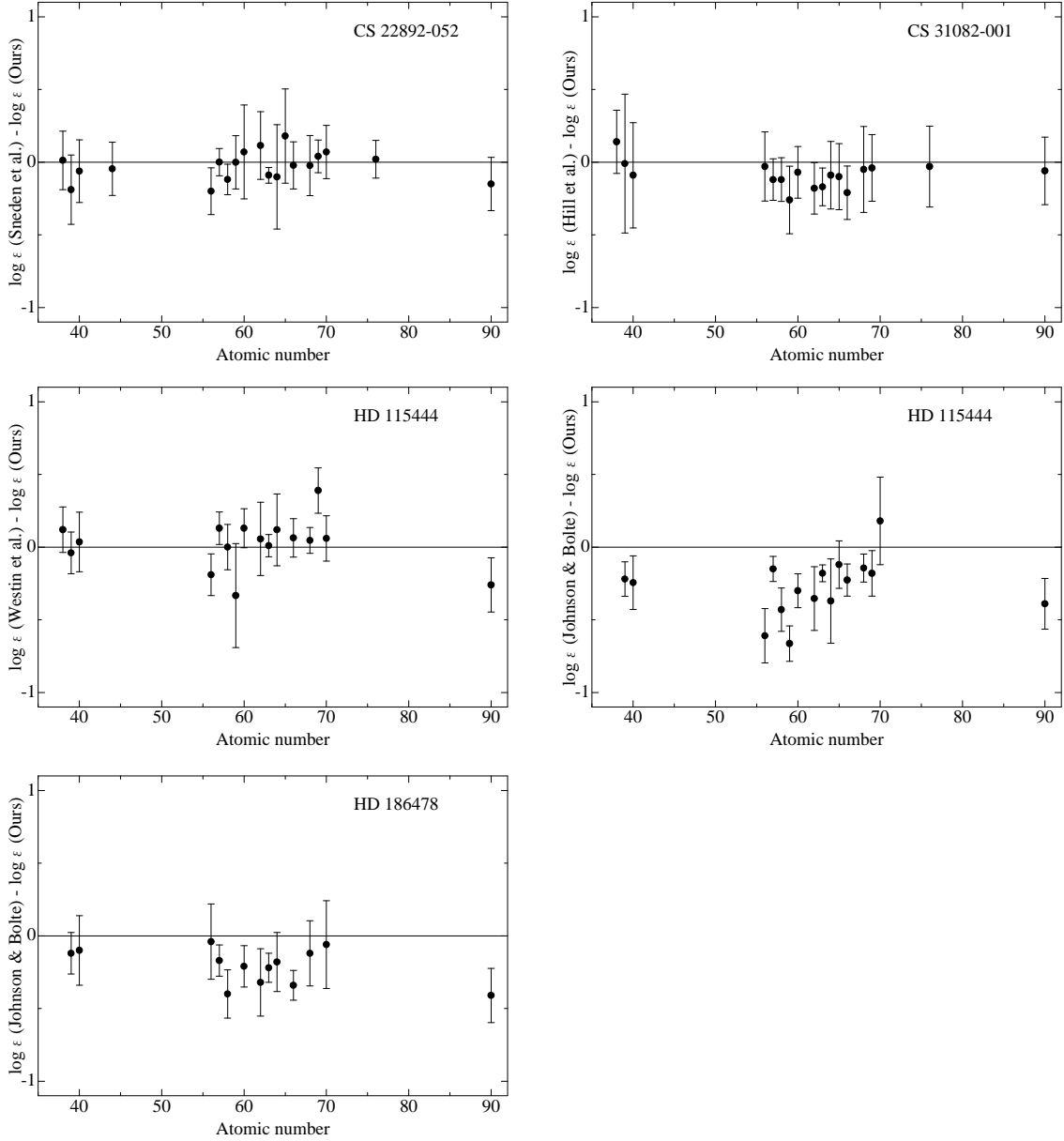}
\caption{The comparisons with the abundances derived by the present
 analysis and previous studies (Sneden et al. 2003 for CS~22892--052;
 Hill et al. 2002 for CS~31082--001; Westin et al. 2000 for HD~115444;
 Johnson \& Bolte 2001 for HD~115444 and HD~186478) as a function of atomic
 number. 
The error bar indicates the sum of the errors in both works.}
\label{fig:de052}
\end{figure}

The abundances of many neutron-capture elements in CS~22892--052 have been
investigated by \citet{sneden96,sneden00,sneden03}. We compare our results with
those of the comprehensive study for this object carried out to date
\citep{sneden03}. Our derived abundances for most elements agree well with their
results; Figure~\ref{fig:de052} shows the difference between our results and
theirs. The Th abundance of CS~22892--052 obtained by our present analysis is
log$\epsilon$(Th) $= -1.42 \pm 0.15$, which is 0.15~dex higher than the Th
abundance derived by their latest analyses ($-1.57 \pm 0.10$). This small
difference is probably due to the differences in adopted $T_{\rm eff}$ and $\log
g$ (see subsection 2.3 and Table 7). We note that the log(Th/Eu) of this object
by our analysis is $-0.56$, which is in good agreement with their result
(log(Th/Eu) = $-0.66$).

The extremely r-process-enhanced, very metal-poor star CS~31082--001 has been
studied in detail by \citet{cayrel01} and \citet{hill02}. Our results for the
abundances of the neutron-capture elements show good agreement with these
previous studies within the stated errors (Figure~\ref{fig:de052}). Although the
species Sm and Eu show small differences, this may be due primarily to
measurement errors, since our HDS spectrum is not of the same quality as the
extremely high S/N spectrum obtained by \citet{hill02}. CS~31082--001 is known
to exhibit unusually strong absorption lines of \ion{Th}{2}. The Th abundance
derived by our analysis is log$\epsilon$(Th) $= -0.92 \pm 0.10$
(log(Th/Eu) $= -0.33 \pm 0.10$ ). These values
agrees very well with those (log$\epsilon$(Th) $= -0.98 \pm$ 0.05;
log(Th/Eu) $= -0.22 \pm 0.12$) reported by Hill et al. (2002).

\citet{westin00} derived the abundances of a number of neutron-capture
elements in HD~115444 from a high-quality spectrum of this star. The abundances
of most neutron-capture elements derived by our analysis agree with theirs
within the errors. An exception is the case of Tm (Figure~\ref{fig:de052}).
Although we investigated a number of possible sources of difficulty, we could
not resolve the likely reason for the difference. We note, however, that our
present result is based on two lines, which agree well with one another, while
the \citet{westin00} analysis is based on only a single line.

Our estimate of the Th abundance in HD~115444, log$\epsilon$(Th) $= -1.97 \pm
0.15$, appears to be somewhat higher than the value obtained by Westin et al.
(2000), log$\epsilon$(Th) $=-2.23 \pm 0.11$. The difference between the two
measurements is at a similar level as the uncertainties in the analysis, so
perhaps this should not be of concern. However, inspection of the synthetic
spectra used in this analysis indicates that this discrepancy could arise
from differences in the line lists that were adopted. In our synthetic
spectrum {\it without} the \ion{Th}{2} line (Figures~\ref{fig:th1}) the
absorption feature at 4019.1~{\AA}, where \ion{Th}{2} exists, is considerably
weaker than that shown in the synthetic spectrum of \citet{westin00} (their
Figure 7). In order to identify the reason for this difference, we also compare
the synthetic spectra for HD~122563, which was studied by both teams. This star
is a very metal-poor giant with similar atmospheric parameters as HD~115444, but
exhibits no Th feature. We find that our synthetic spectrum of HD~122563,
calculated with the same line list as used for HD~115444, reproduces very well
the observed spectrum at 4019.1~{\AA}. On the other hand, the absorption feature
at 4019.1~{\AA} in the synthetic spectrum of \citet{westin00} for this star is
deeper than that in the observed spectrum. Although the discrepancy between the
observed and synthetic spectra seems to be similar to the noise level of their
observed spectra, the spectral feature at 4019.1~{\AA} was also observed in our
very high-quality spectrum of this star. For this reason, we suggest that the
discrepancy between the observed and synthetic spectra found in Figure 7 (lower
panel) of \citet{westin00} is {\it not} due to random errors in the observed
spectrum, but rather arises from an overestimate of the absorption at
4019.1~{\AA} in their calculation. If the absorption at this wavelength is also
overestimated for HD~115444, the contribution of the
\ion{Th}{2} is underestimated. This may explain the difference of
0.2~dex in the Th abundances between the results of \citet{westin00} and our
present analysis.

\citet{johnson01} derived the elemental abundances of HD~115444 and
HD~186478. The Fe abundances derived by these authors are lower by 0.3 dex and
0.1 dex than ours for HD~115444 and HD~186478, respectively
(Figure~\ref{fig:de052}). 
These differences are likely due to differences in the atmospheric
parameters adopted in the two analyses. Their $T_{\rm eff}$ and $\log g$ for
these objects are lower than ours by about 200~K and 0.8~dex, respectively. The
Th abundances derived by \citet{johnson01} for HD~115444 and HD~186478 are
log$\epsilon$(Th) $= -2.36$ and log$\epsilon$(Th) $= -2.26$, respectively. These
values are lower by 0.39 dex and 0.41 dex than our results, respectively. These
differences are well explained by the differences in the adopted $T_{\rm eff}$
and $\log g$. Note that differences of this size do not affect the relative
abundances of neutron-capture elements very significantly. For example, the
differences in the value of log(Th/Eu) become smaller, 0.21 and 0.19
for HD 115444 and HD 186478, respectively, roughly the same size
as the value of the estimated random errors.

\section{Discussion}

The present analysis has shown that our objects indeed possess very low
metallicities ($-3.2 <$ [Fe/H] $< -2.4$). The iron abundances derived by the
present analysis are similar to the values which were estimated by the previous
studies from lower dispersion spectroscopy (Beers et al. 1992; Bonifacio et al.
2000; Allende-Prieto et al. 2000). One exception is the star
CS~22952--015, whose metallicity was estimated to be [Fe/H] $= -3.50$ by
Beers et al. (1992), while our estimate is rather higher, [Fe/H] $= -2.94$.
Another is CS~30306--132, whose metallicity was estimated to be
[Fe/H]$=-3.1$ from lower-dispersion spectroscopy by Beers et al. (2003,
in preparation), while our result is [Fe/H]$=-2.44$. This discrepancy
may be due to the fact that this star has strong CH molecular bands.

The metallicity range of the stars in our sample should be kept in mind, as it
is known that a large scatter in the abundance ratios in many neutron-capture
elements appears when the metallicity drops to [Fe/H] = --2.5 or below
\citep[e.g., ][]{mcwilliam95b}. In this section, we discuss the relative
abundances of the neutron-capture elements, the dispersion of the observed
abundance ratios, and the origin of these elements (subsection 4.1). We discuss
in detail the abundance pattern of neutron-capture elements for the seven stars
in our sample for which Th is detected, and the impact of our new results on the
cosmochronology technique based on Th (subsection 4.2). The abundances of the
$\alpha$- and iron-peak elements are discussed only for comparison purposes.
Details on the behavior of these elements will be presented separately in a
future paper in this series (Honda et al., in preparation).

\subsection{Relative Abundances of the Neutron-Capture Elements}

\subsubsection{The Heavy Neutron-Capture Elements ($56 \leq Z \leq 76$)}

\begin{figure}[p]
\includegraphics[width=8cm]{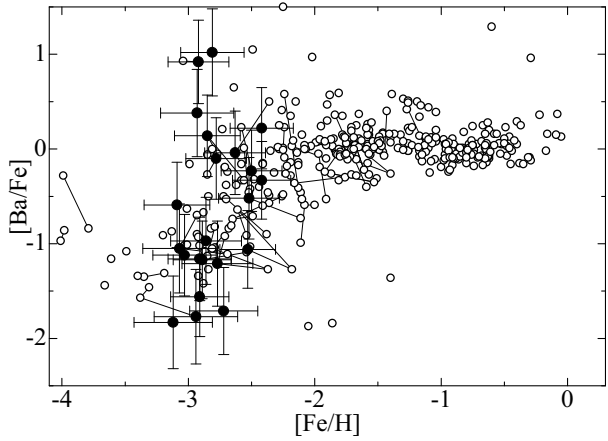}
\caption{[Ba/Fe] values as a function of [Fe/H]. Our results are shown
by filled circles with error bars, while results by previous studies
(compilation of literature data taken from Norris et al. 2001, and
 McWilliam 1998, Burris et al. 2000) are shown by open
 circles. The error bars shown here are random and systematic errors for
 [Fe/H], and rss of random errors [Ba/H] and [Fe/H] for [Ba/Fe].
}
\label{fig:ba}
\end{figure}

Figure \ref{fig:ba} shows the values of [Ba/Fe] as a function of [Fe/H] for all
of the stars in our sample (filled circles with error bars), as well as the
results reported by previous authors for comparison (open circles). Ba
abundances have been reported for stars with lower metallicity than the
abundances of other neutron-capture elements such as Eu. This is because Ba has
strong Ba {\small II} resonance lines (4554 and 4934~{\AA}), which remain
detectable even as the overall level of metallicity decreases. Previous studies
have shown that [Ba/Fe] drops below the solar ratio, on average (e.g., Ryan et
al. 1996; McWilliam 1998), at the lowest metallicities. This trend is thought to
arise because of a change in the primary nucleosynthesis sources for Ba between
stars with [Fe/H] $\la -2.5$ and those with [Fe/H] $\ga -2.5$. Ba in the more
metal-rich stars is believed to originate primarily from the (main) s-process in
low-mass or intermediate-mass stars (1-8 $M_\odot$) with a comparatively small
contribution by the r-process. The production of Ba at the lowest metallicities,
however, is most likely due to the r-process alone, occurring in Type II
supernova explosions of massive and hence rapidly-evolving stars. The
contribution of the s-process to Ba in the early Galaxy is small, because the
time-scale for the evolution of lower mass stars is long, and the ejecta from
these stars contribute only to stars which formed somewhat later, with [Fe/H]
$\ga -2$~\citep[e.g., ][ and references therein]{truran02}.

Our results, shown in Figure \ref{fig:ba}, confirm the existence of an extremely
large scatter (a factor of $\sim$1000 over the entire range) in the Ba
abundances of the most metal-poor stars, even larger than reported in previous
studies. For instance, the standard deviation of the [Ba/Fe] values of our
sample is 0.82~dex, while that found by \citet{mcwilliam98} for 24 objects with
[Fe/H] $<-2.5$ is 0.59~dex. One clear reason for this larger dispersion is that
the detection limit for Ba lines in our work is lower than in previous programs,
thanks to the high quality of the Subaru/HDS spectra. The other reason is likely
due to our selection of candidate neutron-capture-enhanced stars, as mentioned
in Paper I. In spite of this selection bias, the large dispersion in Ba
abundances that exists in the metallicity range of $-3.0 \la$ [Fe/H] $\la -2.5$,
and the possibly decreased scatter at lower iron abundances, provides an important
clue to the sites and mechanisms of astrophysical neutron-capture processes
\citep{wasserburg00}.

\begin{figure}[p]
\includegraphics[width=8cm]{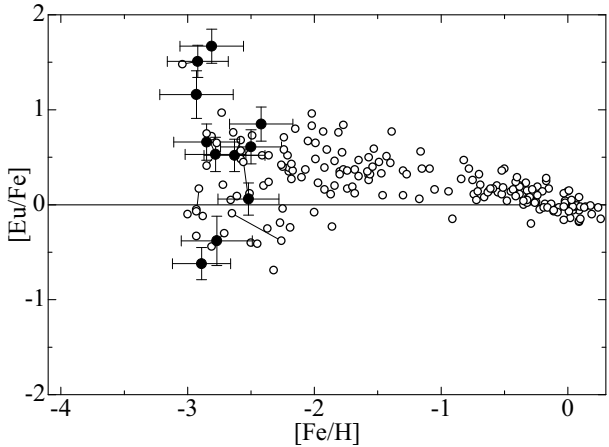}
\caption{The same as Figure 7, but for [Eu/Fe].}
\label{fig:eu}
\end{figure}

Roughly 95\% of the Eu in solar-system material is believed to be produced
by the r-process \citep[e.g.,][]{arlandini99,burris00}, hence this element is
particularly suitable for investigating the characteristics of the r-process in
the early Galaxy. However, measurements of Eu abundances in extremely metal-poor
stars ([Fe/H] $\sim -3.0$) are still quite limited, because there is no strong
Eu line, unlike those of Sr and Ba. We have detected Eu in 11 objects in our
sample; the derived [Eu/Fe] ratios for our sample are shown in Figure
\ref{fig:eu} as a function of [Fe/H] (filled circles), along with the results
obtained by previous studies (open circles). Our results show a large scatter
also in [Eu/Fe] in very metal-poor stars, as has been reported by previous
authors \citep[e.g.,][]{mcwilliam95b,burris00}.

\begin{figure}[p]
\includegraphics[width=8cm]{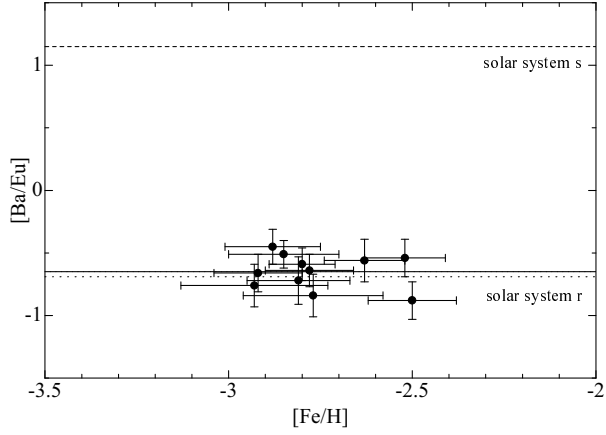}
\caption{[Ba/Eu] as a function of [Fe/H] for our sample. The solid line
indicates mean value of our [Ba/Eu]. The dotted line indicates the
[Ba/Eu] of the r-process component in the solar system, while the
dashed line means that of the s-process-component in the solar
system.}
\label{fig:baeu}
\end{figure}

The ratio [Ba/Eu] is useful for distinguishing the contributions of the r- and
s-processes, because the expected ratios from the two processes are quite
different. Figure \ref{fig:baeu} shows [Ba/Eu] as a function of [Fe/H] for the
11 stars in our sample for which Eu has been detected. The dotted line indicates
the value of [Ba/Eu] of the solar-system r-process component ([Ba/Eu] $= -0.69$,
Arlandini et al. 1999), while the dashed line indicates that of the
s-process component ([Ba/Eu] $= +1.15$, Arlandini et al. 1999). The [Ba/Eu]
ratios exhibited by our stars are clearly associated with the
r-process, rather than the s-process.

Similar results have been reported by previous authors. McWilliam (1998) found
that the mean of the [Ba/Eu] ratios for 10 stars with [Fe/H] $\leq$ --2.4 is
--0.69, consistent with the value expected from pure r-process nucleosynthesis
within the measurement uncertainties. Here we find that the ratio of [Ba/Eu]
matches that of the solar-system r-process component for many stars at low
metallicity. These results suggest that Ba, as well as Eu, is primarily produced
by the r-process during the early history of the Galaxy, and that Ba can be used
as a powerful tool to investigate the behavior of r-process elements in the
early Galaxy, as it remains detectable in stars of metallicities that are far
lower than those in which Eu is detectable.\footnote{A number of stars with
large excesses of s-process elements {\it are} known to exist in this
metallicity range \citep[e.g, ][]{aoki02b}. The chemical compositions of these
stars are interpreted as resulting from s-process nucleosynthesis in
intermediate-mass, evolved stars followed by mass-transfer across the binary system.
These stars do not exist in our sample, presumably because carbon-enhanced
metal-poor stars have been specifically excluded in our sample selection (see
Paper I).}. Though most of the stars shown in Figure \ref{fig:eu} exhibit
[Eu/Fe] $\gtrsim0$, we expect that the [Eu/Fe] values of the stars with low Ba
abundances are lower than zero, and in reality will, once detected, exhibit a
similarly large dispersion as seen in [Ba/Fe]. This prediction should be
confirmed by studies of Eu lines using higher quality spectra for objects with
[Fe/H] $< -3$ (e.g., Ishimaru et al. 2004).

\begin{figure}[p]
\includegraphics[width=8cm]{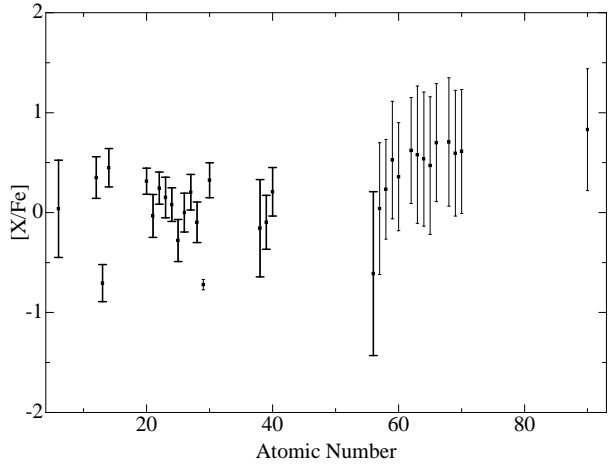}
\caption{
Average of the elemental abundances relative to iron ([X/Fe]) for our
objects as a function of atomic number (dots).
The standard deviation of the abundances are shown by bars. 
The thin bars indicate the standard deviation of the abundances for
the element which is detected in less than 12 objects.
}
\label{fig:scatter}
\end{figure}

Large levels of scatter are also found in the abundance ratios of almost all
neutron-capture elements with {\it Z} $\geq$ 56. Figure \ref{fig:scatter} shows
the average of the elemental abundances relative to iron [X/Fe], and the
standard deviation as a measure of the scatter in the abundances, as a function
of atomic number. The number of objects in which the species is detected depends
upon the element: i.e., neutron-capture elements heavier than Ba are detected only
in stars with excesses of neutron-capture elements. The dispersion of the
abundances for elements which are detected in less than 12 objects is shown by
the thin bars. Even though this limitation will make the dispersion of the
abundances of heavy neutron-capture elements ($Z \ga 57$) smaller, the
dispersion of the abundances of neutron-capture elements is much larger than
found for the $\alpha$- and iron-peak elements.

The scatter in the abundances of heavy neutron-capture elements relative to iron
found in these very metal-poor stars means that the nucleosynthesis site of
iron-peak elements and the r-process elements are quite different, and that the
mixing of the yields from the first supernovae into the ISM is incomplete in the
early stages of the Galaxy. The large scatter of r-process elements appearing at
[Fe/H] $\sim -3$ should provide a constraint on the dominant site of r-process
nucleosynthesis \citep[e.g., ][]{ishimaru99,tsujimoto00}.

We have confirmed that CS~22892--052 and CS~31082--001 are extremely
r-process-rich objects, and that HD~6268, HD~115444 and HD~186478 are moderately
r-process-rich objects. In addition, we have found two new r-process-enhanced
objects, CS~22183--031 and CS~30306-132. CS~22183--031 exhibits a large excess
of r-process elements (e.g., [Eu/Fe] $\gtrsim+1$). Unfortunately, we could not
detect many lines because of the relatively low S/N in our spectrum of
CS~22183-031. We are able to detect numerous lines in CS~30306-132, and found
that this object has a moderate enhancement of r-process elements ([Eu/Fe]
$\sim$ +0.8). In \S 4.2 we describe this object in detail.

\subsubsection{The Light Neutron-Capture Elements ($38 \leq Z \leq 40$)}

\begin{figure}[p]
\includegraphics[width=5cm]{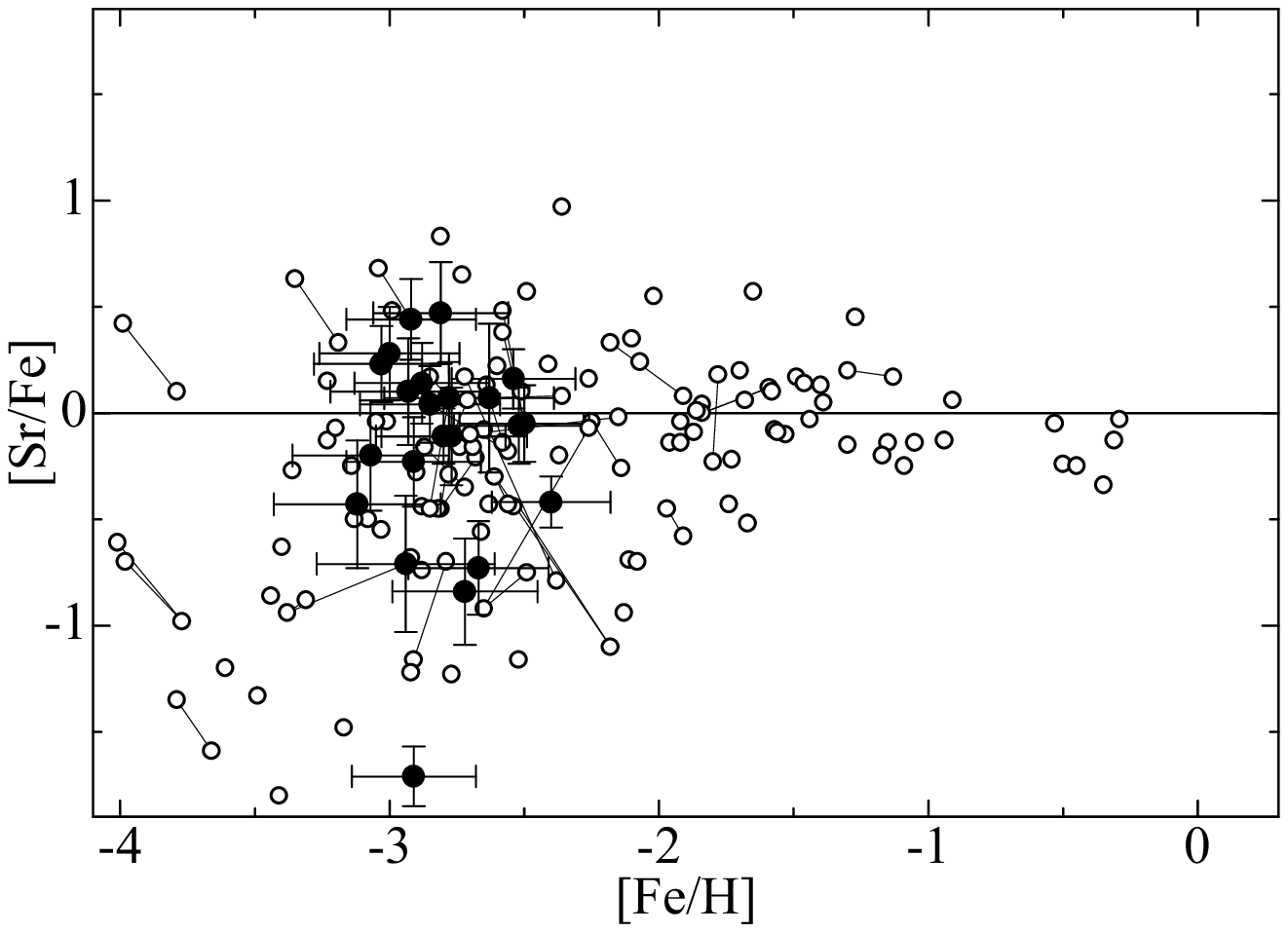}
\includegraphics[width=5cm]{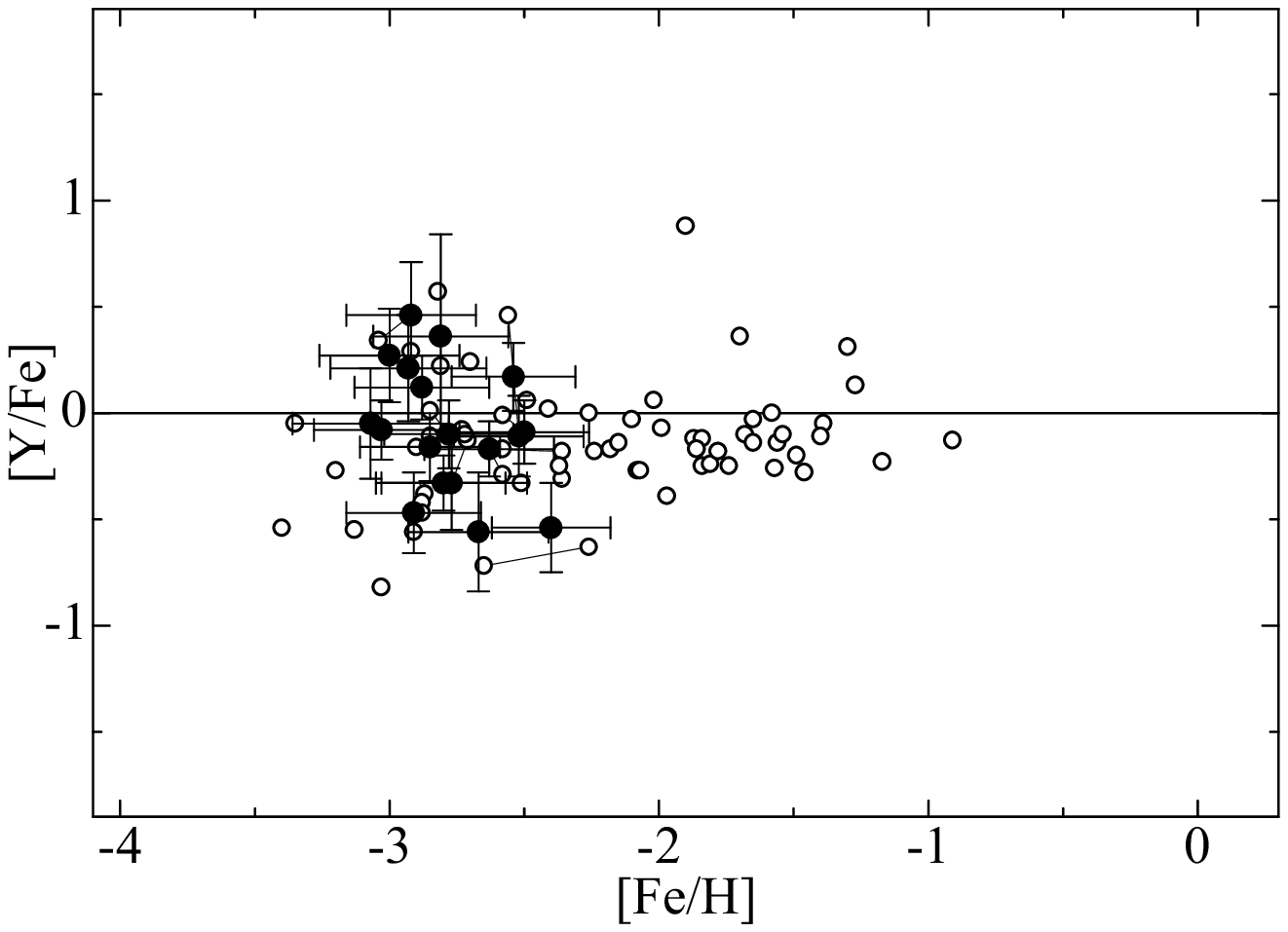}
\includegraphics[width=5cm]{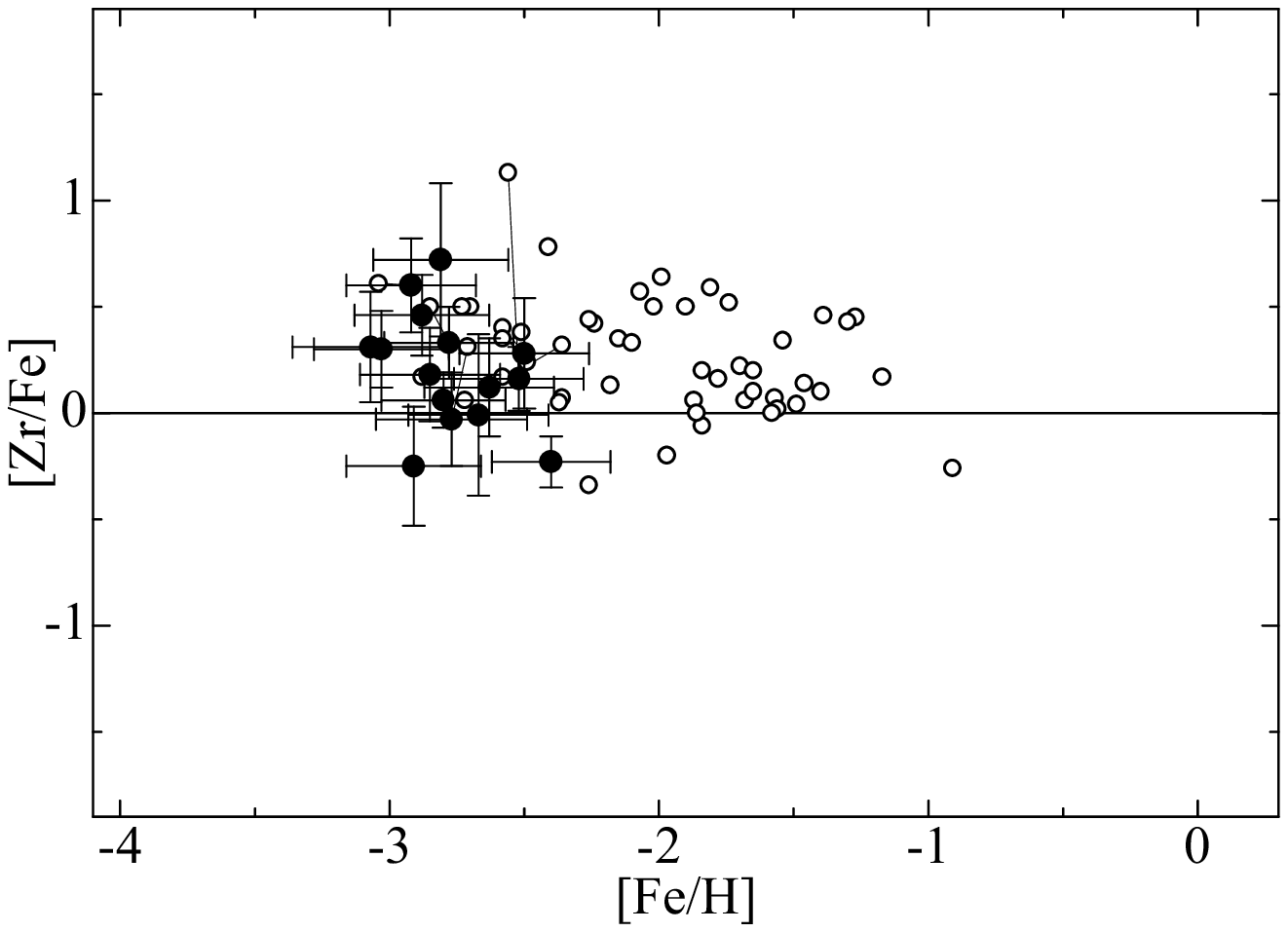}

\caption{The same as Figure 7, but for [Sr/Fe] (upper panel), [Y/Fe]
 (middle panel), and [Zr/Fe] (lower panel).}
\label{fig:lightr}
\end{figure}

Figure \ref{fig:lightr} shows the abundance ratios of [Sr/Fe], [Y/Fe], and
[Zr/Fe] as a function of [Fe/H]. The [Sr/Fe] ratios in our program stars
are distributed over a very wide range, from $-1.7$ to $+0.5$, confirming the large
dispersions in this ratio found by previous studies \citep[e.g.,
][]{mcwilliam95b,burris00}. The stars in our sample appear to exhibit a
rather higher mean [Sr/Fe] ratio than those of previous studies, but this is
likely because of the sample selection, as discussed in the previous subsection.
The scatter that appears in the [Y/Fe] and [Zr/Fe] ratios is smaller than that
in [Sr/Fe]. However, Y and Zr do not have such strong spectral lines as the
\ion{Sr}{2} resonance lines, hence these two elements are not detected in
several objects in our sample. This may account for their smaller dispersion;
clearly, this should be investigated by obtaining higher-quality spectra of
stars that presently have only upper limits on their Y and Zr abundances.

\begin{figure}[p]
\includegraphics[width=8cm]{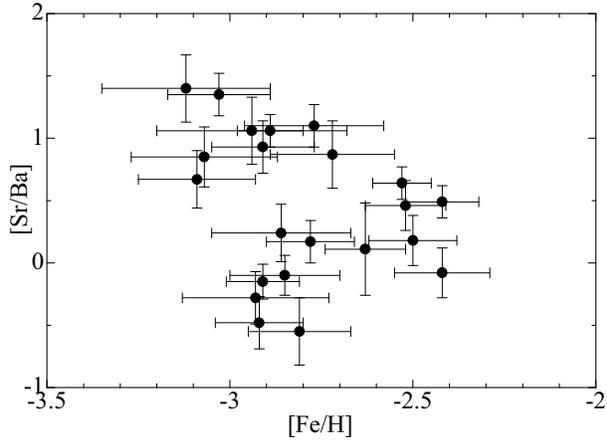}
\caption{The same as Figure 8, but for [Sr/Ba]}
\label{fig:srbafe}
\end{figure}

In order to investigate the reason of the large dispersion in the Sr abundances,
in Figure \ref{fig:srbafe} we plot [Sr/Ba] as a function of [Fe/H]. As discussed
previously by \citet{mcwilliam98}, although the dispersion in [Sr/Ba] at very
low metallicity is rather smaller than that of [Sr/Fe], the range is still
almost 2~dex. This stands in stark contrast to the range of [Ba/Eu] exhibited by
the stars in our sample, all of which have quite similar values. This result
suggests that either (a) the process that contributed significant amounts of Sr
in these metal-deficient stars did not yield similar amounts of Ba, or (b) the
process that produced Ba at very low metallicity yielded a variety of Sr/Ba
ratios.

\begin{figure}[p]
\includegraphics[width=8cm]{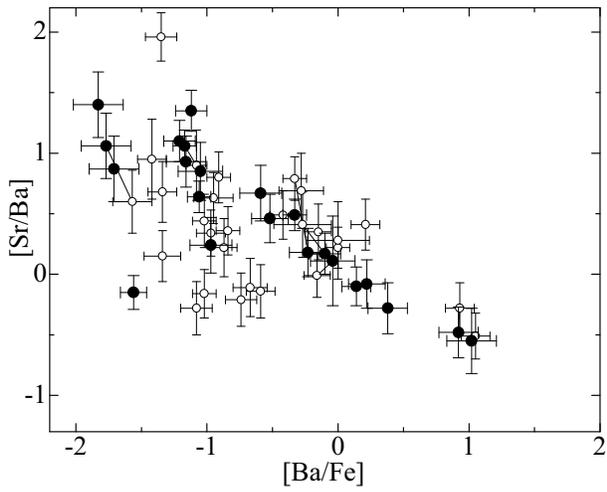}
\caption{[Sr/Ba] as a function of [Ba/Fe] for our sample (filled
circles). The results by previous work (McWilliam et al. 1998) are also shown by
open circles.}
\label{fig:srbabafe}
\end{figure}

To investigate the correlation between Sr and Ba, Figure~\ref{fig:srbabafe}
shows the ratio [Sr/Ba] for the stars in our sample as a function of [Ba/Fe].
Such a diagram was was also shown, for a different set of stars, by
\citet{truran02} and Sneden, Preston, \& Cowan (2003). The sample of Sneden et
al. (2003) includes s-process-rich stars, in which Pb is detected. By way of
contrast, we have excluded stars known to exhibit large excesses of s-process
elements and stars with [Fe/H] $>-2.5$ from this figure, in order to avoid
possible contamination by s-process nucleosynthesis. (A possible contribution of
the so-called weak s-process is discussed below.) One clear result found in this
figure is that the dispersion in [Sr/Ba] decreases with increasing Ba abundance.
This correlation is much clearer in our figure than in Figure 10 of
\citet{truran02}, presumably because their sample includes several rather
metal-rich objects, which could well be affected by the contribution of
s-process nucleosynthesis. Moreover, our new sample of stars has added a number
of objects at the high end of the [Ba/Fe] range, hence this contributes to
making the correlation in this range clearer.

\begin{figure}[p]
\includegraphics[width=8cm]{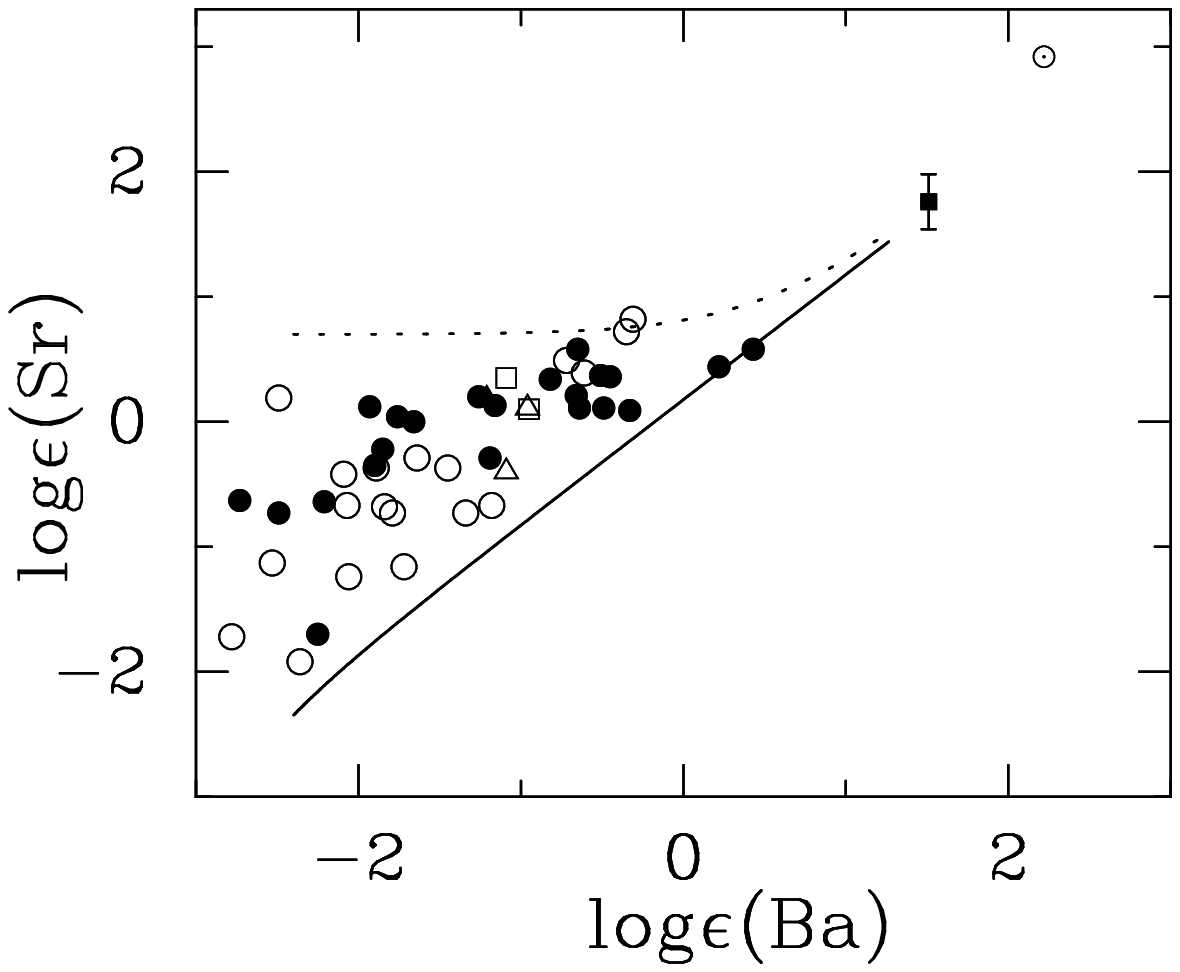}

\caption{Correlation between abundances of Sr and Ba for objects with
[Fe/H]$\lesssim -2.5$ from \citet{mcwilliam95b} (open circles),
\citet{burris00} (squares), \citet{johnson02} (triangles), and the
present work (filled circles). For the object which was studied by
both our study and others, we adopted the result by ours. The objects
which are known to show s-process abundance pattern are excluded from
the sample. The solar values are also plotted. The filled square
indicates the values of the r-process component of solar-system
material. Solid lines show the enrichment of Sr
and Ba assuming the initial abundances of Sr and Ba and a constant
Sr/Ba ratio in the yields of the main r-process. See text for
details.}

\label{fig:srbares}
\end{figure}

Figure \ref{fig:srbares} shows the abundances of Sr and Ba for very metal-poor
stars. For the stars in common between our study and those of others, we have
adopted the abundances derived in the present study. As a result, a total of 46
stars are shown in this figure. It is obvious that the dispersion in the Sr
abundance decreases with increasing Ba abundance. To demonstrate this
quantitatively, we divided the sample into three groups on the Ba abundance
($\log \epsilon$(Ba)$>-1$: 16 stars, $-1\geq \log \epsilon$(Ba)$>-2$: 19
stars, and $-2\geq \log \epsilon$(Ba): 11 stars), then measured the
standard deviation of the Sr abundances for each group. 
The results are 0.25 dex, 0.38 dex, and 0.71~dex, respectively.
Clearly, the dispersion increases with decreasing Ba abundances. While the
standard deviation in the range of $\log \epsilon$(Ba)$>-1$ is of a
similar level as the typical observational errors, in the range of $\log
\epsilon$(Ba)$\leq -2$ the dispersion is significantly larger than the
measurement errors.

We also plot in Figure \ref{fig:srbares} the values of Sr and Ba in solar-system
material (Grevesse, Noels, \& Sauval 1996, the dotted circle), as well
as the solar-system r-process component (the filled square). 
Unfortunately, estimation of the s-process component for Sr (knowledge
of which is required in order to obtain the r-process residual value) is
quite uncertain, because of possible contamination from the so-called
weak component of the s-process \citep{kappeler89}. 
For the r-process component of Sr, we adopted the average of
the values derived by \citet{kappeler89} and by Arlandini et al. (1999, the
estimate from their classical model), and show the difference by an error bar.

A simple model can be constructed for the enrichment of Sr and Ba,
assuming initial abundances of Sr and Ba ($\epsilon_{0}$(Sr) and
$\epsilon_{0}$(Ba)) and a constant Sr/Ba ratio ((Sr/Ba)$_{\rm r}$) in
the yields of the r-process, i.e.,

$\epsilon$(Sr)=$\epsilon_{0}$(Sr)$ + $(Sr/Ba)$_{\rm r} x$

$\epsilon$(Ba)=$\epsilon_{0}$(Ba)$ + x$ .

The two solid lines in Figure \ref{fig:srbares} show cases for different initial
Sr abundances: $\epsilon_{0}$(Sr) $= 3 \times 10^{-3}$ and 5, respectively. 
$\epsilon_{0}$(Ba)$ = 3 \times 10^{-3}$ and (Sr/Ba)$_{\rm r}=1.5$
is assumed for both cases. A glance at this figure shows that the observational
data fill the range between the two lines. Hence, the
distribution of the observed Sr and Ba abundances are quite naturally explained
by the simple assumptions of a large dispersion of Sr abundances at $\log
\epsilon$(Ba)$\sim -2.5$ and enrichment of Ba and Sr with a constant Sr/Ba
ratio. Moreover, if we extend the line representing the enrichment of Sr and Ba,
the Sr and Ba abundance of the r-process component in solar-system material is
also explained.

A similar scenario for Sr and Ba enrichment has already been proposed by
previous studies \citep[e.g., ][]{sneden00,truran02}. In particular,
\citet{truran02} selected several metal-deficient stars with high and low Ba
abundances, and concluded that Ba-poor stars show high Sr/Ba ratios, while the
Sr/Ba ratios of Ba-rich stars are similar to that of the r-process component in
the Solar System. These authors suggested the existence of two processes (sites)
that produce Sr. Our interpretation for the Sr and Ba abundance distributions is
essentially the same as theirs. However, the enrichment of Sr and Ba is seen
much more clearly in the present study, as the result of adding new measurements
and excluding stars of higher metallicity.

The enrichment of Ba in the metallicity range around [Fe/H]$\sim -3$ is
sometimes referred to as the main r-process, which produces heavy (A $>$ 130)
neutron-capture elements \citep{wasserburg00}. Our interpretation for the
variation of the Sr and Ba abundances results in the Sr/Ba ratio produced by
this process, (Sr/Ba)$_{\rm r}$, to be about 1.5. The absence of stars with
Sr/Ba $<1$ in Figure \ref{fig:srbares} means that (Sr/Ba)$_{\rm r}$ is not
significantly smaller than unity. This value is much higher than the Sr/Ba
predicted by recent models of r-process nucleosynthesis (e.g., $\sim 0.03$:
Otsuki, Mathews, \& Kajino 2003). The value derived by these models should be,
however, a lower limit of the Sr/Ba ratio produced by a single site, as these
models deal with only one condition which can produce heavy neutron-capture
elements like Ba. The yields in a real r-process site, such as a Type II
supernova explosion, should be an integration of the results of a variety of
conditions including the cases with low neutron-to-seed ratios, and
produce light neutron-capture elements like Sr. If our interpretation for the
correlation between Sr and Ba abundances is correct, this places a quite strong
constraint on the Sr/Ba ratio produced by the main r-process.

In contrast, the nucleosynthesis process that is responsible for the production
of light neutron-capture elements (e.g., Sr) without producing the heavier
elements (e.g., Ba) is as yet unclear. One possibility is the existence of an
independent nucleosynthesis process which dominantly produces lighter
neutron-capture elements, probably prior to the 'main' r-process which provides
heavier ones (e.g., Truran et al. 2002). For instance, the above observational
results are qualitatively explained by the assumption that different mass ranges
of supernovae are responsible for the light and heavy neutron-capture
elements (e.g., Qian \& Wasserburg 2000). Two different explosion mechanisms,
i.e., prompt and delayed explosion, are proposed for low-mass and high-mass
supernovae, respectively, and numerical simulations of r-process nucleosynthesis
have been made \citep[e.g., ][]{hillebrandt84,woosley94,meyer92,otsuki00,
sumiyoshi01,wanajo03}. Possible r-process nucleosynthesis in
neutron-star mergers may also play a role in the enrichment of the light
neutron-capture elements \citep[e.g.,][]{rosswog99, freiburghaus99}.

On the other hand, another explanation by r-process nucleosynthesis in a single
site may be possible if incomplete mixing of supernova ejecta with interstellar
matter is assumed. For instance, some neutrino-heated wind models predict
significantly large overproduction of light neutron-capture elements in the
early phase ($\sim 1$ second) of the explosion prior to the r-process which
produces the heavier elements (e.g., Woosley et al. 1994). Cameron (2001)
suggested that r-process nucleosynthesis may take place in the jets associated
with gamma-ray bursts, which would be a possible mechanism for production
of an inhomogeneous r-process in a single event. Stars with high Sr and low Ba
abundances may form from the interstellar medium polluted by the ejecta enriched
in light neutron-capture elements, if the ejecta is not well mixed with the
matter ejected in the later phase of the r-process.

An alternative explanation for the enrichment of Sr without Ba may be the
so-called weak s-process, which proposes a neutron-capture process
proposed to occur in core He-burning massive stars \citep[e.g.,
][]{hoffman01}. 
This process is expected to be unimportant in metal-deficient stars,
because $^{22}$Ne is believed to be the neutron source. 
However, if this process {\it can} operate,
it may contribute to light neutron-capture elements in metal-deficient stars.

Further systematic studies of stars with low Ba abundances, especially
of the abundance patterns for elements around Sr, will provide important
information to constrain the nucleosynthesis process responsible for their
production, and their ejection dynamics as well.

\subsection{Distribution of Neutron-Capture Elements}

One important discovery in recent abundance studies of the neutron-capture
elements in metal-deficient stars is that the abundance pattern of heavy
neutron-capture elements are very similar, essentially an exact match, within
observational errors, to that of the r-process component in the Solar System. To
examine details of r-process nucleosynthesis at low metallicity, it is essential
to detect as many elements as possible over the entire range of atomic numbers
({\it Z} = 31 $\sim$ 92). Such a study is possible for very metal-poor stars
with overabundances of neutron-capture elements ([X/Fe] $\ga$ +0.5). In such
stars the absorption lines of neutron-capture elements are relatively strong,
and the blending from lines of other lighter elements is comparatively weak, due
to the overall low metallicity of the star.  

We examine the abundance pattern of neutron-capture elements in detail for the
seven very metal-poor stars in which the Th absorption line (4019 {\AA}) was
detected (see Figures~\ref{fig:th1} and \ref{fig:th2}). All of these stars
exhibit enhancements of their neutron-capture elements. Among them,
CS~31082--001 and CS~22892--052, which have already been studied by Hill et al.
(2002) and Sneden et al. (2003), are the most extreme cases, showing large
enhancements of the neutron-capture elements (e.g., [Eu/Fe]=+1.7 and +1.5,
respectively). The star HD~110184 does not exhibit a remarkable enhancement of
neutron-capture elements relative to iron, but our high S/N spectrum makes it
possible to study Th and other r-process elements in this star as well.

\subsubsection{Abundance Pattern for Elements with 56 $\leq$ {\it Z} $\leq$ 70}

Abundance studies of r-process-enhanced, metal-poor stars to date have shown
that the abundance patterns of the neutron-capture elements with 56 $\leq$ {\it
Z} $\leq$ 70 agree very well with that of the solar-system r-process pattern (e.g.,
Sneden et al. 2003). In order to investigate this phenomenon further, using our
extended sample, we now compare the scaled abundance patterns of our stars with
the abundance distributions in solar-system material.

\begin{figure}[p]
\includegraphics[width=6.5cm]{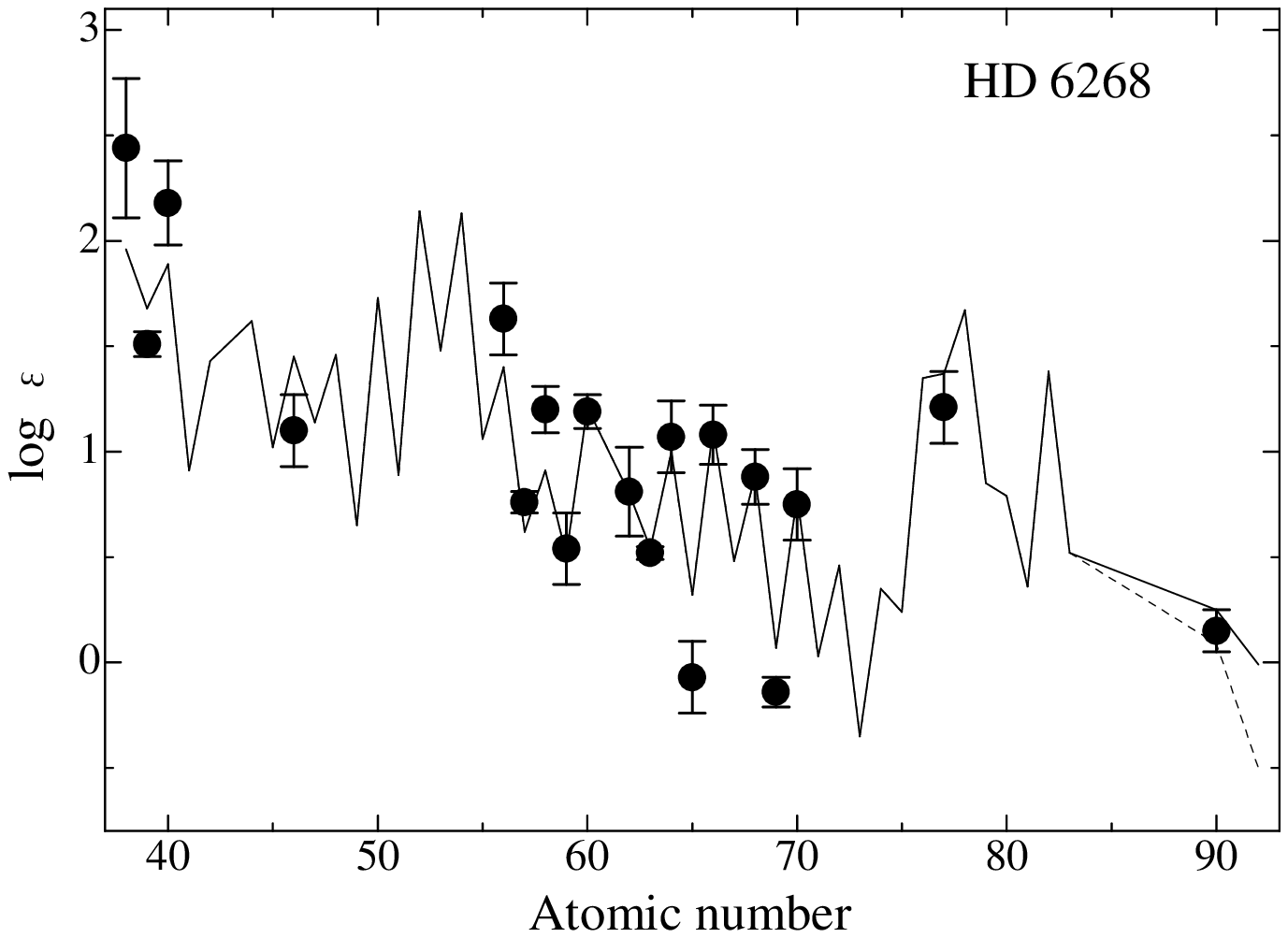}
\includegraphics[width=6.5cm]{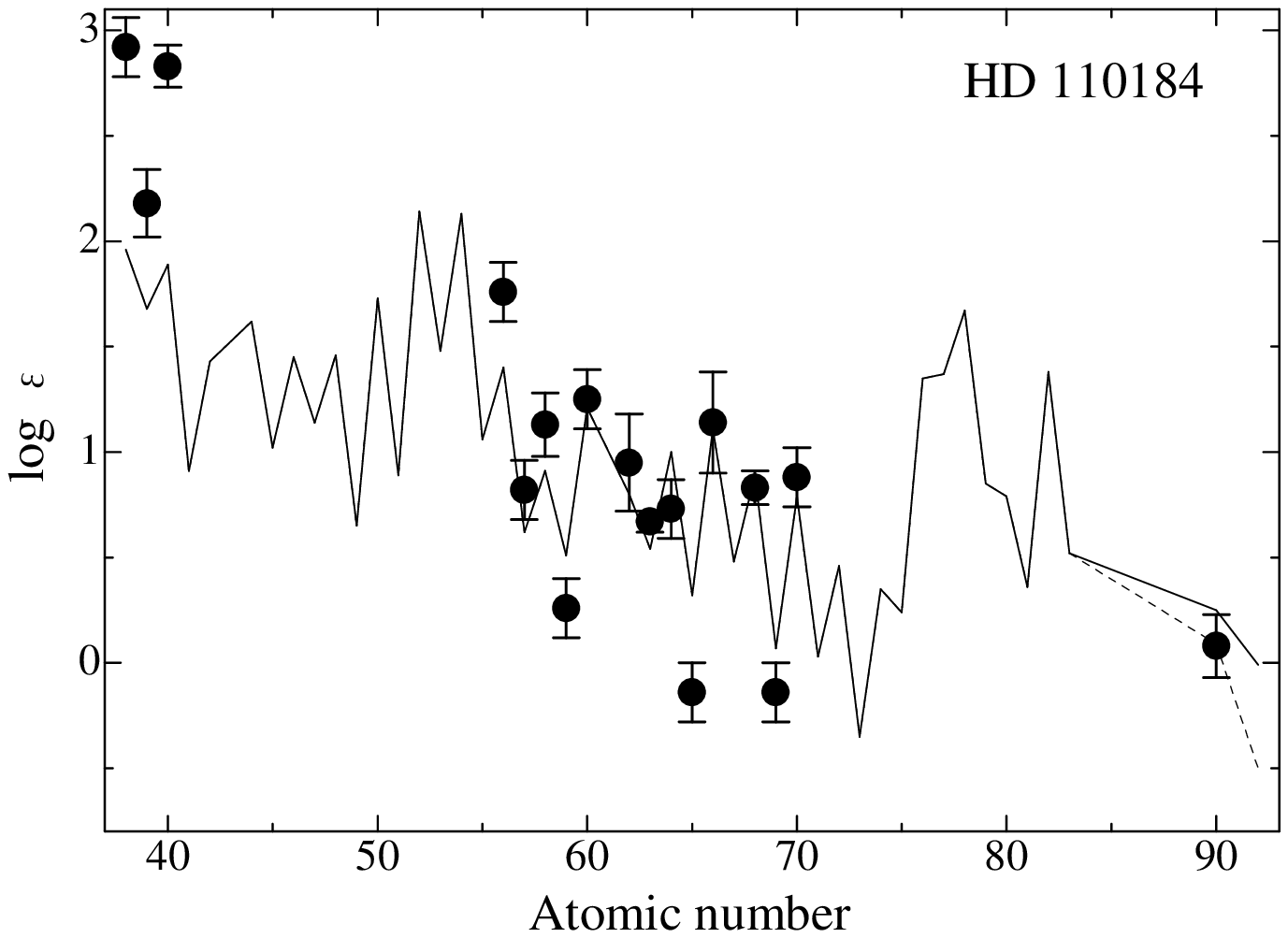}
\includegraphics[width=6.5cm]{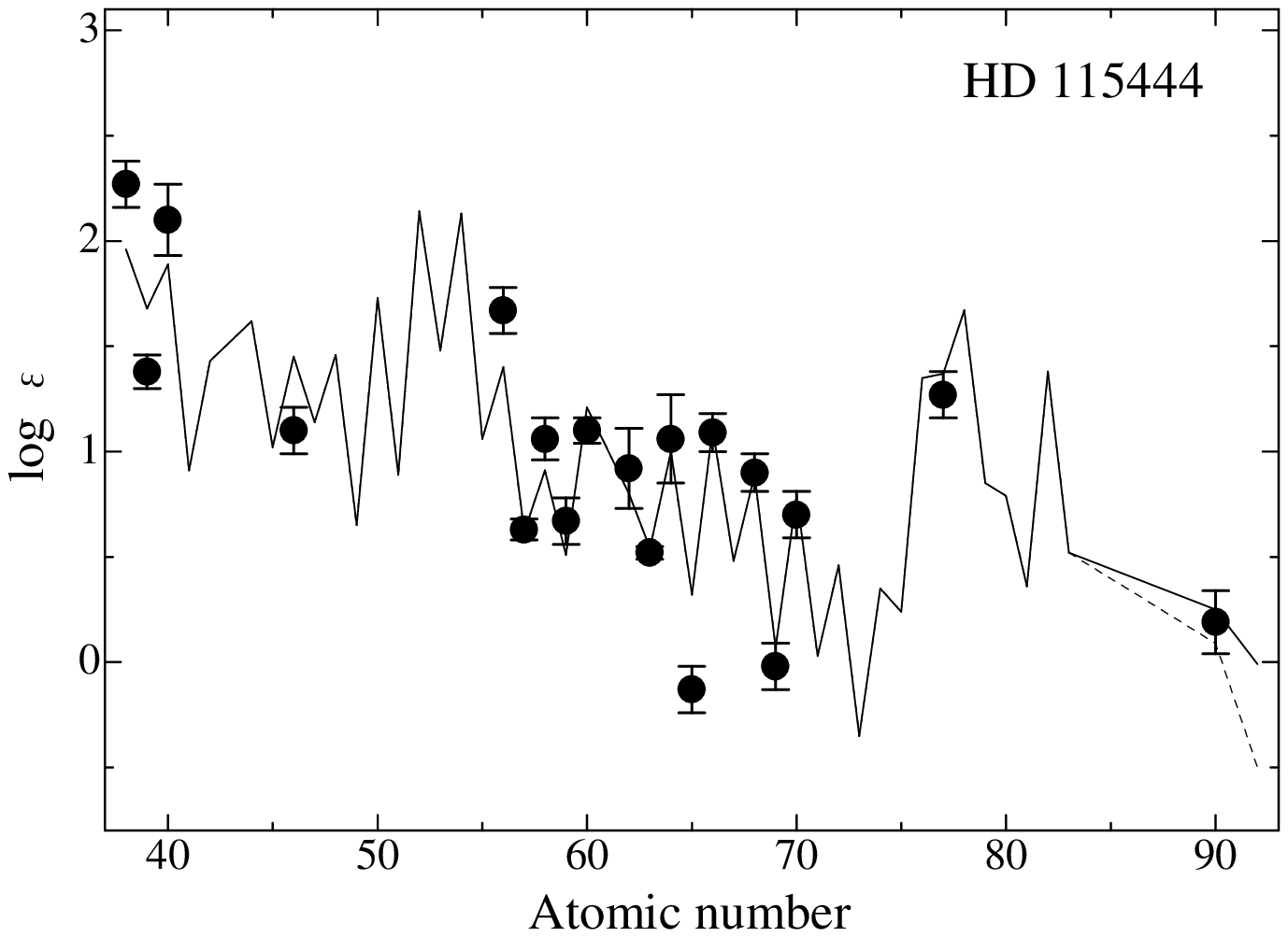}
\includegraphics[width=6.5cm]{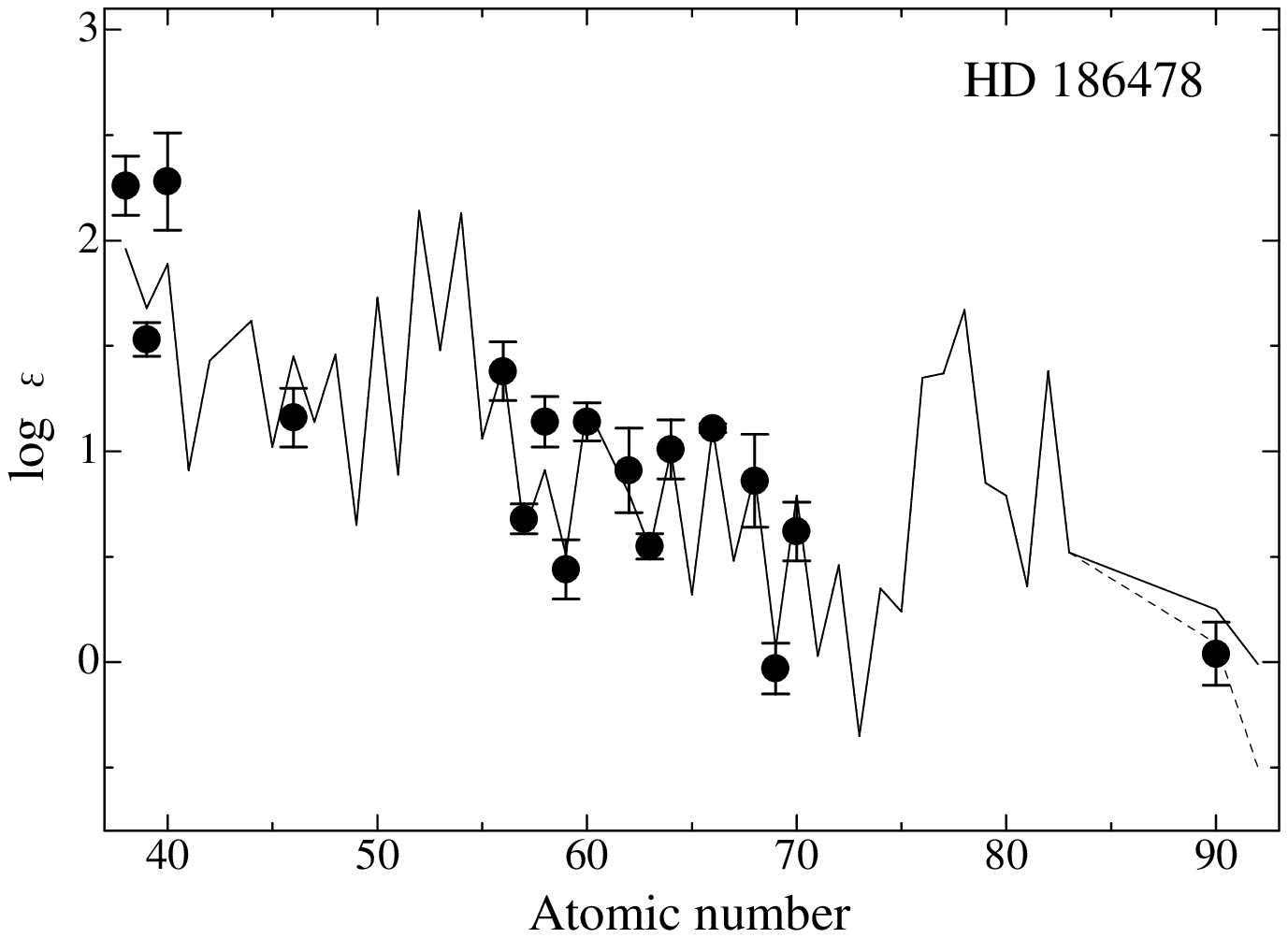}
\includegraphics[width=6.5cm]{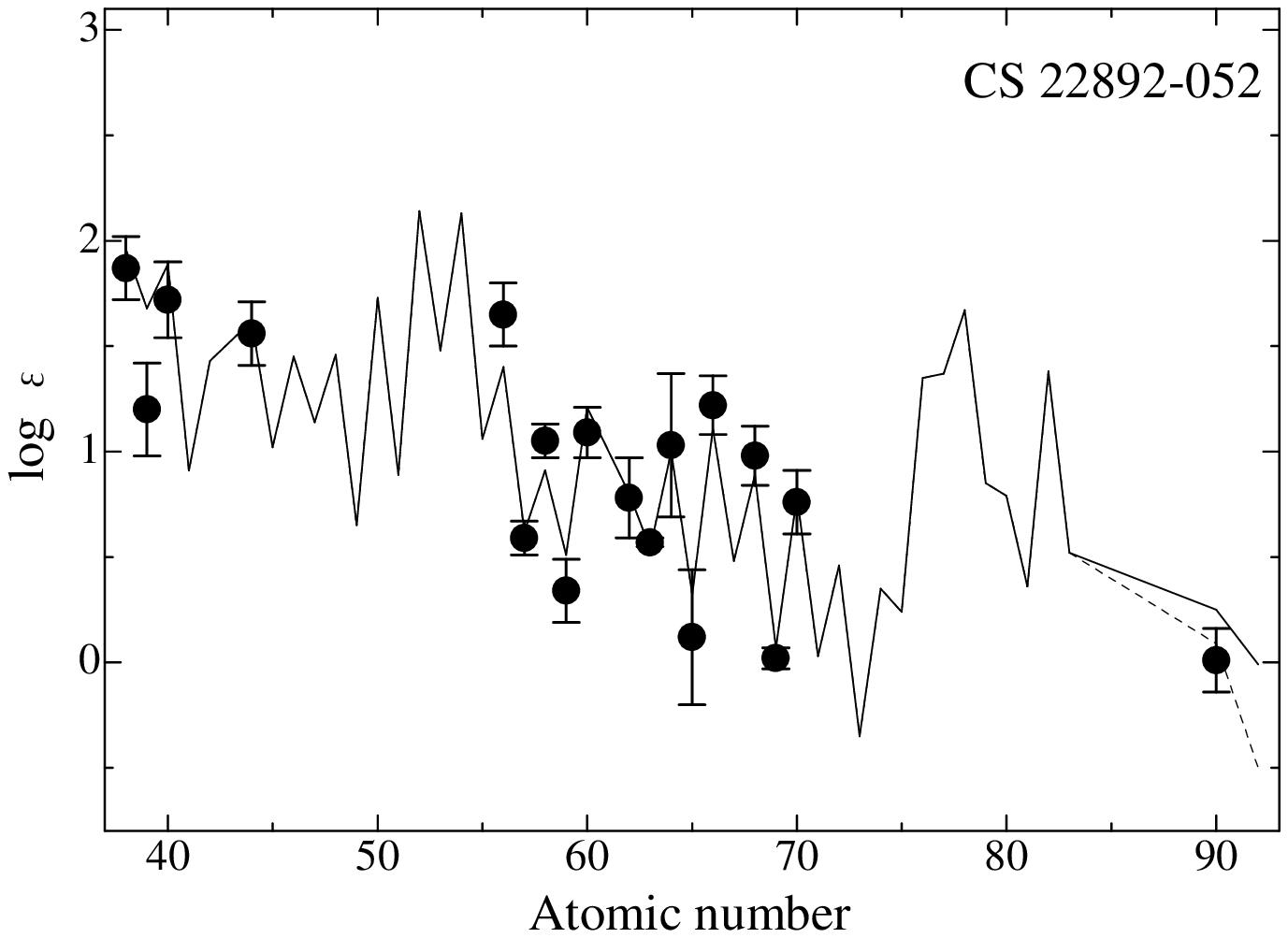}
\includegraphics[width=6.5cm]{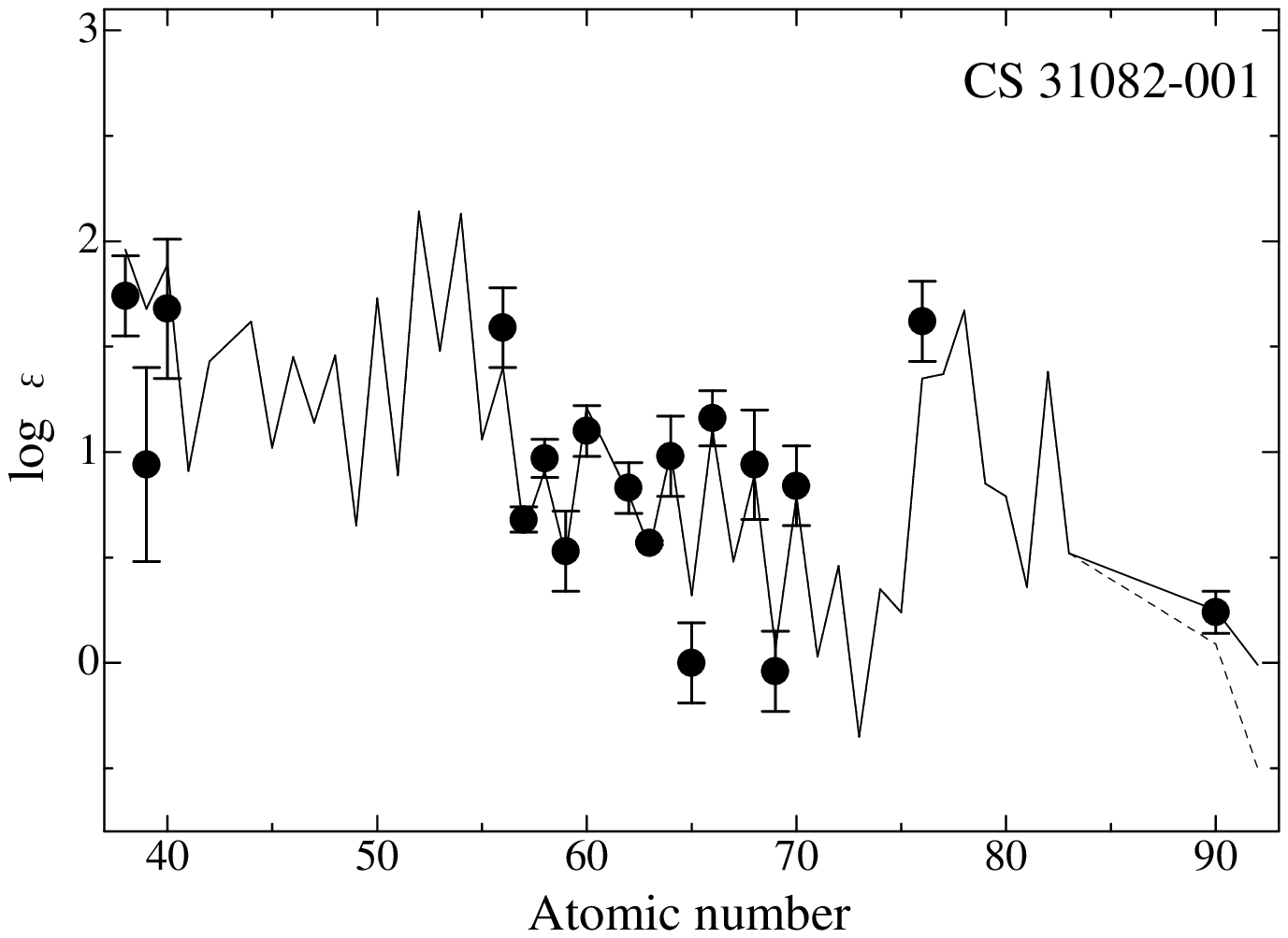}
\includegraphics[width=6.5cm]{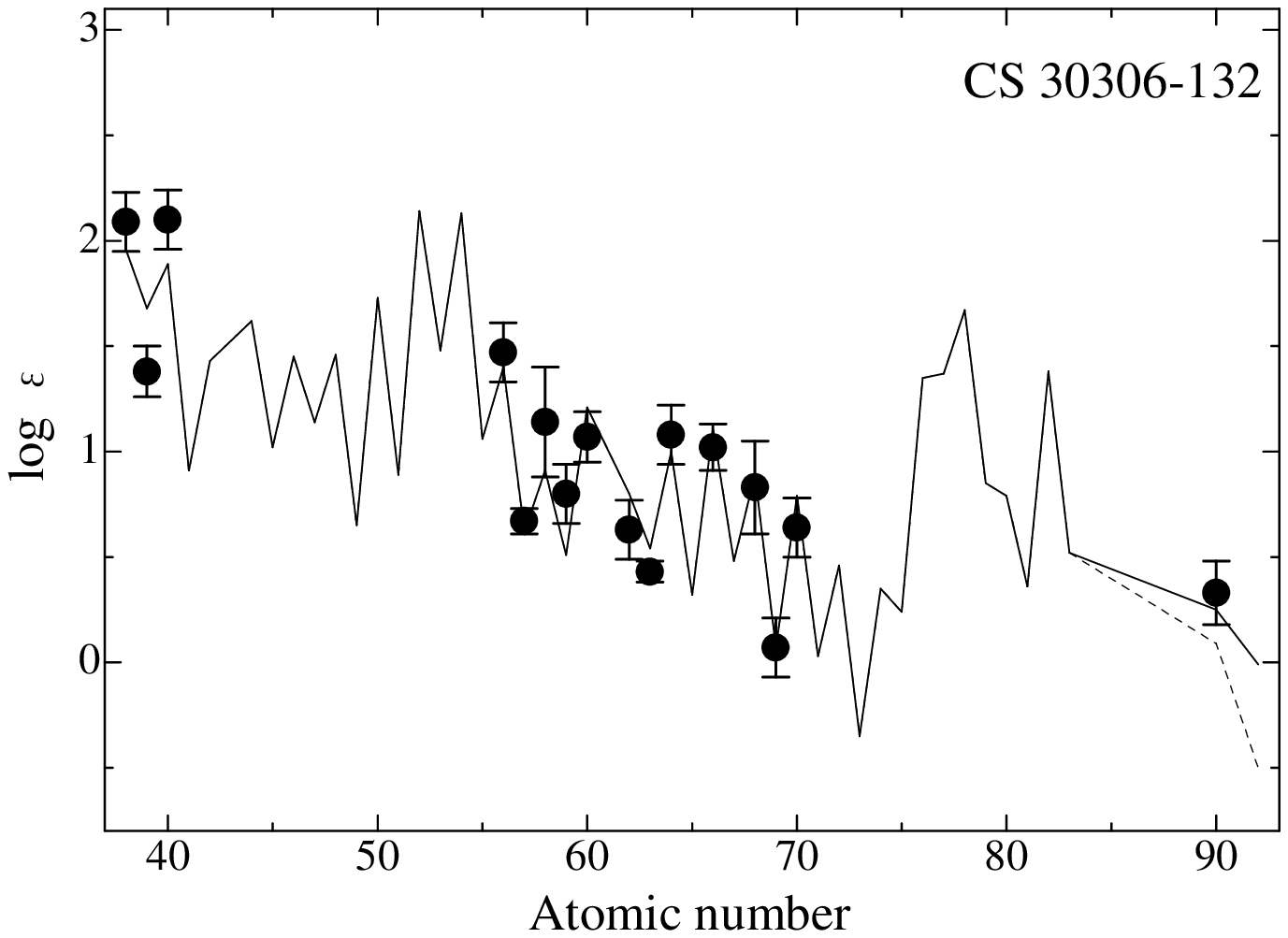}
\caption{Scaled abundance patterns for our seven objects are compared with the
 solar system r-process abundance fractions. For the Th abundance, the
 solid line indicates the initial abundance taken from Cowan et
 al. 1999, while the thin dashed line indicates the present abundance.}
\label{fig:pattern}
\end{figure}

For each of the seven stars with detectable r-process elements, we use the
solar-system r-process abundance pattern as a template to compare the
heavy-element abundances of the stars on a common scale (Figure
\ref{fig:pattern}).  We scale the elemental abundances of our objects to match the
solar-system abundances for elements between Ba and Yb ($56\leq Z\leq 70$). The
logarithmic values of the scaling factor ($\log f$) for individual stars are
listed in the columns labeled 's.s. r-process' in Table \ref{tab:scale}. For this
analysis, we adopt the r-process fraction in the Solar System given by Burris et
al. (2000). The total solar-system abundances were taken from
\citet{grevesse96}.  Figure \ref{fig:pattern} shows that the abundance patterns
of the elements with 56 $\leq Z \leq$ 70 for these very metal-poor stars agree
very well with that of the solar-system r-process component.

\begin{figure}[p]
\includegraphics[width=8cm]{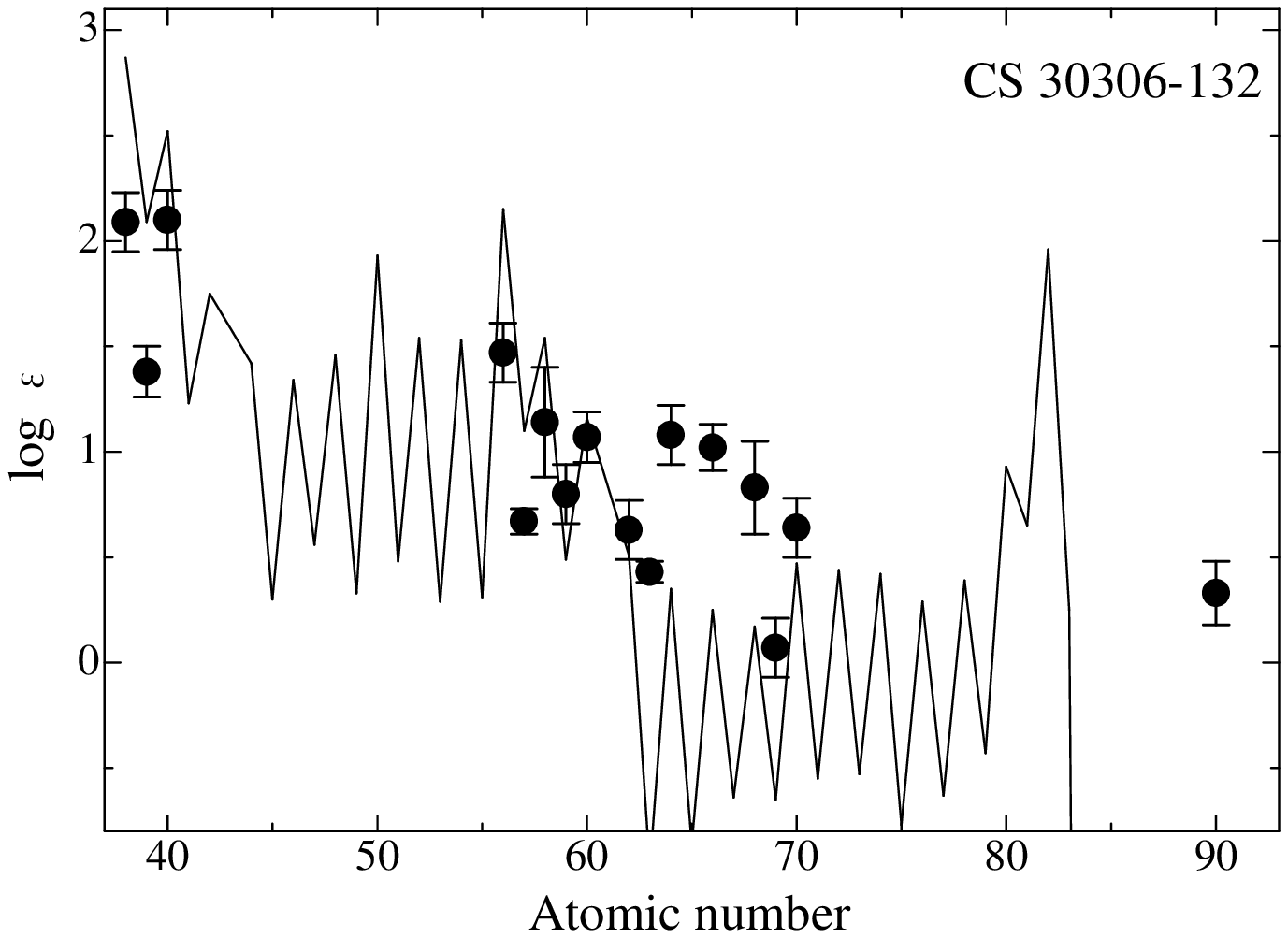}

\includegraphics[width=8cm]{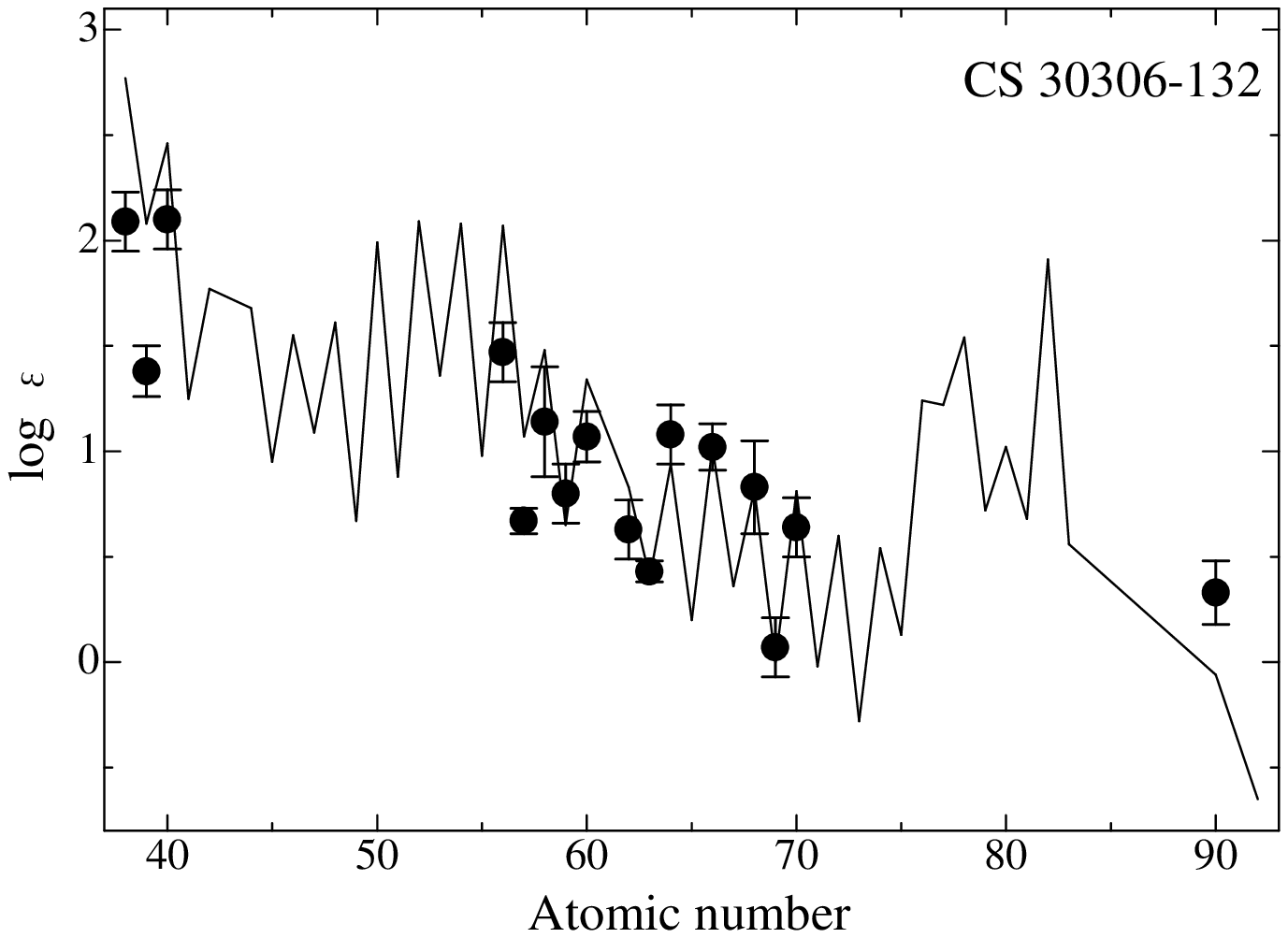}
\caption{Scaled abundance patterns for CS~30306--132 is compared with the
 scaled solar system s-process abundance fractions (upper panel) and the
 scaled solar system abundance (lower panel).}
\label{fig:pattern132}
\end{figure}

We also attempted to compare the abundances of our metal-poor stars with the
solar-system s-process distribution in the same manner. Take CS~30306--132, for
example (Figure~\ref{fig:pattern132}). As expected from the [Ba/Eu] ratios of
these objects, the agreement between the abundance patterns of our objects and
the s-process pattern is poor. We also compare the abundances of our subset of
metal-poor stars with r-process enhancements with the total elemental abundance
distribution of the Solar System (Figure \ref{fig:pattern132}). The logarithmic
values of the scale factor are given in the column labeled 's.s. total' in Table
\ref{tab:scale}. The agreement is clearly better than that in the case for the
abundance pattern of the s-process component alone. This is because the
r-process fraction dominates in the total solar-system abundances of elements
with $62\lesssim Z\lesssim 70$. The scaled abundances of Ba, La, and Ce in our
stars, shown in Figures \ref{fig:pattern}, are less than the solar-system total
abundances of these elements. Presumably, this arises because the s-process
contribution to the abundances of Ba, La, and Ce in the Solar System is larger
than for other elements like Eu (Burris et al. 2000). From these comparisons, we
conclude that the abundance patterns of heavy neutron-capture elements in our
objects agree best with the r-process component in the Solar System.

In Table \ref{tab:scale} we also provide the standard deviation (1-$\sigma$) of
the difference between the scaled abundances of each star and the total
solar-system abundances, as well as its r-process component. These standard
deviations can be taken as indicators of the level of agreement between the
abundance pattern of each star and the pattern of the solar-system abundances.
The average of the observational errors for elements with $56\leq Z\leq 70$
($\sigma_{\rm obs}$) are also provided in Table 8. Comparisons of the observed
1-$\sigma$ values with $\sigma_{\rm obs}$ indicates good agreement between the scaled
abundance pattern of our objects with that of the solar-system r-process
component. We note that the 1-$\sigma$ value for HD~110184 is slightly larger than
those for other objects, and than its $\sigma_{\rm obs}$. One reason for this
result may be that the r-process enhancement of this star (e.g., [Eu/Fe] = 0.06)
is small, and a small contribution by the s-process may affect the
abundance pattern of this star.

The excellent agreement between the abundance pattern of the heavy
neutron-capture elements in very metal-deficient stars and that of the
solar-system r-process component has already been reported for several very
metal-poor stars by previous studies \citep[e.g., ][]{sneden96,sneden00,
westin00,johnson01,hill02,cowan02, sneden03}, and is sometimes referred to as the
'universality' of r-process nucleosynthesis. The apparent universality is
ascribed to the fact that the predicted abundance patterns of elements with
$56\leq Z\leq 70$ appear to be rather insensitive to variations in the
parameters of the current r-process models (e.g., the entropy/baryon ratio).
The nucleosynthesis paths in this mass range (i.e., $A \sim$ 150) are quite
similar among the r-process models that are used to predict the production of the
actinide nuclei, even though the abundances of the actinides show an apparent
variation in some stars \citep[e.g., ][]{wanajo02,otsuki03}.

\subsubsection{The Radioactive Element Th and the Impact on
Cosmochronometry}

As mentioned in \S1, the actinide Th is heavier than the elements at the third
abundance peak produced by the r-process (Os, Ir, Pb, etc.), and is a key
element for the understanding of this nucleosynthesis process. Application of
the abundances of this element to cosmochronometry has also been discussed
extensively in recent studies of very metal-deficient stars.

The abundances of Th measured for the seven stars in our sample are presented in
Figure \ref{fig:pattern}. For the Th abundance, the solid line here shows the
{\it initial} abundance of this radioactive element, as estimated by Cowan et
al. (1999), rather than the {\it present} Th abundance, which is shown by
the dashed line in this figure. Since these very metal-poor stars are
believed to be born in the early Galaxy, the Th abundances of our sample are
expected to be lower than the value shown by the dashed line, if we assume
that the abundance patterns of heavy neutron-capture elements, including Th,
produced by the r-process, is indeed universal. This is in fact found for the stars
HD~110184, HD~115444, HD~186478, and CS~22892--052. However, the Th abundances
of the other three stars are {\it higher} than would be expected from this
logic. 

\begin{figure}
\begin{center}
\includegraphics[width=12cm]{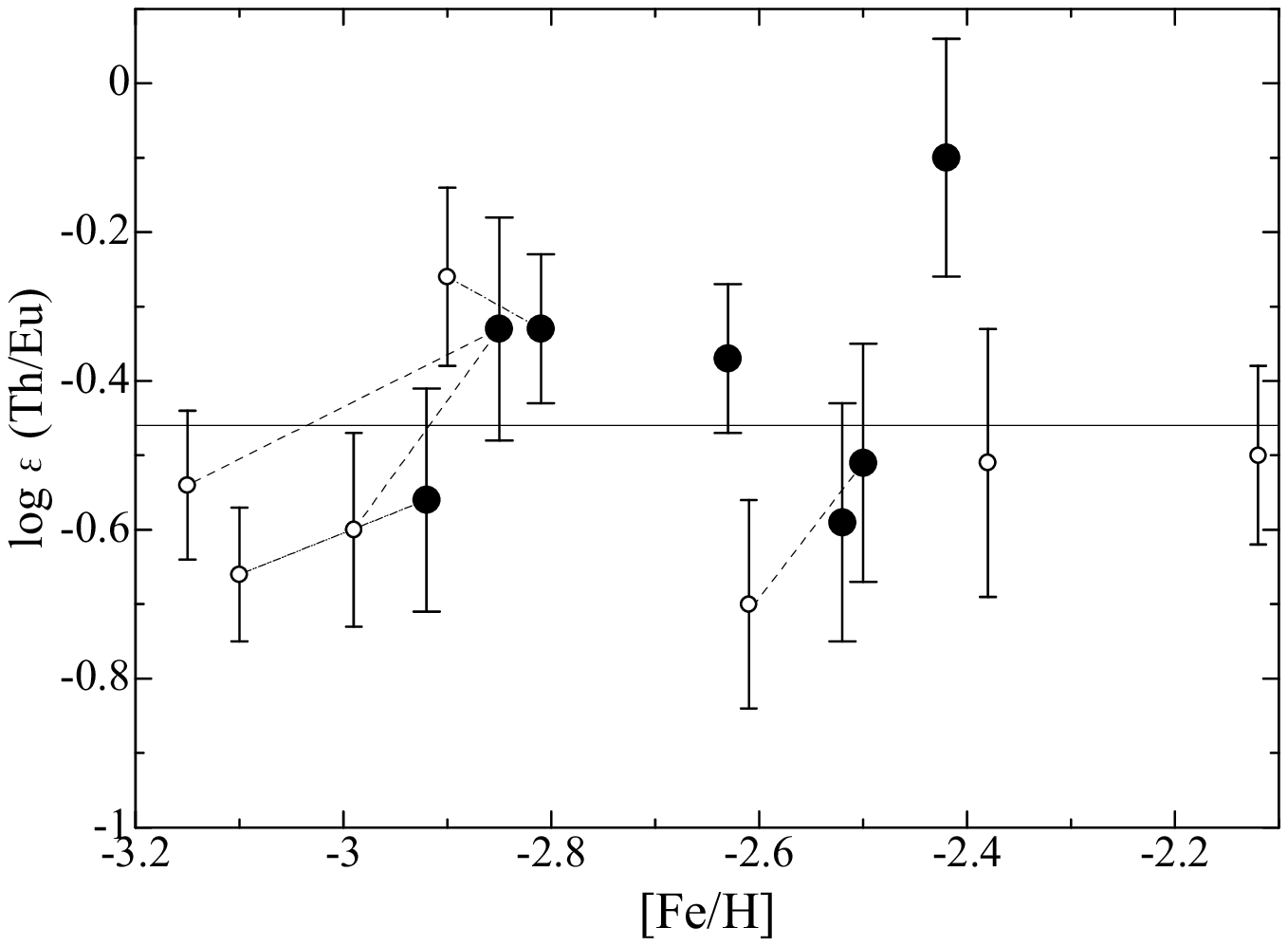}
\caption{Th/Eu ratios as a function of [Fe/H]. The filled circles are
 our sample, while open circles are the results by previous studies
\citep{sneden00,westin00,johnson01,hill02,cowan02}. The value of
solar-system ($-0.46$) is shown by the solid line.}
\label{fig:theu}
\end{center}
\end{figure}

In order to investigate this issue more clearly, we show the abundance ratios
between Th and the stable r-process element Eu ($\log$(Th/Eu) = $\log
\epsilon$(Th) $- \log \epsilon$(Eu)) in Figure \ref{fig:theu}, where we plot our
results, and the results of previous studies \citep{westin00,sneden00,johnson01,
hill02,cowan02}. The average and standard deviation of the values of our seven
stars are $-0.40$ dex and 0.17~dex, respectively. The standard deviation is as
large as, or slightly larger than, the typical observation errors, which were
estimated from the root-sum-square of the random errors of Th and Eu abundances
(error bars in Figure \ref{fig:theu}).

We have found that CS~31082--001 and CS~30306--132 have clearly higher Th/Eu
ratios than the well studied star CS~22892--052 (Figure \ref{fig:theu}). In
Figure \ref{fig:th2}, the observed spectra around the \ion{Th}{2} line are
shown. The \ion{Nd}{2} 4018.6~{\AA} line exists blueward of the
\ion{Th}{2} line.  Since the atmospheric parameters in these giants are quite similar, the
ratio of the line strengths between \ion{Th}{2} and \ion{Nd}{2} directly
reflects the Th/Nd abundance ratio. In CS~22892--052, the \ion{Th}{2} line is as
strong as the \ion{Nd}{2} line. The ratios of the line strengths in HD~6268 and
HD~115444 are rather similar to that of CS~22892--052. In contrast, in
CS~31082--001 the \ion{Th}{2} line is significantly stronger than the
\ion{Nd}{2} line. In addition, in CS~30306--132, the \ion{Th}{2} is clearly
detected, while there is no evidence for the presence of the \ion{Nd}{2} line.
These results indicate that the Th/Nd ratios in CS~31082--001 and CS~30306--132
are significantly higher than in the other three stars. As discussed in detail
in subsection 4.2.1, Nd in these objects is regarded as a product of the
r-process, and can be taken as representative of the stable r-process elements.
Therefore, the above inspection suggests the existence of some dispersion in the
abundance ratios between Th and the other neutron-capture elements ($56
\leq Z \leq 70$).

A similar result was already reported for CS~31082--001 by \citet{cayrel01} and
\citet{hill02}, as mentioned in \S 1. Our present study shows that this object
is not unique, but that there is at least one other similar object that shares
the same ``actinide boost,'' CS~30306--132, though its enhancement factor of
r-process elements is much smaller than that of CS~31082--001. We might expect
that, in the near future, as additional r-process-enhanced metal-poor stars are
identified, such behaviors will be seen in additional stars.

If the conventional Th/Eu chronometer (Cowan et al, 1999) is simply applied to
the Th/Eu abundance ratios, the ages of a few stars, those exhibiting actinide
boosts, are estimated to be shorter than that of the present age of the Sun. In
particular, the derived ages of CS~31082--001 and CS~30306--132 are as low as
zero. Even with a level of uncertainty of as much as 5 Gyr in the age
estimation, the ages derived from the above method appear unrealistic. We
note that the average of the Th/Eu ratios of our seven stars ($<\log$(Th/Eu)
$>=-0.42$) is quite similar to the value of solar-system material ($\log$(Th/Eu)
$=-0.46$ \citep{grevesse96}). That is, the average of the ages derived from
application of the Th/Eu chronometer is similar to the age of the Sun, and hence
is also unrealistic. 

The observations seem to suggest that some very metal-poor stars
had {\it higher} initial Th abundances than expected from the solar-system
r-process abundance pattern. In other words, even though the abundance pattern
of the elements with $56\leq Z\leq 70$ agrees with the abundance pattern of the
solar system r-process component, the initial abundance ratios of the heaviest
elements, like Th, to the lighter ones (e.g., Eu) are not necessarily the same as
those expected from the r-process component in the Solar System. In order to apply
the abundance ratios between Th and other stable elements as
chronometers, estimates of the initial abundance ratios for these
elements are essential, hence a deeper understanding of the r-process
nucleosynthesis for wider mass ranges is necessary.

One possible alternative cosmochronometer is the U/Th ratio, recently applied
by Cayrel et al. (2001) and Hill et al. (2002) for the extremely
r-process-enhanced, metal-poor star CS~31082--001. Since $^{232}$Th and
$^{238}$U are neighboring actinide nuclei, their production rate is expected to
be quite similar. This justifies the assumption that the initial abundance ratio
is the same as in the initial solar-system abundance ratio, as shown by recent
theoretical calculations
\citep[e.g., ][]{wanajo02,otsuki03}. At present, U/Th is expected to be the best
available chronometer.

We would like to point out that, even though we conclude from our analysis that
there exists a real scatter in the abundance ratios between Th and other
neutron-capture elements with $Z \sim 60$, the level of this scatter is at
most a factor of three, much smaller than the ratios between light ($Z \sim 40$)
and heavy ($Z \sim 60$) neutron capture elements, which are as large as a
factor of 10 (see Figure \ref{fig:pattern}). Recent models of the r-process
showed that the abundance ratios between the elements at the second and the
third r-process peaks are quite sensitive to the parameters in the calculation
\citep[e.g.,][]{hoffman97,otsuki00,wanajo02,otsuki03}. The small
dispersion of Th/Eu ratios found in our stars, as well as in other
previously studied stars, should be an important constraint on modeling
the r-process, as we are beginning to place limits on the possible range
over within which the initial production ratio can fall.
In order to derive a clear conclusion, it is required to systematically
study the Th abundances for a larger sample of very metal-poor stars, based on
high S/N spectra, which would enable one to detect the Th line even in
stars with lower Th abundances.

\section{Summary}

We have conducted detailed abundance analyses for 22 very metal-poor stars
based on high-resolution, high S/N near-UV-blue spectra obtained with
Subaru/HDS. Our sample of stars covers the metallicity range $-3.2 <$ [Fe/H]
$< -2.4$. This paper reports the results of abundance analyses concentrating
in particular the r-process elements. The main conclusions of this
study are as follows:

(1) We have identified a new highly r-process element-enhanced, metal-poor star,
CS~22183--031, a giant with [Fe/H] = $-2.93$ and [Eu/Fe] = +1.2.  The lower S/N
of its spectrum, however, prevented Th from being clearly detected.  
We also identified a new, moderately r-process-enhanced, metal-poor
star, CS~30306--132, a giant with [Fe/H] = $-2.42$ and [Eu/Fe] = +0.85, in which
Th was detected.  

(2) We have confirmed the large star-to-star scatter in the abundances of
neutron-capture elements relative to iron observed for stars with [Fe/H] $<
-2.5$. The abundance pattern of the heavy neutron-capture elements ($56\leq
Z\lesssim 72$) in seven r-process-enhanced, metal-poor stars are, however, quite
similar to that of the r-process component in solar-system material. The Ba to
Eu ratios in 11 metal-poor stars, including these seven objects, are nearly
equal to that of the solar system r-process component. These results prove that
heavy neutron-capture elements in very metal-poor stars are primarily
synthesized by the r-process.

(3) We have investigated the correlation between Sr and Ba abundances in our
sample, and in other very metal-poor stars ([Fe/H] $<-2.5$) studied by previous
authors, and conclude that the dispersion of the Sr abundances clearly decreases
with increasing Ba abundance. This trend suggests the existence of two
nucleosynthesis processes, one of which produces Sr with very small production
of Ba, and the other which produces Sr and Ba with Sr/Ba $\gtrsim 1$. The
so-called main r-process might be associated with the latter process, while that
responsible for the former is yet unidentified.

(4) The Th/Eu abundance ratios (log(Th/Eu)) measured for the seven
r-process-enhanced stars range from $-0.10$ to $-0.59$. We have confirmed the
high Th/Eu ratio found for the extremely r-process-enhanced star CS~31082--001
\citep{hill02}; the newly discovered moderately r-process-enhanced star
CS~30306-132 exhibits a similar (high) Th/Eu ratio as CS~31082--001. Since
these very metal-poor stars are believed to be formed in the early Galaxy, this
result means that a small dispersion appears in the abundance ratios between Th
and rare-earth elements, such as Eu, in very metal-poor stars. In order to apply
the Th/Eu ratios to estimates of stellar ages, further understanding of Th
production by r-process nucleosynthesis is required.

Clearly, important information on the nature of r-process nucleosynthesis is
being obtained from abundance studies of very metal-poor stars, especially for
those with enhancements of r-process elements. Further observational studies of
these objects, based on spectra of even higher quality, will provide strong
constraints on element production via r-process nucleosynthesis in the early
Galaxy.

In this study we have discovered a new, moderately r-process-enhanced,
metal-poor star, CS~30306--132, which has a similar (high) Th/Eu ratio to
CS~31082--001. We have also identified a new highly r-process element-enhanced,
metal-poor star, CS~22183--031, a giant with [Fe/H] = $-2.93$ and [Eu/Fe] =
+1.2.

\acknowledgments

We would like to thank Yoichi Takeda for development of the SPTOOL
software.
We would like to thank Chris Sneden for providing useful atomic line lists.
We also thank an anonymous referee for many helpful comments.
This publication makes use of data products from the Two Micron All Sky Survey,
which is a joint project of the University of Massachusetts and the Infrared
Processing and Analysis Center/California Institute of Technology, funded by the
National Aeronautics and Space Administration and the National Science
Foundation.  We have also made use of the SIMBAD database, maintained by the
CDS, Strasbourg, France.  This work has supported in part by
Grants-in-Aid for Scientific Research 12047233 and 13640313 of the
Ministry of Education, Science, Sports and Culture of Japan.
T.C.B. acknowledges partial support for this work from
grants AST 00-98508 and AST 00-98549 awarded by the U.S. National
Science Foundation.

\clearpage

\begin{deluxetable}{lp{7mm}p{7mm}p{7mm}p{7mm}p{7mm}p{7mm}crrrrrr}
\tabletypesize{\small}
\tablewidth{0pt}
\tablecaption{PHOTOMETRIC DATA OF THE PROGRAM STARS \label{tab:photo}}
\startdata
\tableline
\tableline
Object & $B$ & $V$ & $V_{\rm 0}$ & $J$ & $H$ & $K$ & $B-V$ & $E(B-V)$ &
 $(B-V)_{\rm 0}$ & $V_{\rm 0}-K$ \\
\tableline
HD~4306 & 9.71 & 9.08 & 9.08 & 7.42 & 6.98 & 6.82 & 0.63 & 0.00 & 0.63 & 2.26\\
HD~6268 & 8.89 & 8.10 & 8.10 & 6.34 & 5.84 & 5.71 & 0.79 & 0.00 & 0.79 & 2.39 \\
HD~88609 & 9.52 & 8.59 & 8.59 & 6.67 & 6.13 & 6.01 & 0.93 & 0.00 & 0.93 & 2.58 \\
HD~110184 & 9.48 & 8.31 & 8.31 & 6.13 & 5.51 & 5.35 & 1.17 & 0.00 & 1.17 & 2.96 \\
HD~115444 & 9.75 & 8.97 & 8.97 & 7.16 & 6.70 & 6.61 & 0.78 & 0.00 & 0.78 & 2.36 \\
HD~122563 & 7.11 & 6.20 & 6.20 & 4.79 & 4.03 & 3.73 & 0.91 & 0.00 & 0.91 & 2.47 \\
HD~126587 & 9.88 & 9.15 & 8.84 & 7.26 & 6.78 & 6.67 & 0.73 & 0.10 & 0.63 & 2.17 \\
HD~140283 & 7.70 & 7.21 & 7.18 & 6.01 & 5.70 & 5.59 & 0.49 & 0.01 & 0.48 & 1.59 \\
HD~186478 & 10.08 & 9.18 & 8.81 & 7.12 & 6.60 & 6.44 & 0.90 & 0.12 & 0.78 & 2.37 \\
BS~16082--129 & 14.22 & 13.55 & 13.49 & 11.86 & 11.40 & 11.31 & 0.67 & 0.02 & 0.65 & 2.18 \\
BS~16085--050 & 12.89 & 12.15 & 12.09 & 10.47 & 10.01 & 9.94 & 0.74 & 0.02 & 0.72 & 2.15 \\
BS~16469--075 & 14.19 & 13.42 & 13.36 & 11.75 & 11.26 & 11.20 & 0.77 & 0.02 & 0.75 & 2.16 \\
BS~16920--017 & 14.64 & 13.88 & 13.85 & 12.17 & 11.68 & 11.59 & 0.76 & 0.01 & 0.75 & 2.26 \\
BS~16928--053 & 14.32 & 13.47 & 13.44 & 11.66 & 11.14 & 11.04 & 0.85 & 0.01 & 0.84 & 2.40 \\
BS~16929--005 & 14.23 & 13.61 & 13.58 & 12.17 & 11.75 & 11.67 & 0.62 & 0.01 & 0.61 & 1.91 \\
BS~17583--100 & 12.88 & 12.37 & 12.06 & 11.10 & 10.69 & 10.66 & 0.51 & 0.10 & 0.41 & 1.40 \\
CS~22169--035 & 13.80 & 12.88 & 12.76 & 11.00 & 10.48 & 10.35 & 0.92 & 0.04 & 0.88 & 2.40 \\
CS~22183--031 & 14.27 & 13.62 & 13.50 & 12.11 & 11.65 & 11.58 & 0.65 & 0.04 & 0.61 & 1.91 \\
CS~22892--052 & 13.96 & 13.18 & 13.18 & 11.30 & 10.85 & 10.93 & 0.78 & 0.00 & 0.78 & 2.25 \\
CS~22952--015 & 14.05 & 13.27 & 13.15 & 11.49 & 11.02 & 10.92 & 0.78 & 0.04 & 0.74 & 2.23 \\
CS~30306--132 & 13.61 & 12.81 & 12.78 & 11.52 & 11.00 & 10.75 & 0.80 & 0.01 & 0.79 & 2.03 \\
CS~31082--001 & 12.44 & 11.67 & 11.67 & 10.05 & 9.61 & 9.46 & 0.77 & 0.00 & 0.77 & 2.21 \\
\tableline
\enddata
\end{deluxetable}

\begin{deluxetable}{lccccccccc}  
\tablewidth{0pt}
\tablecaption{ATMOSPHERIC PARAMETERS \label{tab:param}}
\startdata
\tableline
\tableline
Object & $T_{\rm eff}(V-K)$ & $T_{\rm eff}(B-V)$ & $T_{\rm eff}$(adopted) & $\xi$ & $\log g$ & [Fe/H] & $\sigma$ & N \\\hline
\tableline
HD~4306  & 4814 & 5058 & 4810 & 1.6 & 1.8 &  --2.89  & 0.09 & 81 \\
HD~6268  & 4679 & 4683 & 4600 & 2.1 & 1.0 &  --2.63  & 0.11 &  82\\
HD~88609  & 4512 & 4467 & 4550 & 2.4 & 1.1 &  --3.07  & 0.20 &  70\\
HD~110184  & 4220 & 4225 & 4240 & 2.1 & 0.3 &  --2.52  & 0.11 &  47\\
HD~115444  & 4707 & 4702 & 4720 & 1.7 & 1.5 &  --2.85  & 0.15 &  76\\
HD~122563  & 4599 & 4493 & 4570 & 2.2 & 1.1 &  --2.77  & 0.19 &  84\\
HD~126587  & 4919 & 5058 & 4960 & 1.8 & 2.1 &  --2.78  & 0.12 &  80\\
HD~140283  & 5633 & 5585 & 5630 & 1.4 & 3.5 &  --2.53  & 0.08 &  78\\
HD~186478  & 4700 & 4702 & 4720 & 2.2 & 1.6 &  --2.50  & 0.12 &  78\\
BS~16082--129  & 4898 & 5002 & 4900 & 1.6 & 1.8 &  --2.86  & 0.15 &  76\\
BS~16085--050  & 4943 & 4827 & 4950 & 1.8 & 1.8 &  --2.91  & 0.10 &  82\\
BS~16469--075  & 4950 & 4762 & 4880 & 1.4 & 2.0 &  --3.03  & 0.14 &  79\\
BS~16920--017  & 4839 & 4762 & 4760 & 1.4 & 1.2 &  --3.12  & 0.23 &  69\\
BS~16928--053  & 4681 & 4596 & 4590 & 1.6 & 0.9 &  --2.91  & 0.14 &  80\\
BS~16929--005  & 5267 & 5118 & 5270 & 1.3 & 2.7 &  --3.09  & 0.15 &  63\\
BS~17583--100  & 5934 & 5902 & 5930 & 1.4 & 4.0 &  --2.42  & 0.10 &  55\\
CS~22169--035  & 4667 & 4535 & 4670 & 1.9 & 1.3 &  --2.72  & 0.17 &  35\\
CS~22183--031  & 5247 & 5118 & 5270 & 1.2 & 2.8 &  --2.93  & 0.20 &  49\\
CS~22952--015  & 4879 & 4783 & 4850 & 1.5 & 1.5 &  --2.94  & 0.26 &  57\\
CS~30306--132  & 5105 & 4683 & 5110 & 1.8 & 2.5 &  --2.42  & 0.13 &  93\\
CS~22892--052  & 4778 & 4702 & 4790 & 1.8 & 1.6 &  --2.92  & 0.14 &  82\\
CS~31082--001  & 4825 & 4721 & 4790 & 1.9 & 1.8 &  --2.81  & 0.12 &  69\\
\tableline
\enddata
\end{deluxetable}

\begin{deluxetable}
{@{}c@{}c@{\extracolsep{\fill}}c@{\extracolsep{\fill}}c@{\extracolsep{\fill}}c@{\extracolsep{\fill}}c@{\extracolsep{\fill}}c@{\extracolsep{\fill}}c@{\extracolsep{\fill}}c@{\extracolsep{\fill}}c@{\extracolsep{\fill}}c@{\extracolsep{\fill}}c@{\extracolsep{\fill}}c@{\extracolsep{\fill}}c@{\extracolsep{\fill}}c@{\extracolsep{\fill}}c@{\extracolsep{\fill}}c@{\extracolsep{\fill}}c@{\extracolsep{\fill}}c@{\extracolsep{\fill}}c}
\tabletypesize{\footnotesize}
\tablewidth{0pt}
\tablecaption{RESULTS \label{tab:res1}}
\startdata
\tableline
\tableline
\multicolumn{1}{l}{} & \multicolumn{1}{l}{} & \multicolumn{4}{c}{HD~4306} & \multicolumn{4}{c}{HD~6268} & \multicolumn{4}{c}{HD~88609} & \multicolumn{4}{c}{HD~110184} \\
\tableline
& &  \multicolumn{1}{c}{log$\epsilon$}  & \multicolumn{1}{c}{[X/Fe]} & \multicolumn{1}{c}{$\sigma$} & \multicolumn{1}{c}{n} & \multicolumn{1}{c}{log$\epsilon$}& \multicolumn{1}{c}{[X/Fe]} & \multicolumn{1}{c}{$\sigma$} & \multicolumn{1}{c}{n} & \multicolumn{1}{c}{log$\epsilon$}& \multicolumn{1}{c}{[X/Fe]} & \multicolumn{1}{c}{$\sigma$} & \multicolumn{1}{c}{n} & \multicolumn{1}{c}{log$\epsilon$}& \multicolumn{1}{c}{[X/Fe]} & \multicolumn{1}{c}{$\sigma$} & \multicolumn{1}{c}{n} \\\tableline
  &  Fe/H  & 4.62 &  --2.89  & 0.09 & 81 & 4.88 &  --2.63  & 0.11 & 82 & 4.44 &  --3.07  & 0.2 & 70 & 4.99 &  --2.52  & 0.11 & 47 \\
  &  $^{12}$C/$^{13}$C  &  &  $>$20  &    &    &  & 4 &    &    &  &  $>$3  &    &   &  & 5 &   &   \\
6 &  C  & 5.78 & 0.11 &    &    & 5.26 &  --0.67  &    &    & 4.98 &  --0.51  &    &    & 5.37 &  --0.67  &    &    \\
12 &  Mg   & 5.25 & 0.56 & 0.17 & 4 & 5.44 & 0.49 & 0.25 & 4 & 4.94 & 0.43 & 0.28 & 4 & 5.25 & 0.19 & 0.16 & 2 \\
13 &  Al   & 3.02 &  --0.56  & 0.09 & 1 & 3.27 &  --0.57  & 0.17 & 1 & 2.65 &  --0.75  & 0.17 & 1 & 3.11 &  --0.84  & 0.14 & 1 \\
14 &  Si   & 5.26 & 0.60 & 0.09 & 1 & 5.46 & 0.54 & 0.17 & 1 & 5.06 & 0.58 & 0.17 & 1 & 5.52 & 0.49 & 0.14 & 1 \\
20 &  Ca   & 3.97 & 0.50 & 0.01 & 3 & 4.09 & 0.36 & 0.11 & 3 & 3.66 & 0.37 & 0.08 & 4 & 4.13 & 0.29 & 0.08 & 2 \\
21 &  Sc   & 0.45 & 0.17 & 0.03 & 3 & 0.51 &  --0.03  & 0.03 & 3 & 0.14 & 0.04 & 0.07 & 3 & 0.59 &  --0.06  & 0.08 & 2 \\
22 &  Ti   & 2.55 & 0.42 & 0.09 & 21 & 2.62 & 0.23 & 0.10 & 12 & 2.22 & 0.27 & 0.12 & 17 & 2.58 & 0.08 & 0.12 & 21 \\
23 &  V    & 1.24 & 0.13 & 0.17 & 3 & 1.42 & 0.05 & 0.22 & 3 & 1.34 & 0.41 & 0.17 & 2 &  &    &    &    \\
24 &  Cr   & 2.97 & 0.19 & 0.09 & 2 & 3.20 & 0.16 & 0.17 & 2 & 2.73 & 0.13 & 0.17 & 2 & 3.03 &  --0.12  & 0.14 & 1 \\
25 &  Mn   & 2.08 &  --0.42  & 0.13 & 6 & 2.49 &  --0.27  & 0.17 & 6 & 1.91 &  --0.41  & 0.25 & 6 & 2.67 &  --0.20  & 0.19 & 4 \\
27 &  Co   & 2.32 & 0.29 & 0.09 & 2 & 2.48 & 0.19 & 0.17 & 2 & 1.77 &  --0.08  & 0.17 & 2 & 2.28 &  --0.12  & 0.14 & 2 \\
28 &  Ni   & 3.33 &  --0.03  & 0.09 & 2 & 3.40 &  --0.22  & 0.17 & 2 & 2.75 &  --0.43  & 0.17 & 2 & 3.54 &  --0.19  & 0.14 & 2 \\
29 &  Cu   &  &    &    &    &  &    &    &    & 0.39 &  --0.75  & 0.17 & 1 & 1.02 &  --0.67  & 0.14 & 1 \\
38 &  Sr   &  --0.08 &  --0.11  & 0.09 & 1 & 0.36 & 0.07 & 0.33 & 3 &  --0.35 &  --0.20  & 0.17 & 2 & 0.34 &  --0.06  & 0.14 & 2 \\
39 &  Y    &  --0.99 &  --0.33  & 0.18 & 6 &  --0.57 &  --0.17  & 0.06 & 8 &  --0.89 &  --0.05  & 0.17 & 5 &  --0.40 &  --0.11  & 0.16 & 5 \\
40 &  Zr   &  --0.22 & 0.06 & 0.09 & 2 & 0.10 & 0.12 & 0.20 & 4 &  --0.15 & 0.31 & 0.17 & 3 & 0.25 & 0.16 & 0.10 & 3 \\
45 &  Ru   &  &    &    &    &  &    &    &    &  &    &    &    &  &    &    &    \\
46 &  Pd   &  &    &    &    &  --0.98 &  --0.05  & 0.17 & 1 &  &    &    &    &  &    &    &    \\
56 &  Ba   &  --1.84 &  --1.17  & 0.09 & 2 &  --0.45 &  --0.04  & 0.17 & 2 &  --1.90 &  --1.05  & 0.17 & 2 &  --0.82 &  --0.52  & 0.14 & 2 \\
57 &  La   &  --2.55 &  --0.88  & 0.09 & 1 &  --1.32 & 0.09 & 0.05 & 5 &  &    &    &    &  --1.76 &  --0.46  & 0.14 & 5 \\
58 &  Ce   &  &    &    &    &  --0.88 & 0.12 & 0.11 & 8 &  &    &    &    &  --1.45 &  --0.56  & 0.14 & 7 \\
59 &  Pr   &  &    &    &    &  --1.54 & 0.29 & 0.17 & 2 &  &    &    &    &  --2.32 &  --0.60  & 0.14 & 1 \\
60 &  Nd   &  &    &    &    &  --0.89 & 0.25 & 0.08 & 8 &  &    &    &    &  --1.33 &  --0.30 & 0.14 & 7 \\
62 &  Sm   &  &    &    &    &  --1.27 & 0.38 & 0.21 & 3 &  &    &    &    &  --1.63 &  --0.09  & 0.23 & 4 \\
63 &  Eu   &  --2.96 &  --0.62  & 0.09 & 2 &  --1.56 & 0.52 & 0.03 & 3 &  &    &    &    &  --1.91 & 0.06 & 0.05 & 3 \\
64 &  Gd   &  &    &    &    &  --1.01 & 0.53 & 0.17 & 2 &  &    &    &    &  --1.85 &  --0.42  & 0.14 & 1 \\
65 &  Tb   &  &    &    &    &  --2.15 & 0.13 & 0.17 & 1 &  &    &    &    &  --2.72 &  --0.55 & 0.14 & 2 \\
66 &  Dy   &  &    &    &    &  --1.00 & 0.46 & 0.14 & 5 &  &    &    &    &  --1.44 &  --0.09  & 0.24 & 4 \\
68 &  Er   &  &    &    &    &  --1.20 & 0.46 & 0.13 & 3 &  &    &    &    &  --1.75 &  --0.20  & 0.08 & 3 \\
69 &  Tm   &  &    &    &    &  --2.22 & 0.26 & 0.07 & 3 &  &    &    &    &  --2.72 &  --0.35  & 0.14 & 2 \\
70 &  Yb   &  &    &    &    &  --1.33 & 0.34 & 0.17 & 1 &  &    &    &    &  --1.70 &  --0.14 & 0.14 & 1 \\
76 &  Os   &  &    &    &    &  &    &    &    &  &    &    &    &  &    &    &    \\
77 &  Ir   &  &    &    &    &  --0.87 & 0.39 & 0.17 & 1 &  &    &    &    &  &    &    &    \\
90 &  Th   &  &    &    &    &  --1.93 & 0.61 & 0.10 & 1 &  &    &    &    &  --2.50 &  --0.07  & 0.15 & 1 \\
92 &  U   &  &    &    &    &  --2.63 &  $<$0.50  &    & 1 &  &    &    &    &  --2.52 &  $<$0.50  &    & 1 \\
\tableline
\enddata
~\\

[Fe/H] and $^{12}$C/$^{13}$C are given for the first and second lines, respectively
\end{deluxetable}

\begin{deluxetable}
{@{}c@{}c@{\extracolsep{\fill}}c@{\extracolsep{\fill}}c@{\extracolsep{\fill}}c@{\extracolsep{\fill}}c@{\extracolsep{\fill}}c@{\extracolsep{\fill}}c@{\extracolsep{\fill}}c@{\extracolsep{\fill}}c@{\extracolsep{\fill}}c@{\extracolsep{\fill}}c@{\extracolsep{\fill}}c@{\extracolsep{\fill}}c@{\extracolsep{\fill}}c@{\extracolsep{\fill}}c@{\extracolsep{\fill}}c@{\extracolsep{\fill}}c@{\extracolsep{\fill}}c@{\extracolsep{\fill}}c}
\tabletypesize{\footnotesize}
\tablewidth{0pt}
\tablecaption{RESULTS \label{tab:res2}}
\startdata
\tableline
\tableline
\multicolumn{1}{l}{} & \multicolumn{1}{l}{} & \multicolumn{4}{c}{HD~115444} & \multicolumn{4}{c}{HD~122563} & \multicolumn{4}{c}{HD~126587} & \multicolumn{4}{c}{HD~140283} \\
\tableline
& &  \multicolumn{1}{c}{log$\epsilon$}  & \multicolumn{1}{c}{[X/Fe]} & \multicolumn{1}{c}{$\sigma$} & \multicolumn{1}{c}{n} & \multicolumn{1}{c}{log$\epsilon$}& \multicolumn{1}{c}{[X/Fe]} & \multicolumn{1}{c}{$\sigma$} & \multicolumn{1}{c}{n} & \multicolumn{1}{c}{log$\epsilon$}& \multicolumn{1}{c}{[X/Fe]} & \multicolumn{1}{c}{$\sigma$} & \multicolumn{1}{c}{n} & \multicolumn{1}{c}{log$\epsilon$}& \multicolumn{1}{c}{[X/Fe]} & \multicolumn{1}{c}{$\sigma$} & \multicolumn{1}{c}{n} \\\tableline
  &  Fe/H  & 4.66 &  --2.85  & 0.15 & 76 & 4.74 &  --2.77  & 0.19 & 84 & &  --2.78  & 0.12 & 80 & 4.98  &  --2.53  & 0.08 & 78 \\
  &  $^{12}$C/$^{13}$C  &  & 7 &    &    &  & 5 &    &     &  &  $>$20  &    &    &  &  $>$20  &    &    \\
6 &  C  & 5.30 &  --0.41  &    &    & 5.38 &  --0.41  &    &     & 5.97 & 0.19 &    &    & 6.31 & 0.28 &    &    \\
12 &  Mg   & 5.12 & 0.39 & 0.03 & 4 & 5.35 & 0.54 & 0.06 & 6 & 5.22 & 0.42 & 0.1 & 6 & 5.26 & 0.21 & 0.13 & 5 \\
13 &  Al   & 3.14 &  --0.48  & 0.11 & 1 & 3.30 &  --0.40  & 0.12 & 1 & 2.98 &  --0.71  & 0.12 & 1 & 2.92 &  --1.02  & 0.09 & 1 \\
14 &  Si   & 5.05 & 0.35 & 0.11 & 1 & 5.22 & 0.44 & 0.12 & 1 & 5.33 & 0.56 & 0.12 & 1 & 5.26 & 0.24 & 0.09 & 1 \\
20 &  Ca   & 3.78 & 0.27 & 0.06 & 4 & 3.89 & 0.30 & 0.10 & 4 & 3.94 & 0.36 & 0.11 & 4 & 4.08 & 0.25 & 0.04 & 4 \\
21 &  Sc   & 0.30 &  --0.02  & 0.00 & 3 & 0.40 & 0.00 & 0.07 & 3 & 0.52 & 0.13 & 0.03 & 2 & 0.66 & 0.02 & 0.04 & 3 \\
22 &  Ti   & 2.47 & 0.30 & 0.10 & 26 & 2.45 & 0.20 & 0.12 & 27 &  & 0.41 & 0.29 & 26 & 2.72 & 0.23 & 0.08 & 22 \\
23 &  V   & 1.10 &  --0.05  & 0.15 & 4 & 1.28 & 0.05 & 0.13 & 4 & 1.45 & 0.23 & 0.12 & 2 & 1.55 & 0.07 & 0.08 & 3 \\
24 &  Cr   & 2.88 & 0.06 & 0.17 & 3 & 3.05 & 0.15 & 0.15 & 3 & 3.36 & 0.47 & 0.15 & 3 & 3.35 & 0.21 & 0.15 & 3 \\
25 &  Mn   & 2.01 &  --0.53  & 0.12 & 6 & 2.46 &  --0.16  & 0.09 & 6 & 2.21 &  --0.40  & 0.11 & 6 & 2.59 &  --0.27  & 0.10 & 6 \\
27 &  Co   & 2.26 & 0.19 & 0.12 & 3 & 2.33 & 0.18 & 0.16 & 3 & 2.44 & 0.3 & 0.13 & 3 & 2.65 & 0.26 & 0.03 & 4 \\
28 &  Ni   & 3.29 &  --0.11  & 0.11 & 2 & 3.33 &  --0.15  & 0.12 & 2 & 3.35 &  --0.12  & 0.12 & 2 & 3.76 & 0.04 & 0.09 & 2 \\
29 &  Cu   &  &    &    &    & 0.66 &  --0.78  & 0.12 & 1 &  &    &    &    &  &    &    &    \\
38 &  Sr   & 0.11 & 0.04 & 0.11 & 2 & 0.04 &  --0.11  & 0.12 & 1 & 0.21 & 0.07 & 0.12 & 1 &  --0.03 &  --0.42  & 0.09 & 1 \\
39 &  Y   &  --0.78 &  --0.16  & 0.08 & 7 &  --0.87 &  --0.33  & 0.12 & 6 &  --0.65 &  --0.10  & 0.1 & 5 &  --0.84 &  --0.54  & 0.18 & 3 \\
40 &  Zr   &  --0.06 & 0.18 & 0.17 & 4 &  --0.19 &  --0.03  & 0.11 & 3 & 0.16 & 0.33 & 0.12 & 2 &  --0.15 &  --0.23  & 0.09 & 2 \\
45 &  Ru   &  &    &    &    &  &    &    &     &  &    &    &    &  &    &    &    \\
46 &  Pd   &  --1.06 & 0.09 & 0.11 & 1 &  &    &    &     &  &    &    &    &  &    &    &    \\
56 &  Ba   &  --0.49 & 0.14 & 0.11 & 2 &  --1.76 &  --1.21  & 0.12 & 2 &  --0.66 &  --0.10  & 0.12 & 2 &  --1.37 &  --1.06  & 0.09 & 2 \\
57 &  La   &  --1.53 & 0.10 & 0.05 & 5 &  --2.44 &  --0.89  & 0.12 & 5 &  --1.59 &  --0.03  & 0.07 & 4 &  &    &    &    \\
58 &  Ce   &  --1.10 & 0.12 & 0.10 & 8 &  &    &    &     &  --1.08 & 0.07 & 0.12 & 5 &  &    &    &    \\
59 &  Pr   &  --1.49 & 0.56 & 0.11 & 2 &  &    &    &     &  --1.29 & 0.69 & 0.12 & 1 &  &    &    &    \\
60 &  Nd   &  --1.06 & 0.30 & 0.06 & 7 &  --1.95 &  --0.67 & 0.12 & 1 &  --1.07 & 0.22 & 0.19 & 6 &  &    &    &    \\
62 &  Sm   &  --1.24 & 0.63 & 0.19 & 4 &  &    &    &     &  --1.27 & 0.53 & 0.12 & 2 &  &    &    &    \\
63 &  Eu   &  --1.64 & 0.66 & 0.03 & 3 &  --2.60 &  --0.38  & 0.12 & 1 &  --1.70 & 0.53 & 0.04 & 3 &  &    &    &    \\
64 &  Gd   &  --1.10 & 0.66 & 0.21 & 3 &  --2.31 &  --0.63  & 0.12 & 1 &  --1.30 & 0.39 & 0.12 & 1 &  &    &    &    \\
65 &  Tb   &  --2.29 & 0.21 & 0.11 & 1 &  &    &    &     &  &    &    &    &  &    &    &    \\
66 &  Dy   &  --1.07 & 0.61 & 0.09 & 5 &  &    &    &     &  --1.19 & 0.42 & 0.12 & 2 &  &    &    &    \\
68 &  Er   &  --1.26 & 0.62 & 0.09 & 4 &  &    &    &     &  &    &    &    &  &    &    &    \\
69 &  Tm   &  --2.18 & 0.52 & 0.11 & 2 &  &    &    &     &  &    &    &    &  &    &    &    \\
70 &  Yb   &  --1.46 & 0.43 & 0.11 & 1 &  &    &    &     &  &    &    &    &  &    &    &    \\
76 &  Os   &  &    &    &    &  &    &    &     &  &    &    &    &  &    &    &    \\
77 &  Ir   & --0.89 & 0.59 & 0.11 & 1 &  &    &    &     &  &    &    &    &  &    &    &    \\
90 &  Th   & --1.97 & 0.79 & 0.15 & 1 &  &    &    &     &  &    &    &    &  &    &    &    \\
92 &  U   & --2.35 &  $<$1.00  &    & 1 &  &    &    &     &  &    &    &    &  &    &    &    \\
\tableline
\enddata
~\\

\end{deluxetable}

\begin{deluxetable}
{@{}c@{}c@{\extracolsep{\fill}}c@{\extracolsep{\fill}}c@{\extracolsep{\fill}}c@{\extracolsep{\fill}}c@{\extracolsep{\fill}}c@{\extracolsep{\fill}}c@{\extracolsep{\fill}}c@{\extracolsep{\fill}}c@{\extracolsep{\fill}}c@{\extracolsep{\fill}}c@{\extracolsep{\fill}}c@{\extracolsep{\fill}}c@{\extracolsep{\fill}}c@{\extracolsep{\fill}}c@{\extracolsep{\fill}}c@{\extracolsep{\fill}}c@{\extracolsep{\fill}}c@{\extracolsep{\fill}}c}
\tabletypesize{\footnotesize}
\tablewidth{0pt}
\tablecaption{RESULTS \label{tab:res3}}
\startdata
\tableline
\tableline
\multicolumn{1}{l}{} & \multicolumn{1}{l}{} & \multicolumn{4}{c}{HD~186478} & \multicolumn{4}{c}{BS~16082--129} & \multicolumn{4}{c}{BS~16085--050} & \multicolumn{4}{c}{BS~16469--075} \\
\tableline
& &  \multicolumn{1}{c}{log$\epsilon$}  & \multicolumn{1}{c}{[X/Fe]} & \multicolumn{1}{c}{$\sigma$} & \multicolumn{1}{c}{n} & \multicolumn{1}{c}{log$\epsilon$}& \multicolumn{1}{c}{[X/Fe]} & \multicolumn{1}{c}{$\sigma$} & \multicolumn{1}{c}{n} & \multicolumn{1}{c}{log$\epsilon$}& \multicolumn{1}{c}{[X/Fe]} & \multicolumn{1}{c}{$\sigma$} & \multicolumn{1}{c}{n} & \multicolumn{1}{c}{log$\epsilon$}& \multicolumn{1}{c}{[X/Fe]} & \multicolumn{1}{c}{$\sigma$} & \multicolumn{1}{c}{n} \\\tableline
  &  Fe/H  & 5.01 &  --2.50  & 0.12 & 78 & 4.65 &  --2.86  & 0.15 & 76 & 4.60 &  --2.91  & 0.10 & 82 & 4.48 &  --3.03  & 0.14 & 79 \\
  &  $^{12}$C/$^{13}$C  &  & 6 &    &    &  &  $>$7  &    &   &  &  $>$3   &   &    &  &  $>$5  &    &     \\
6 &  C  & 5.81 &  --0.25  &    &    & 5.99 & 0.29 &    &    &  &  $<$--0.52  &    &    & 5.74 & 0.21 &    &     \\
12 &  Mg   & 5.50 & 0.42 & 0.15 & 4 & 5.00 & 0.28 & 0.15 & 7 & 5.28 & 0.61 & 0.17 & 7 & 4.91 & 0.36 & 0.24 & 4 \\
13 &  Al   & 3.36 &  --0.61  & 0.14 & 1 & 2.67 &  --0.94  & 0.16 & 1 & 3.03 &  --0.53  & 0.10 & 1 & 2.67 &  --0.77  & 0.12 & 1 \\
14 &  Si   & 5.48 & 0.43 & 0.14 & 1 & 5.05 & 0.36 & 0.16 & 1 & 5.57 & 0.93 & 0.10 & 1 & 4.92 & 0.40 & 0.12 & 1 \\
20 &  Ca   & 4.25 & 0.39 & 0.13 & 4 & 3.81 & 0.31 & 0.04 & 4 & 3.83 & 0.38 & 0.03 & 4 & 3.63 & 0.30 & 0.16 & 4 \\
21 &  Sc   & 0.70 & 0.03 & 0.09 & 3 & 0.30 & --0.01 & 0.06 & 2 & 0.56 & 0.30 & 0.07 & 2 & 0.18 & 0.04 & 0.05 & 2 \\
22 &  Ti   & 2.81 & 0.29 & 0.12 & 24 & 2.40 & 0.24 & 0.12 & 22 & 2.33 & 0.22 & 0.09 & 24 & 2.22 & 0.23 & 0.10 & 21 \\
23 &  V   & 1.64 & 0.14 & 0.15 & 3 & 1.16 & 0.02 & 0.16 & 2 & 1.34 & 0.25 & 0.10 & 2 & 1.04 & 0.07 & 0.12 & 2 \\
24 &  Cr   & 3.34 & 0.17 & 0.14 & 2 & 2.81 & 0.00 & 0.16 & 3 & 2.82 & 0.06 & 0.10 & 2 &  &    &    &     \\
25 &  Mn   & 2.65 &  --0.24  & 0.11 & 6 & 2.22 &  --0.31  & 0.10 & 6 & 2.47 &  --0.01  & 0.15 & 6 & 1.83 &  --0.53  & 0.15 & 4 \\
27 &  Co   & 2.50 & 0.08 & 0.14 & 2 & 2.38 & 0.32 & 0.13 & 3 & 2.29 & 0.28 & 0.09 & 3 & 2.23 & 0.34 & 0.07 & 3 \\
28 &  Ni   & 3.42 &  --0.33  & 0.14 & 2 & 3.20 &  --0.19  & 0.16 & 2 & 3.54 & 0.20 & 0.10 & 2 & 3.24 & 0.02 & 0.12 & 2 \\
29 &  Cu   & 1.03 &  --0.68  & 0.14 & 1 &  &    &    &    &  &    &    &    &  &    &    &     \\
38 &  Sr   & 0.37 &  --0.05  & 0.14 & 2 &  --0.67 &  --0.73  & 0.16 & 1 &  --1.70 &  --1.71  & 0.10 & 1 & 0.12 & 0.23 & 0.12 & 1 \\
39 &  Y   &  --0.36 &  --0.09  & 0.08 & 5 &  --1.19 &  --0.56  & 0.25 & 4 &  &    &    &    & --0.88 &  --0.08  & 0.00 & 4 \\
40 &  Zr   & 0.39 & 0.28 & 0.23 & 3 &  --0.26 &  --0.01  & 0.36 & 4 &  &  &  &    &  --0.12 & 0.30 & 0.12 & 2 \\
45 &  Ru   &  &    &    &    &  &    &    &    &  &    &    &    &  &    &    &     \\
46 &  Pd   &  --0.73 & 0.07 & 0.14 & 1 &  &    &    &    &  &    &    &    &  &    &    &     \\
56 &  Ba   &  --0.51 &  --0.23  & 0.14 & 2 &  --1.61 &  --0.97  & 0.16 & 2 &  &  --1.56  & 0.10 & 2 & --1.93 &  --1.12  & 0.12 & 2 \\
57 &  La   &  --1.21 & 0.07 & 0.06 & 5 &  &    &    &    &  &    &    &    &  &    &    &     \\
58 &  Ce   &  --0.75 & 0.12 & 0.12 & 8 &  &    &    &    &  &    &    &    &  &    &    &     \\
59 &  Pr   &  --1.45 & 0.25 & 0.14 & 1 &  &    &    &    &  &    &    &    &  &    &    &     \\
60 &  Nd   &  --0.75 & 0.26 & 0.09 & 6 &  &    &    &    &  &    &    &    &  &    &    &     \\
62 &  Sm   &  --0.98 & 0.54 & 0.20 & 4 &  &    &    &    &  &    &    &    &  &    &    &     \\
63 &  Eu   &  --1.34 & 0.61 & 0.06 & 3 &  &    &    &    &  &    &    &    &  &    &    &     \\
64 &  Gd   &  --0.88 & 0.53 & 0.14 & 2 &  &    &    &    &  &    &    &    &  &    &    &     \\
65 &  Tb   &  &    &    &    &  &    &    &    &  &    &    &    &  &    &    &     \\
66 &  Dy   &  --0.78 & 0.55 & 0.02 & 4 &  &    &    &    &  &    &    &    &  &    &    &     \\
68 &  Er   &  --1.03 & 0.50 & 0.22 & 4 &  &    &    &    &  &    &    &    &  &    &    &     \\
69 &  Tm   &  --1.92 & 0.43 & 0.12 & 3 &  &    &    &    &  &    &    &    &  &    &    &     \\
70 &  Yb   &  --1.27 & 0.27 & 0.14 & 1 &  &    &    &    &  &    &    &    &  &    &    &     \\
76 &  Os   &  &    &    &    &  &    &    &    &  &    &    &    &  &    &    &     \\
77 &  Ir   &  &    &    &    &  &    &    &    &  &    &    &    &  &    &    &     \\
90 &  Th   &  --1.85 & 0.56 & 0.15 & 1 &  &    &    &    &  &    &    &    &  &    &    &     \\
92 &  U   &  --2.00 &  $<$1.00  &    & 1 &  &    &    &    &  &    &    &    &  &    &    &     \\
\tableline
\enddata
~\\

\end{deluxetable}

\begin{deluxetable}
{@{}c@{}c@{\extracolsep{\fill}}c@{\extracolsep{\fill}}c@{\extracolsep{\fill}}c@{\extracolsep{\fill}}c@{\extracolsep{\fill}}c@{\extracolsep{\fill}}c@{\extracolsep{\fill}}c@{\extracolsep{\fill}}c@{\extracolsep{\fill}}c@{\extracolsep{\fill}}c@{\extracolsep{\fill}}c@{\extracolsep{\fill}}c@{\extracolsep{\fill}}c@{\extracolsep{\fill}}c@{\extracolsep{\fill}}c@{\extracolsep{\fill}}c@{\extracolsep{\fill}}c@{\extracolsep{\fill}}c}
\tabletypesize{\footnotesize}
\tablewidth{0pt}
\tablecaption{RESULTS \label{tab:res4}}
\startdata
\tableline
\tableline
\multicolumn{1}{l}{} & \multicolumn{1}{l}{} & \multicolumn{4}{c}{BS~16920--017} & \multicolumn{4}{c}{BS~16928--053} & \multicolumn{4}{c}{BS~16929--005} & \multicolumn{4}{c}{BS~17583--100} \\
\tableline
& &  \multicolumn{1}{c}{log$\epsilon$}  & \multicolumn{1}{c}{[X/Fe]} & \multicolumn{1}{c}{$\sigma$} & \multicolumn{1}{c}{n} & \multicolumn{1}{c}{log$\epsilon$}& \multicolumn{1}{c}{[X/Fe]} & \multicolumn{1}{c}{$\sigma$} & \multicolumn{1}{c}{n} & \multicolumn{1}{c}{log$\epsilon$}& \multicolumn{1}{c}{[X/Fe]} & \multicolumn{1}{c}{$\sigma$} & \multicolumn{1}{c}{n} & \multicolumn{1}{c}{log$\epsilon$}& \multicolumn{1}{c}{[X/Fe]} & \multicolumn{1}{c}{$\sigma$} & \multicolumn{1}{c}{n} \\\tableline
  &  Fe/H  & 4.39 &  --3.12  & 0.23 & 69 & 4.60 &  --2.91  & 0.14 & 80 & 4.52 &  --3.09  & 0.15 & 63 & 5.09 &  --2.42  & 0.10 & 55 \\
  &  $^{12}$C/$^{13}$C  &  &  $>$3  &    &    &  & $>$5   &    &    &  &  $>$7  &    &    &  & $>$3   &    &    \\
6 &  C  &  &  $<$--0.07  &    &    & 5.42 & --0.23 &    &    & 6.39 & 0.92 &    &    & 6.67 & 0.53 &    &    \\
12 &  Mg   & 4.67 & 0.21 & 0.28 & 6 & 5.06 & 0.39 & 0.35 & 4 & 4.87 & 0.38 & 0.32 & 4 & 5.43 & 0.27 & 0.26 & 5 \\
13 &  Al   & 2.64 &  --0.71  & 0.19 & 1 & 2.74 &  --0.83  & 0.15 & 1 & 2.53 &  --0.85  & 0.16 & 1 & 3.13 &  --0.92  & 0.09 & 1 \\
14 &  Si   & 4.42 &  --0.01  & 0.19 & 1 & 4.93 & 0.29 & 0.15 & 1 & 4.84 & 0.38 & 0.16 & 1 & 5.45 & 0.32 & 0.09 & 1 \\
20 &  Ca   & 3.21 &  --0.03  & 0.18 & 4 & 3.76 & 0.31 & 0.07 & 4 & 3.73 & 0.46 & 0.17 & 4 & 4.22 & 0.28 & 0.07 & 6 \\
21 &  Sc   &  --0.23 &  --0.28  & 0.07 & 2 & --0.04 &  --0.30  & 0.13 & 2 &  --0.45 &  --0.53  & 0.16 & 1 & 0.92 & 0.17 & 0.07 & 2 \\
22 &  Ti   & 2.24 & 0.34 & 0.13 & 23 & 2.24 & 0.13 & 0.09 & 23 & 2.35 & 0.42 & 0.12 & 16 & 2.99 & 0.39 & 0.07 & 17 \\
23 &  V   & 1.14 & 0.26 & 0.19 & 1 & 0.93 &  --0.16  & 0.15 & 2 &  &    &    &    & 1.92 & 0.34 & 0.09 & 1 \\
24 &  Cr   & 2.53 &  --0.02  & 0.19 & 2 & 2.68 &  --0.08  & 0.15 & 2 &  &    &    &    & 3.02 &  --0.23  & 0.09 &    \\
25 &  Mn   & 2.50 & 0.23 & 0.20 & 6 & 2.24 &  --0.24  & 0.09 & 6 & 1.56 &  --0.74  & 0.16 & 2 & 2.49 &  --0.48  & 0.06 & 3 \\
27 &  Co   & 2.07 & 0.27 & 0.23 & 3 & 2.14 & 0.13 & 0.11 & 3 & 2.19 & 0.36 & 0.06 & 3 & 2.89 & 0.39 & 0.04 & 3 \\
28 &  Ni   & 3.51 & 0.38 & 0.19 & 2 & 3.09 &  --0.25  & 0.15 & 2 & 3.07 &  --0.09  & 0.16 & 2 & 3.88 & 0.05 & 0.09 & 2 \\
29 &  Cu   &  &    &    &    &  &    &    &    &  &    &    &    &  &    &    &    \\
38 &  Sr   &  --0.63 &  --0.43  & 0.19 & 1 &  --0.22 &  --0.23  & 0.15 & 1 & 0.11 & 0.28 & 0.16 & 2 & 0.66 & 0.16 & 0.09 & 1 \\
39 &  Y   &  &    &    &    &  --1.15 &  --0.47  & 0.13 & 5 &  --0.70 & 0.16 & 0.16 & 2 &  --0.02 & 0.17 & 0.12 & 3 \\
40 &  Zr   &  &    &    &    &  --0.55 &  --0.25  & 0.25 & 3 &  &    &    &    &  &    &    &    \\
45 &  Ru   &  &    &    &    &  &    &    &    &  &    &    &    &  &    &    &    \\
46 &  Pd   &  &    &    &    &  &    &    &    &  &    &    &    &  &    &    &    \\
56 &  Ba   &  --2.73 &  --1.83  & 0.19 & 2 &  --1.85 &  --1.16  & 0.15 & 2 &  --1.46 &  --0.59  & 0.16 & 2 &  --0.53 &  --0.33  & 0.09 & 2 \\
\tableline
\enddata
~\\

\end{deluxetable}

\begin{deluxetable}
{@{}c@{}c@{\extracolsep{\fill}}c@{\extracolsep{\fill}}c@{\extracolsep{\fill}}c@{\extracolsep{\fill}}c@{\extracolsep{\fill}}c@{\extracolsep{\fill}}c@{\extracolsep{\fill}}c@{\extracolsep{\fill}}c@{\extracolsep{\fill}}c@{\extracolsep{\fill}}c@{\extracolsep{\fill}}c@{\extracolsep{\fill}}c@{\extracolsep{\fill}}c@{\extracolsep{\fill}}c@{\extracolsep{\fill}}c@{\extracolsep{\fill}}c@{\extracolsep{\fill}}c@{\extracolsep{\fill}}c}
\tabletypesize{\footnotesize}
\tablewidth{0pt}
\tablecaption{RESULTS \label{tab:res5}}
\startdata
\tableline
\tableline
\multicolumn{1}{l}{} & \multicolumn{1}{l}{} & \multicolumn{4}{c}{CS~22169--035} & \multicolumn{4}{c}{CS~22183--031} & \multicolumn{4}{c}{CS~22952--015} & \multicolumn{4}{c}{CS~30306--132} \\
\tableline
& &  \multicolumn{1}{c}{log$\epsilon$}  & \multicolumn{1}{c}{[X/Fe]} & \multicolumn{1}{c}{$\sigma$} & \multicolumn{1}{c}{n} & \multicolumn{1}{c}{log$\epsilon$}& \multicolumn{1}{c}{[X/Fe]} & \multicolumn{1}{c}{$\sigma$} & \multicolumn{1}{c}{n} & \multicolumn{1}{c}{log$\epsilon$}& \multicolumn{1}{c}{[X/Fe]} & \multicolumn{1}{c}{$\sigma$} & \multicolumn{1}{c}{n} & \multicolumn{1}{c}{log$\epsilon$}& \multicolumn{1}{c}{[X/Fe]} & \multicolumn{1}{c}{$\sigma$} & \multicolumn{1}{c}{n} \\\tableline
  &  Fe/H  & 4.79 &  --2.72  & 0.17 & 35 & 4.58 &  --2.93  & 0.20 & 49 & 4.57 &  --2.94  & 0.26 &    & 5.09 &  --2.42  & 0.13 &    \\
  &  $^{12}$C/$^{13}$C  &  &  $>$3  &    &    &  & $>$3   &    &     &  &     &    &    &  &  $>$20   &    &    \\
6 &  C  & 5.59 &  --0.25  &    &    & 6.05 & 0.42 &    &     & 5.40 &  --0.22  &    &    & 6.48 & 0.34 &    &    \\
12 &  Mg   & 4.80 &  --0.06  & 0.50 & 4 & 5.07 & 0.42 & 0.32 & 5 & 4.45 &  --0.19  & 0.21 & 5 & 5.49 & 0.33 & 0.19 & 7 \\
13 &  Al   & 2.64 &  --1.11  & 0.19 & 1 & 2.98 &  --0.56  & 0.15 & 1 & 2.85 &  --0.68  & 0.19 & 1 & 3.41 &  --0.64  & 0.14 & 1 \\
14 &  Si   & 5.15 & 0.32 & 0.19 & 1 & 5.36 & 0.74 & 0.15 & 1 & 5.00 & 0.39 & 0.19 & 1 & 5.64 & 0.51 & 0.14 & 1 \\
20 &  Ca   &  &    &    &    & 3.72 & 0.29 & 0.15 & 4 & 3.45 & 0.03 & 0.16 & 2 & 4.28 & 0.34 & 0.03 & 3 \\
21 &  Sc   & 0.12 &  --0.33  & 0.05 & 2 & 0.54 & 0.30 & 0.05 & 2 &  --0.13 &  --0.36  & 0.06 & 2 & 0.87 & 0.12 & 0.14 & 2 \\
22 &  Ti   & 2.17 &  --0.13  & 0.06 & 13 & 2.47 & 0.38 & 0.03 & 16 & 1.96 &  --0.12  & 0.06 & 13 & 2.95 & 0.35 & 0.03 & 25 \\
23 &  V   &  &    &    &    & 1.75 & 0.68 & 0.15 & 1 & 1.03 &  --0.03  & 0.19 & 1 & 1.79 & 0.21 & 0.14 & 2 \\
24 &  Cr   &  &    &    &    &  &    &    &     &  &    &    &    & 3.31 & 0.06 & 0.14 & 2 \\
25 &  Mn   & 2.76 & 0.09 & 0.19 & 2 & 2.01 &  --0.45  & 0.27 & 3 & 2.11 &  --0.34  & 0.39 & 3 & 2.84 &  --0.13  & 0.23 & 6 \\
27 &  Co   & 1.98 &  --0.22  & 0.15 & 3 & 2.36 & 0.37 & 0.11 & 4 & 2.04 & 0.06 & 0.19 & 2 & 2.74 & 0.24 & 0.14 & 3 \\
28 &  Ni   & 3.12 &  --0.41  & 0.19 & 2 & 3.44 & 0.12 & 0.15 & 2 & 3.20 &  --0.11  & 0.19 & 2 & 3.57 &  --0.26  & 0.14 & 2 \\
29 &  Cu   &  &    &    &    &  &    &    &     &  &    &    &    &  &    &    &    \\
38 &  Sr   &  --0.64 &  --0.84  & 0.19 & 2 & 0.09 & 0.10 & 0.15 & 1 &  --0.73 &  --0.71  & 0.19 & 2 & 0.64 & 0.14 & 0.14 & 2 \\
39 &  Y   &  &    &    &    &  --0.49 & 0.21 & 0.15 & 2 &  &    &    &    &  --0.07 & 0.12 & 0.12 & 6 \\
40 &  Zr   &  &    &    &    &  &    &    &     &  &    &    &    & 0.65 & 0.46 & 0.14 & 4 \\
45 &  Ru   &  &    &    &    &  &    &    &     &  &    &    &    &  &    &    &    \\
46 &  Pd   &  &    &    &    &  &    &    &     &  &    &    &    &  &    &    &    \\
56 &  Ba   &  --2.21 &  --1.71  & 0.19 & 2 &  --0.33 & 0.38 & 0.15 & 2 &  --2.49 &  --1.77  & 0.19 & 2 & 0.02 & 0.22 & 0.14 & 2 \\
57 &  La   &  &    &    &    &  &    &    &     &  &    &    &    &  --0.78 & 0.42 & 0.06 & 5 \\
58 &  Ce   &  &    &    &    &  &    &    &     &  &    &    &    &  --0.31 & 0.48 & 0.26 & 5 \\
59 &  Pr   &  &    &    &    &  &    &    &     &  &    &    &    &  --0.65 & 0.97 & 0.14 & 1 \\
60 &  Nd   &  &    &    &    &  &    &    &     &  &    &    &    &  --0.38 & 0.55 & 0.12 & 8 \\
62 &  Sm   &  &    &    &    &  &    &    &     &  &    &    &    &  --0.82 & 0.62 & 0.14 & 2 \\
63 &  Eu   &  &    &    &    &  --1.22 & 1.16 & 0.08 & 3 &  &    &    &    &  --1.02 & 0.85 & 0.05 & 3 \\
64 &  Gd   &  &  &  &  &  &  &  &  &  &    &    &    &  --0.37 & 0.96 & 0.14 & 2 \\
65 &  Tb   &  &  &  &  &  &  &  &  &  &    &    &    &  &    &    &    \\
66 &  Dy   &  &  &  &  &  &  &  &  &  &    &    &    &  --0.43 & 0.82 & 0.11 & 3 \\
68 &  Er   &  &  &  &  &  &  &  &  &  &    &    &    &  --0.62 & 0.83 & 0.22 & 4 \\
69 &  Tm   &  &  &  &  &  &  &  &  &  &    &    &    &  --1.38 & 0.89 & 0.14 & 2 \\
70 &  Yb   &  &  &  &  &  &  &  &  &  &    &    &    &  --0.81 & 0.65 & 0.14 & 1 \\
76 &  Os   &  &  &  &  &  &  &  &  &  &    &    &    &  &    &    &    \\
77 &  Ir   &  &  &  &  &  &  &  &  &  &    &    &    &  &    &    &    \\
90 &  Th   &  &  &  &  &  &  &  &  &  &    &    &    &  --1.12 & 1.21 & 0.15 & 1 \\
92 &  U    &  &  &  &  &  &  &  &  &  &    &    &    &  --1.42 &  $<$1.50  &     & 1 \\
\tableline
\enddata
~\\

\end{deluxetable}

\begin{deluxetable}
{@{}c@{}c@{\extracolsep{\fill}}c@{\extracolsep{\fill}}c@{\extracolsep{\fill}}c@{\extracolsep{\fill}}c@{\extracolsep{\fill}}c@{\extracolsep{\fill}}c@{\extracolsep{\fill}}c@{\extracolsep{\fill}}c@{\extracolsep{\fill}}c@{\extracolsep{\fill}}c@{\extracolsep{\fill}}c@{\extracolsep{\fill}}c@{\extracolsep{\fill}}c@{\extracolsep{\fill}}c@{\extracolsep{\fill}}c@{\extracolsep{\fill}}c@{\extracolsep{\fill}}c@{\extracolsep{\fill}}c}
\tabletypesize{\footnotesize}
\tablewidth{0pt}
\tablecaption{RESULTS \label{tab:res6}}
\startdata
\tableline
\tableline
\multicolumn{1}{l}{} & \multicolumn{1}{l}{} & \multicolumn{4}{c}{CS~31082--001} & \multicolumn{4}{c}{CS~22892--052} \\
\tableline
& &  \multicolumn{1}{c}{log$\epsilon$}  & \multicolumn{1}{c}{[X/Fe]} & \multicolumn{1}{c}{$\sigma$} & \multicolumn{1}{c}{n} & \multicolumn{1}{c}{log$\epsilon$}& \multicolumn{1}{c}{[X/Fe]} & \multicolumn{1}{c}{$\sigma$} & \multicolumn{1}{c}{n} \\\tableline
  &  Fe/H  & 4.70 &  --2.81  & 0.12 &    & 4.59 &  --2.92  & 0.14 &     \\
  &  $^{12}$C/$^{13}$C  &  &  $>$10  &    &    &  & 20 &    &     \\
6 &  C  & 5.84 & 0.09 &    &    & 6.55 & 0.91 &    &     \\
12 &  Mg   & 5.43 & 0.66 & 0.11 & 4 & 4.93 & 0.27 & 0.17 & 7 \\
13 &  Al   & 3.05 &  --0.61  & 0.19 & 1 & 2.94 &  --0.61  & 0.15 & 1 \\
14 &  Si   & 5.36 & 0.62 & 0.19 & 1 & 4.87 & 0.24 & 0.15 & 1 \\
20 &  Ca   & 3.97 & 0.42 & 0.19 & 1 & 3.70 & 0.26 & 0.09 & 4 \\
21 &  Sc   & 0.19 &  --0.17  & 0.19 & 1 & 0.17 &  --0.08  & 0.01 & 2 \\
22 &  Ti   & 2.55 & 0.34 & 0.05 & 15 & 2.18 & 0.08 & 0.03 & 19 \\
23 &  V   & 1.47 & 0.28 & 0.19 & 2 & 1.07 &  --0.01  & 0.15 & 1 \\
24 &  Cr   &  &    &    &    &  &    &    &     \\
25 &  Mn   & 2.41 &  --0.17  & 0.24 & 5 & 2.14 &  --0.33  & 0.12 & 4 \\
27 &  Co   & 2.49 & 0.38 & 0.13 & 3 & 2.12 & 0.12 & 0.05 & 3 \\
28 &  Ni   & 3.64 & 0.20 & 0.19 & 1 & 3.05 &  --0.28  & 0.15 & 2 \\
29 &  Cu   &  &    &    &    &  &    &    &     \\
38 &  Sr   & 0.58 & 0.47 & 0.19 & 2 & 0.44 & 0.44 & 0.15 & 2 \\
39 &  Y   &  --0.22 & 0.36 & 0.46 & 6 &  --0.23 & 0.46 & 0.22 & 7 \\
40 &  Zr   & 0.52 & 0.72 & 0.33 & 4 & 0.29 & 0.60 & 0.18 & 3 \\
45 &  Ru   &  &    &    &    & 0.13 & 1.22 & 0.15 & 1 \\
46 &  Pd   &  &    &    &    &  &    &    &     \\
56 &  Ba   & 0.43 & 1.02 & 0.19 & 2 & 0.22 & 0.92 & 0.15 & 2 \\
57 &  La   &  --0.48 & 1.11 & 0.06 & 6 &  --0.84 & 0.86 & 0.08 & 5 \\
58 &  Ce   &  --0.19 & 0.99 & 0.09 & 7 &  --0.38 & 0.91 & 0.08 & 7 \\
59 &  Pr   &  --0.63 & 1.38 & 0.19 & 1 &  --1.09 & 1.03 & 0.15 & 1 \\
60 &  Nd   &  --0.06 & 1.26 & 0.12 & 9 &  --0.34 & 1.09 & 0.12 & 9 \\
62 &  Sm   &  --0.33 & 1.50 & 0.12 & 4 &  --0.65 & 1.29 & 0.19 & 4 \\
63 &  Eu   &  --0.59 & 1.67 & 0.01 & 3 &  --0.86 & 1.51 & 0.02 & 3 \\
64 &  Gd   &  --0.18 & 1.54 & 0.19 & 2 &  --0.40 & 1.43 & 0.34 & 3 \\
65 &  Tb   &  --1.16 & 1.30 & 0.19 & 2 &  --1.31 & 1.26 & 0.32 & 3 \\
66 &  Dy   & 0.00 & 1.64 & 0.13 & 5 &  --0.21 & 1.54 & 0.14 & 5 \\
68 &  Er   &  --0.22 & 1.62 & 0.26 & 4 &  --0.45 & 1.50 & 0.14 & 5 \\
69 &  Tm   &  --1.20 & 1.46 & 0.19 & 3 &  --1.41 & 1.36 & 0.05 & 4 \\
70 &  Yb   &  --0.32 & 1.54 & 0.19 & 1 &  --0.67 & 1.29 & 0.15 & 1 \\
76 &  Os   & 0.46 & 1.90 & 0.19 & 1 &  &    &    &     \\
77 &  Ir   &  &    &    &    &  &    &    &     \\
90 &  Th   &  --0.92 & 1.80 & 0.10 & 1 &  --1.42 & 1.41 & 0.15 & 1 \\
92 &  U    &  --1.96 &  $<$1.35  &    & 1 & --1.92 &  $<$1.50  &        & 1 \\
\tableline
\enddata
~\\

\end{deluxetable}

\clearpage
\begin{deluxetable}{lcccccccccccc} 
\tablewidth{0pt}
\tablecaption{ERROR ESTIMATES FOR HD~115444 \label{tab:error}}
\startdata
\tableline
\tableline
 species & \multicolumn{2}{c}{$\Delta T_{\rm eff}$} && \multicolumn{2}{c}{$\Delta \log g$} && \multicolumn{2}{c}{$\Delta$ [Fe/H]} && \multicolumn{2}{c}{$\Delta \xi$} & r.s.s. \\
\cline{2-3}  \cline{5-6}    \cline{8-9}    \cline{11-12}   
 & $+100K$ & $-100K$ && $-0.3$ & $+0.3$ && $+0.5$ & $-0.5$ && $-0.5$ & $+0.5$ & \\
\tableline
\ion{Mg}{1} & +0.12 & $-$0.13 && +0.08 & $-$0.08 && $-$0.03 & +0.01 && 0.07 & $-$0.09 & 0.17 \\
\ion{Al}{1}& +0.11 & $-$0.13 && +0.06 & $-$0.06 && $-$0.07 & +0.03 && 0.31 & $-$0.33 & 0.36 \\
\ion{Si}{1} & +0.12 & $-$0.12 && +0.02 & $-$0.01 && $-$0.03 & +0.01 && 0.12 & $-$0.09 & 0.15 \\
\ion{Ca}{1} & +0.08 & $-$0.08 && +0.02 & $-$0.01 && $-$0.02 & +0.01 && +0.11 & $-$0.08 & 0.11 \\
\ion{Sc}{2} & +0.06 & $-$0.06 && $-$0.09 & +0.10 && +0.02 & +0.00 && +0.12 & $-$0.07 & 0.13 \\
\ion{Ti}{1} & +0.13 & $-$0.14 && +0.02 & $-$0.02 && $-$0.03 & +0.01 && +0.08 & $-$0.05 & 0.14 \\
\ion{Ti}{2} & +0.05 & $-$0.05 && $-$0.08 & +0.09 && +0.00 & +0.01 && +0.26 & $-$0.19 & 0.21 \\
\ion{V}{1} & +0.14 & $-$0.14 && +0.03 & $-$0.01 && $-$0.02 & +0.00 && +0.02 & $-$0.01 & 0.14 \\
\ion{V}{2} & +0.05 & $-$0.04 && $-$0.09 & +0.10 && +0.02 & $-$0.01 && +0.03 & $-$0.02 & 0.12 \\
\ion{Cr}{1} & +0.14 & $-$0.16 && +0.04 & $-$0.04 && $-$0.07 & +0.03 && +0.44 & $-$0.31 & 0.35 \\
\ion{Cr}{2} & +0.00 & +0.01 && $-$0.11 & +0.11 && +0.01 & 0.00 && +0.01 & $-$0.01 & 0.11 \\
\ion{Mn}{1} & +0.13 & $-$0.15 && +0.03 & $-$0.03 && $-$0.05 & +0.02 && +0.23 & $-$0.17 & 0.22 \\
\ion{Fe}{1} & +0.12 & $-$0.14 && +0.05 & $-$0.04 && $-$0.04 & +0.02 && +0.17 & $-$0.16 & 0.21 \\
\ion{Co}{1} & +0.13 & $-$0.15 && +0.03 & $-$0.03 && $-$0.05 & +0.01 && +0.36 & $-$0.20 & 0.25 \\
\ion{Ni}{1} & +0.14 & $-$0.18 && +0.05 & $-$0.05 && $-$0.09 & +0.03 && +0.46 & $-$0.42 & 0.46 \\
\ion{Sr}{2} & +0.05 & $-$0.07 && $-$0.06 & +0.05 && $-$0.01 & +0.02 && +0.16 & $-$0.22 & 0.23 \\
\ion{Y}{2} & +0.07 & $-$0.07 && $-$0.09 & +0.09 && +0.02 & 0.00 && +0.15 & $-$0.08 & 0.14 \\
\ion{Zr}{2} & +0.06 & $-$0.07 && $-$0.09 & +0.09 && +0.02 & $-$0.01 && +0.05 & $-$0.03 & 0.12 \\
\ion{Ba}{2} & +0.11 & $-$0.12 && $-$0.07 & +0.06 && $-$0.03 & +0.04 && +0.46 & $-$0.37 & 0.39 \\
\ion{La}{2} & +0.08 & $-$0.08 && $-$0.09 & +0.10 && +0.02 & $-$0.01 && +0.04 & $-$0.02 & 0.13 \\
\ion{Pr}{2} & +0.08 & $-$0.08 && $-$0.09 & +0.10 && +0.03 & $-$0.01 && +0.02 & $-$0.01 & 0.13 \\
\ion{Nd}{2} & +0.08 & $-$0.08 && $-$0.09 & +0.10 && +0.03 & $-$0.01 && +0.03 & $-$0.02 & 0.13 \\
\ion{Sm}{2} & +0.09 & $-$0.08 && $-$0.09 & +0.11 && +0.03 & $-$0.01 && +0.01 & +0.00 & 0.14 \\
\ion{Eu}{2} & +0.08 & $-$0.09 && $-$0.09 & +0.09 && +0.02 & $-$0.02 && +0.01 & $-$0.01 & 0.14 \\
\ion{Gd}{2} & +0.08 & $-$0.09 && $-$0.09 & +0.10 && +0.02 & $-$0.01 && +0.05 & $-$0.03 & 0.13 \\
\ion{Dy}{2} & +0.08 & $-$0.08 && $-$0.09 & +0.10 && +0.03 & $-$0.01 && +0.02 & $-$0.01 & 0.13 \\
\ion{Th}{2} & +0.10 & $-$0.09 && $-$0.10 & +0.10 && +0.00 & $-$0.02 && +0.00 & +0.00 & 0.14 \\
\tableline
\enddata
\end{deluxetable}

\clearpage
\begin{deluxetable}{lcclrclr}
\tablewidth{0pt}
\tablecaption{LOGARITHM OF ABUNDANCE SCALE-FACTOR \label{tab:scale}}
\startdata
\tableline
\tableline
 Object name &  \multicolumn{2}{c}{s.s. r-process} & & \multicolumn{2}{c}{s.s. total} & $\sigma_{\rm obs}$ \\
\cline{2-3} \cline{5-6}  
 & $\log f$ & $\sigma$ & & $\log f$ & $\sigma$ &  \\
\tableline 
HD~6268    & 2.08 & 0.18 & & 2.34 & 0.18 & 0.13 \\
HD~110184  & 2.58 & 0.24 & & 2.84 & 0.22 & 0.15 \\
HD~115444  & 2.16 & 0.18 & & 2.42 & 0.22 & 0.11 \\
HD~186478  & 1.89 & 0.10 & & 2.17 & 0.25 & 0.12 \\
CS~22892--052 & 1.43 & 0.13 & & 1.69 & 0.24 & 0.15 \\
CS~30306--132 & 1.45 & 0.15 & & 1.73 & 0.24 & 0.14 \\
CS~31082--001 & 1.16 & 0.12 & & 1.42 & 0.23 & 0.15 \\
\tableline
\enddata
\end{deluxetable}

\clearpage

\clearpage

\clearpage

\clearpage

\clearpage


\clearpage

\clearpage

\clearpage

\end{document}